\documentclass[aps,prb,reprint,showpacs,superscriptaddress,longbibliography]{revtex4-2}
\usepackage{mathtools,amssymb,graphicx,units}

\usepackage[plainpages=false,pdfpagelabels,colorlinks=true,linkcolor=red,urlcolor=blue,citecolor=blue,pdftitle={Title},pdfauthor={},pdfdisplaydoctitle=true,pdfduplex=DuplexFlipLongEdge]{hyperref}
\usepackage{amsmath,graphicx,units}
\usepackage[table,xcdraw]{xcolor}
\usepackage{verbatim}
\usepackage[english]{babel}
\usepackage{color}
\usepackage{float}

\begin{document}

\title{Algorithmic overlaps as thermodynamic variables:
from local to cluster Monte Carlo dynamics in critical phenomena}

\author{Ian Pil\'{e}}
\affiliation{HSE University, 101000 Moscow, Russia}
\author{Youjin Deng}
\affiliation{Hefei National Laboratory, University of Science and Technology of China, Hefei 230088, China}
\author{Lev Shchur}
\affiliation{HSE University, 101000 Moscow, Russia}
\affiliation{Osipyan Institute for Solid State Physics, 142432 Chernogolovka, Russia}
\date{\today}
\begin{abstract}
We investigate the spatial overlap of successive spin configurations in Markov chain Monte Carlo simulations using the local Metropolis algorithm and the Swendsen-Wang and Wolff cluster algorithms. We examine the dynamics of these algorithms for models in different universality classes: Ising model, Potts model with three components, and four-state Potts model. The overlap of two successive Wolff clusters reflects critical behavior and can be used as an order parameter for the algorithm's dynamics. In the case of the Swendsen-Wang algorithm, similar behavior is demonstrated by the variance of the overlap of two consecutive lattice configurations, which behaves like an order parameter. Nothing similar is observed for the Metropolis algorithm, where the dynamics in the critical region are determined by the spin-flip frequency, which is equivalent to the acceptance rate. Thus, the critical behavior of the Wolff cluster overlap and of the variance of the configuration overlap in the Swendsen-Wang algorithm are naturally related to the critical behavior of geometric objects---Fortuin-Kasteleyn clusters. Interestingly, in all cases the geometric quantity---the configuration overlap or its variance---reflects the thermodynamics of the phase transition.

\end{abstract}

\maketitle

\section*{Introduction}

Monte Carlo methods based on Markov chains are among the most versatile tools in computational statistical physics: they provide access to thermodynamic averages in systems whose configuration spaces are far too large to probe exhaustively~\cite{binder1986,landau2005}. Over decades, a diverse ecosystem of Monte Carlo strategies has developed, ranging from elementary local single-spin flips to cluster updates and, more recently, to generalized-ensemble and replica-exchange techniques. This evolution has been shaped by two concurrent pressures. On the one hand, increasingly subtle physical questions---such as the nature of critical fluctuations, the role of disorder, or the structure of frustrated states---demand algorithms capable of sampling rare events and long-ranged correlations efficiently. On the other hand, the architecture of modern computing platforms has shifted dramatically: multi-core CPUs, graphics processors, and rack-scale distributed clusters now dictate that algorithms must not only be statistically sound but also exploit massive parallelism and memory locality~\cite{weigel2012,bisson2025,lin2017}.

A notable advance in this landscape was the introduction of nonlocal update schemes such as the Swendsen--Wang and Wolff algorithms~\cite{swendsen1987,wolff1989}. By identifying and flipping clusters of collectively aligned spins, these methods bypass the microscopic rearrangements that plague local Metropolis updates and thereby suppress the critical slowing down near phase transitions. The underlying ideas, rooted in Fortuin--Kasteleyn percolation~\cite{fortuin1972a, fortuin1972b}, have since been generalized to a wide range of models and even to quantum Monte Carlo variants~\cite{evertz2003}. At the same time, cluster algorithms themselves have been redesigned for high-performance computing: data layouts and cluster-construction procedures have been reformulated to exploit GPUs, communication-efficient graph decompositions, and task-parallel growth strategies~\cite{weigel2012,bisson2025,lin2017}. The result is an algorithmic toolkit in which local and cluster updates are not competitors but complementary approaches whose statistical performance and computational footprint can be tuned to the geometry and scale of the problem under study.

In parallel with algorithmic advances in local and cluster updates, there has been remarkable progress in generalized-ensemble techniques. Multicanonical sampling~\cite{berg1992}, Wang--Landau sampling~\cite{wang2001}, the broad-histogram method~\cite{oliveira1996}, and parallel tempering and replica exchange~\cite{hukushima1996,earl2005} were all devised to overcome rough energy landscapes and large free-energy barriers that challenge canonical simulations. A distinctive feature of these approaches is the elevation of algorithmic quantities---such as histogram flatness, swap probabilities, and transition statistics---to observables that guide and monitor the simulation. Instead of serving merely as diagnostic outputs, these quantities become objects that encode physical information about the sampling process itself. In broad-histogram and Wang--Landau schemes, for example, the evolution of the density-of-states estimator and its modification factors can be exploited to infer dynamical exponents of the underlying Markov chain~\cite{trebst2006,belardinelli2007}, while transfer-matrix formulations in energy space provide stringent convergence criteria for density-of-states estimation~\cite{barash2017b}.

Among generalized-ensemble methods, population annealing provides an effective approach to systems with complex free-energy landscapes~\cite{machta2010}. The method promotes population weights, resampling factors, and inter-temperature histogram overlap to central algorithmic observables that both control and diagnose equilibration, conceptually combining elements of sequential Monte Carlo and replica-based tempering. Its statistical properties, temperature-step optimization via histogram overlap, and convergence behavior have been analyzed in detail~\cite{weigel2021}, and efficient GPU implementations have been developed to leverage modern hardware~\cite{barash2017a}. Extensions to microcanonical and equilibrium microcanonical variants further broaden its applicability, particularly for first-order phase transitions and tricritical systems~\cite{rose2019,mozolenko2024a,mozolenko2024b}.

Evidence has accumulated that such intrinsic Monte Carlo observables exhibit a systematic and model-specific dependence on thermodynamic variables. In local Metropolis updates, the mean acceptance rate varies smoothly with temperature and can, in simple settings, be expressed analytically in terms of the internal energy~\cite{lev2019}. In replica-exchange simulations, the temperature dependence of exchange probabilities between adjacent replicas provides a sensitive probe of ensemble overlap~\cite{trebst2006}. These examples suggest a broader principle: algorithm-specific observables can play the role of thermodynamic variables, reflecting both equilibrium properties and dynamical constraints.

A complementary line of investigation has focused on intrinsic observables tracked along the trajectory of local single-spin-flip dynamics, which is the setting in which the Metropolis and Glauber algorithms operate. Derrida, Bray, and Godr\`eche~\cite{derrida1994persistence} introduced in this context the persistence probability $r_t$, defined as the fraction of spins that have never flipped between the beginning of the simulation and time $t$, and showed for the one-dimensional Ising and $q$-state Potts models under zero-temperature Glauber dynamics that $r_t$ decays algebraically, $r_t \sim t^{-\theta(q)}$, with a non-trivial $q$-dependent exponent ($\theta \simeq 0.37$ for $q=2$, $\theta \simeq 0.53$ for $q=3$, and $\theta \to 1$ as $q\to\infty$). Stauffer~\cite{stauffer1994persistence} extended these measurements up to five dimensions, reporting $d$-dependent exponents in one to three dimensions, and providing evidence that above four dimensions a finite fraction of spins remains permanently frozen along the trajectory of the dynamics. The notion was subsequently generalized to non-zero temperature and to critical dynamics, where the so-called global persistence exponent at $T_c$ acts as an independent dynamical critical exponent~\cite{majumdar1996persistence}. The persistence probability is a cumulative one-time quantity, since it forbids any spin flip throughout the entire interval $[0,t]$, whereas the configuration overlap $U_n$ discussed in the present work is a two-time correlator that compares only the configurations at $t$ and $t+n$. Despite this structural difference, both classes of observables share the same spirit: they are geometric, configuration-level functionals defined directly on the trajectory of the algorithm.

In this work we examine three canonical Monte Carlo update schemes---the local Metropolis algorithm, the multi-cluster Swendsen--Wang method, and the single-cluster Wolff algorithm---through the lens of intrinsic algorithmic observables. For each update we define an algorithmic overlap: a short-time correlation between configurations, or, in the Wolff case, a geometric overlap between successive update clusters. We study its mean and variance as functions of temperature $T$, energy per spin $\epsilon$, and system size $L$ for the two-dimensional Ising model, three-state Potts model and four-state Potts model. In particular, we demonstrate that fluctuations of the Wolff cluster overlap locate the phase transition and exhibit a power-law dependence at criticality. Analogous features arise for the two-step overlap in both Metropolis and Swendsen--Wang updates.

%%%%%%%%%%%%%%%%%%%%%%%%%%%%%%%%%%%%%%%%%%%%%%%%%%%%%%%%%%%%%%%%%%%
\section*{Models and Update Algorithms}

We study three prototypical lattice spin systems---the ferromagnetic Ising model ($q=2$), three-state Potts model ($q=3$) and four-state Potts model ($q = 4$) ---defined on a two-dimensional square lattice of linear size $L$ with periodic boundary conditions. Each site $i$ carries a discrete variable $s_i \in \{1, \ldots, q\}$. 

For the Ising case ($q=2$), the spin variables are conventionally written as $s_i = \pm 1$, and the Hamiltonian takes the standard form
\begin{equation}
\mathcal{H}_{\text{Ising}} = - J \sum_{\langle ij\rangle} s_i s_j ,
\label{eq:H_Ising}
\end{equation}
where the sum runs over nearest-neighbor pairs and $J>0$ denotes the ferromagnetic coupling (set to unity in what follows).

For the three-state Potts model ($q=3$), the spin variables take values $s_i \in \{1,2,3\}$, and the Hamiltonian is conventionally written as
\begin{equation}
\mathcal{H}_{\text{Potts}} = - J \sum_{\langle ij\rangle} \delta_{s_i, s_j},
\label{eq:H_Potts}
\end{equation}
with $\delta_{s_i,s_j}$ the Kronecker delta enforcing equal-state interactions between nearest neighbors.

The four-state Potts model ($q=4$) is governed by the same Hamiltonian~(\ref{eq:H_Potts}), with the spin variables now taking values $s_i \in \{1,2,3,4\}$.

The exact critical temperatures are $T_c^{(\text{Ising})} = 2/\ln(1+\sqrt{2})\approx 2.269185$~\cite{isingTc} and $T_c^{(3)} = 1/\ln(1+\sqrt{3}) \approx 0.994973$~\cite{pottsTc}; for the four-state Potts model the corresponding value is $T_c^{(4)} = 1/\ln(1+\sqrt{4}) = 1/\ln 3 \approx 0.910239$~\cite{pottsTc}.

Simulations were performed for lattice sizes $L=16$--$1024$ ($N=L^2$ spins), corresponding to system sizes up to $2^{20}$ degrees of freedom. 

\subsection*{Local Metropolis updates}

The Metropolis algorithm updates the configuration by proposing single-spin flips and accepting them with a probability that depends on the associated energy change. For a given spin configuration $\{s_i\}$ and inverse temperature $\beta = 1/T$, a Monte Carlo sweep consists of $N$ proposals; each proposal selects a site $i$ and proposes a new state $s_i'$, which for the Ising case is $-s_i$. For the $q$-state Potts model (with $q=3$ or $q=4$), the proposed new state $s_i'$ is drawn uniformly at random from the $q-1$ states different from the current $s_i$. The energy difference
\begin{equation}
\Delta E = E(\{s_i'\}) - E(\{s_i\})
\label{eq:energy difference}
\end{equation}
is computed from the local neighborhood of $i$, and the move is accepted with probability
\begin{equation}
P_{\text{acc}} = \min\left(1, e^{-\beta \Delta E}\right).
\label{eq:Metropolis_accept}
\end{equation}

We define one Monte Carlo time unit as a full sweep of $L^2$ Metropolis proposals. In our simulations we perform an initial equilibration of $N_{\text{therm}}$ sweeps, followed by $N_{\text{meas}}$ sweeps during which measurements are taken after every sweep or every few sweeps, as detailed below.

Besides conventional thermodynamic quantities such as the energy per spin $\epsilon = E/N$, for the Metropolis updates one can monitor two intrinsic algorithmic observables:
\begin{enumerate}
    \item The mean acceptance rate $\langle P_{\text{acc}}\rangle$, averaged over all trial moves in a sweep, which is known to behave as a thermodynamic function of temperature in a number of models~\cite{lev2019}.
    \item An $n$-step \emph{overlap} between configurations separated by $n$ successive sweeps, defined in Eq.~\eqref{eq:overlap_general} below with $\Delta t = n$ sweeps.
\end{enumerate}

Alongside these algorithmic observables, we compute standard thermodynamic quantities from the energy time series. The heat capacity is obtained from the fluctuation--dissipation relation,
\begin{equation}
C = \frac{N}{T^2}
\left(
\langle \epsilon^2 \rangle - \langle \epsilon \rangle^2
\right),
\label{eq:heat_capacity}
\end{equation}
and serves as a reference against which the singular behavior of the overlap fluctuations may be compared~\cite{lev2019}.

\subsection*{Swendsen--Wang Cluster updates}

The Swendsen--Wang (SW) algorithm is a multi-cluster update scheme based on the Fortuin--Kasteleyn representation~\cite{swendsen1987}. For a given configuration $\{s_i\}$, bonds are activated between like-spin neighbors with probability
\begin{equation}
p_{\text{bond}} =
\begin{cases}
1 - e^{-2\beta J}, & q = 2,\\[4pt]
1 - e^{-\beta J},  & q = 3 \text{ or } q=4,
\end{cases}
\label{eq:bond_probability}
\end{equation}
producing a set of FK clusters. Each cluster is then flipped collectively. This update satisfies detailed balance and significantly reduces critical slowing down relative to local updates.

In our implementation we perform $N_{\text{therm}}$ SW updates for equilibration, followed by $N_{\text{meas}}$ measurement steps. For each measurement we apply $n$ successive SW updates and record the $n$-step overlap defined in Eq.~\eqref{eq:overlap_general} with $\Delta t = n$ cluster updates.

\subsection*{Wolff Single-Cluster updates} 
\label{Wolff-Single-Cluster-updates}

The Wolff algorithm~\cite{wolff1989} creates and flips one FK cluster for each update. Starting from a randomly selected initial spin, neighboring spins in the same state are added to the cluster with probability $p_{\text{bond}}$, specified in Eq.~\eqref{eq:bond_probability}. After constructing the cluster at Monte Carlo time $t$, it is completely flipped for the Ising model or randomly flipped to one of the other two states for the three-state Potts model, and randomly recolored to one of the other three states for the four-state Potts model.

For Wolff updates, we introduce a geometric overlap between successive clusters, defined as follows:
\begin{equation}
U^{(\text{W})}_{n} = \frac{1}{N} |C^{(t)} \cap C^{(t+n)}|,
\label{eq:wolff_overlap}
\end{equation}
where $C^{(t)}$ is the set of spin coordinates in the Wolff cluster at time $t$, and $C^{(t+n)}$ is the set of spin coordinates in the Wolff cluster at time $t+n$. Thus, the value of $U^{(\text{W})}_{n}$ can vary from 0, if clusters $C^{(t)}$ and $C^{(t+n)}$ have no common coordinates, to 1, if clusters $C^{(t)}$ and $C^{(t+n)}$ completely coincide. Obviously, complete coincidence of clusters occurs at zero temperature with probability 1.

\subsection*{Algorithmic Overlaps as Thermodynamic Observables}

For local Metropolis updates and Swendsen--Wang cluster updates we quantify the short-time relation between configurations by the \emph{configuration overlap}

\begin{equation}
U_{n}(t) = \frac{1}{N}\sum_{i=1}^{N} \delta\!\left(s_i^{(t)}, s_i^{(t+n)}\right),
\label{eq:overlap_general}
\end{equation}
\noindent where $\delta(\cdot,\cdot)$ is the Kronecker delta and $n$ denotes the number of passes through the lattice. The same definition applies uniformly to the Ising model and to the $q=3$ and $q=4$ Potts models; only the random baseline value $1/q$ to which $U_n$ tends in the fully disordered limit changes with $q$.

 It is important to note the difference between the expression~(\ref{eq:wolff_overlap}) for the Wolff algorithm and the expression~(\ref{eq:overlap_general}). In the former case, we seek the intersection of spins in two Wolff clusters, while in the latter case it is the intersection of {\em all} spins in the lattice. These expressions reflect a significant difference between the algorithms, and the behavior of these functions, as we will see below, is also significantly different.

For our simulations we measure the mean values and variances of the corresponding $n$-step overlaps. Just as for the Swendsen--Wang case, we apply $n$ successive updates for each measurement and record the $n$-step overlap defined in Eq.~\eqref{eq:overlap_general} with $\Delta t = n$ cluster updates. Then the measured mean and variance of the overlap become
\begin{equation}
U_n=\frac{1}{N_{meas}} \sum_{t=N_{therm}+1}^{N_{therm}+N_{meas}} U_n(t),
\label{eq:mean_overlap}
\end{equation}
and 
\begin{equation}
\mathrm{Var}(U_n)=\frac{1}{N_{meas}} \sum_{t=N_{therm}+1}^{N_{therm}+N_{meas}} (U_n(t) - U_n)^2.
\label{eq:var_overlap}
\end{equation}

% {\color{red} Ian: Please add a procedure for averaging over a thermodynamic ensemble. For example, $U_n=\frac{1}{N_{meas}} \sum_{t=1}^{N_{meas}} U_n(t)$}
 \section*{Fortuin--Kasteleyn Clusters and Edwards--Sokal Measure}
\label{sec:ES-FK}

Both cluster algorithms above generate samples from the joint spin–bond distribution introduced by Edwards and Sokal~\cite{edwards1988,sokal1997},
which is essentially the algorithmic version of the Fortuin–Kasteleyn random-cluster representation.~\cite{fortuin1972a,fortuin1972b}. The Swendsen--Wang and Wolff algorithms are built on top of the FK representation, and the algorithmic overlaps we measure are observables defined on the same statistical ensemble that underlies FK percolation. This construction applies uniformly to all three models considered here ($q=2$, $q=3$, and $q=4$), and the marginal four-state case fits into the same Edwards--Sokal framework without modification of the formulas below.

On each edge $e=\langle ij\rangle$ of the lattice one can introduce an auxiliary bond variable $\omega_e\in\{0,1\}$ and define the Edwards--Sokal joint distribution
\begin{equation}
\mu_{\mathrm{ES}}(\{s_i\},\{\omega_e\})
= \frac{1}{Z_{\mathrm{ES}}}
\prod_{e=\langle ij\rangle}
\Big[(1-p)\,\delta_{\omega_e,0}
+ p\,\delta_{\omega_e,1}\,\delta_{s_i,s_j}\Big],
\label{eq:ES_measure}
\end{equation}
with $p$ as in Eq.~(\ref{eq:bond_probability}). The corresponding
partition function is
\begin{equation}
Z_{\mathrm{ES}}
= \sum_{\{s_i\}}\sum_{\{\omega_e\}}
\prod_{e=\langle ij\rangle}
\Big[(1-p)\,\delta_{\omega_e,0}
+ p\,\delta_{\omega_e,1}\,\delta_{s_i,s_j}\Big].
\label{eq:ES_Z}
\end{equation}
Marginalizing over $\{\omega_e\}$ in Eq.~(\ref{eq:ES_Z}) one can reproduce the Potts partition function; marginalizing over $\{s_i\}$ yields the random-cluster partition function
\begin{equation}
Z_{\mathrm{RC}}
= \sum_{\{\omega_e\}}
p^{|\omega|}(1-p)^{|\mathcal{E}|-|\omega|}\,
q^{\,k(\omega)},
\label{eq:Z_RC}
\end{equation}
where $|\omega|$ is the number of occupied bonds and $k(\omega)$ is the number of connected components of $\omega$.
Equations~(\ref{eq:ES_measure})--(\ref{eq:Z_RC}) explicitly show the equivalence between the Potts model and the random-cluster model: both are marginals of the same joint measure $\mu_{\mathrm{ES}}$. At $q=4$ the random-cluster weight $q^{k(\omega)}$ takes its largest value within the continuous-transition regime, and it is precisely this factor that drives the logarithmic corrections observed in the four-state Potts critical behavior.

The two conditionals of $\mu_{\mathrm{ES}}$ are the elementary moves of the cluster algorithms. Conditioned on $\{s_i\}$, each edge between aligned neighbors is independently occupied with probability $p$ and otherwise vacant; this is the FK bond-assignment rule of Eq.~(\ref{eq:bond_probability}). Conditioned on $\{\omega_e\}$, each FK cluster is assigned a uniformly random color in $\{1,\dots,q\}$. Iterating both conditionals globally yields the Swendsen--Wang update rule~\cite{swendsen1987}. If one restricts the recoloring to the single FK cluster containing a randomly chosen seed site yields the Wolff update rule~\cite{wolff1989}. Both chains have the same stationary law $\mu_{\mathrm{ES}}$ and generate joint trajectories $(s^{(t)},\omega^{(t)})_{t\ge0}$ in the enlarged configuration space.
The algorithmic overlaps of Eqs.~(\ref{eq:wolff_overlap})
and~(\ref{eq:overlap_general}) are therefore functionals of these FK trajectories. Illustrative snapshots of how these overlaps look on the lattice in the disordered, ordered, and critical regimes are collected in Appendix~\ref{app:snapshots}.

\subsection*{Algorithmic overlaps as two-time FK observables}

Along the Wolff chain, the cluster $C^{(t)}=C(\omega^{(t)},i_0^{(t)})$ is the FK cluster of $\omega^{(t)}$ rooted at a uniform seed $i_0^{(t)}$. The geometric overlap Eq.~(\ref{eq:wolff_overlap}) is the sum of indicators
\begin{multline}
U_n^{(\mathrm{W})}(t)
= \frac{1}{N}\sum_{j=1}^{N}
\mathbf{1}\!\big[\,j\in C(\omega^{(t)},i_0^{(t)})\,\big]\\
\times\,\mathbf{1}\!\big[\,j\in C(\omega^{(t+n)},i_0^{(t+n)})\,\big],
\label{eq:Uw_indicator}
\end{multline}
and its stationary expectation reads
\begin{multline}
\big\langle U_n^{(\mathrm{W})}\big\rangle
= \frac{1}{N}\sum_{j=1}^{N}
\Pr_{\mu_{\mathrm{ES}}}\!\Big[\,j\in C(\omega^{(t)},i_0^{(t)})\\
\wedge\; j\in C(\omega^{(t+n)},i_0^{(t+n)})\,\Big],
\label{eq:Uw_expectation}
\end{multline}
where the joint probability is taken along the Wolff trajectory. This identity is exact, but it does not factorize into static FK quantities. The naive factorization would replace the joint by the product of two independent FK seed-cluster probabilities. By the
Ruge--Wagner cluster-distribution scaling for the Wolff
algorithm~\cite{ruge1992}, the probability that a fixed site $j$ belongs to the seed cluster satisfies 
\begin{equation}
\Pr[j\in C]\big|_{T_c}
=\langle|C|\rangle/N\sim L^{-\beta/\nu}
\end{equation}
which would yield

\begin{equation}
\langle U_n^{(\mathrm{W})}\rangle\big|_{T_c}\sim L^{-2\beta/\nu}.
\end{equation}

For the 2D Ising, three-state Potts and four-state Potts models this would predict
$2\beta/\nu = 1/4$, $4/15$ and again $1/4$ respectively, in clear disagreement with our measured value $\psi^{(W)}\approx 0.42$ for all three models
(Figs.~\ref{fig:wolff-ising-fss} and~\ref{fig:wolff-potts-fss}). The discrepancy is structural rather than numerical: along the Wolff chain the seed $i_0^{(t+n)}$ is correlated with the previously flipped cluster $C^{(t)}$ through the intermediate spin updates, and $\omega^{(t+n)}$ is correlated with $\omega^{(t)}$ through the conditioning spin field. The Wolff overlap is therefore a genuine
two-time functional of the Edwards--Sokal measure, governed by the joint geometry of $\omega^{(t)}$ and $\omega^{(t+n)}$ along the Markov chain rather than by a product of one-time FK probabilities.
The numerical coincidence of $\psi^{(W)}$ between
universality classes --- which would be inexplicable if $\psi^{(W)}$ tracked the static $\beta/\nu$, since these differ between Ising and three-state Potts --- is then natural: it reflects the universal geometry of how two successive Wolff clusters intersect at criticality, a property of the dynamics on the random-cluster ensemble. The four-state Potts model, sharing the static value $\beta/\nu = 1/16$ with the Ising case but belonging to a different universality class with logarithmic corrections, would provide a stringent independent check of this dynamics-driven (rather than static-FK-driven) universality of $\psi^{(W)}$.

For Swendsen--Wang, the geometric content of the configuration overlap is different. Conditioned on $\omega^{(t)}$, every FK cluster $C_\alpha$ is assigned an independent uniform color in $\{1,\dots,q\}$, so the spins $s_i^{(t)}$ and $s_j^{(t)}$ are perfectly correlated when $i,j$ lie in the same FK cluster and independent otherwise. After one SW step $\omega^{(t+1)}$ is resampled from
$\mu_{\mathrm{ES}}(\,\cdot\,|\,s^{(t)})$, so it depends on
$\omega^{(t)}$ only through the spin field that survives the
recoloring. Writing $\mathrm{Var}(U_n^{(\mathrm{SW})})$ as a double sum over sites and using these conditional independences, the variance reduces to a sum of two-time FK connectivity probabilities,
\begin{align}
&\mathrm{Var}\!\big(U_n^{(\mathrm{SW})}\big)
= \frac{1}{N^2}\sum_{i,j}\Big[
\Pr\nolimits_{\!\mu_{\mathrm{ES}}}\!\big[\,i\leftrightarrow j
\nonumber\\
&\quad\text{ in }\omega^{(t)}\text{ and in }\omega^{(t+n)}\,\big]
- \tfrac{1}{q^2}
\Big] + O(1/N),
\label{eq:Usw_variance}
\end{align}
evaluated along the SW trajectory (for the derivation look at \cite{}). The connectivity event
$\{i\leftrightarrow j\text{ in both }\omega^{(t)}\text{ and }
\omega^{(t+n)}\}$ is again a two-time FK observable, whose decay is controlled by the SW transfer operator on the random-cluster space rather than by static FK scaling alone. The measured exponents
$\mu^{(SW)}=0.348(4)$ for Ising and $0.266(3)$ for $q=3$ Potts differ both from each other and from $\mu^{(W)}$, consistent with the fact that the $q$-dependence enters Eq.~(\ref{eq:Usw_variance}) both through the FK weight $q^{k(\omega)}$ in $Z_{\mathrm{RC}}$ and through the recoloring factor $1/q^2$, and that the SW and Wolff
transfer operators have qualitatively different spectra on the random-cluster space~\cite{du2006}. For the four-state Potts model, the recoloring factor in Eq.~(\ref{eq:Usw_variance}) becomes $1/q^2 = 1/16$, and the spectral structure of the SW transfer operator on the random-cluster space is modified by the marginal nature of $q=4$. A monotone $q$-dependence $\mu^{(SW)}(q=2) > \mu^{(SW)}(q=3) > \mu^{(SW)}(q=4)$ is therefore natural, although a quantitative prediction requires the full spectral data of the SW transfer matrix.

\section*{2-step overlaps}
\label{sec:results}

The Ising model and the three-component Potts model were simulated on a square lattice with sizes ranging from $L=64$ to $L=1024$ using three methods: the Metropolis method~\cite{landau2005}, the Swendsen--Wang clustering method~\cite{swendsen1987}, and the single-cluster Wolff method~\cite{wolff1989}. The following observables were calculated: the energy per spin $\epsilon$, the heat capacity, expr.~(\ref{eq:heat_capacity}), and the algorithmic intersections, exprs.~(\ref{eq:wolff_overlap}) and~(\ref{eq:overlap_general}).

For all three update schemes and all three models, the default simulation statistics were $N_{\text{therm}} = 200\,000$ thermalization steps followed by $N_{\text{meas}} = 1\,000\,000$ measurement steps, where one step is a Metropolis sweep of $N=L^2$ proposed spin flips, a Swendsen--Wang multi-cluster update, or a Wolff single-cluster update, depending on the algorithm. In order to control finite-size and statistical-quality effects in the regimes where they are expected to be most pronounced, the statistics were increased to $N_{\text{therm}} = 400\,000$ and $N_{\text{meas}} = 1\,500\,000$ in the following three cases: the largest lattice size $L=1536$ used in the Wolff finite-size-scaling analysis of the Ising model at $T_c$ (Fig.~\ref{fig:wolff-ising-fss}), the Swendsen--Wang simulations of the three-state Potts model for $L \ge 768$; and all simulations of the four-state Potts model for $L \ge 768$. The same $(N_{\text{therm}},N_{\text{meas}})$ pairs are used both for the mean values, Eq.~(\ref{eq:mean_overlap}), and for the variances, Eq.~(\ref{eq:var_overlap}), of all algorithmic and thermodynamic observables reported below.

\subsection{Overlap in Wolff updates}

In this subsection, we present the results of simulations of the Ising model and the three-component Potts model using the single-cluster Wolff algorithm~\cite{wolff1989}. The simulations were performed over a wide range of temperatures $T$. A single step of the Wolff algorithm consists of computing a Wolff cluster, which is built around a randomly selected spin on the lattice; the cluster is built by attaching a neighboring spin to it with probability $p_{bond}$, as specified in expression~(\ref{eq:bond_probability}). All spins in the cluster take on new values. This yields the configuration of all $N$ lattice spins at simulation step $t$. 

Upon reaching the number of steps $N_{therm}$ in the Wolff algorithm, we calculate the values of the internal energy and heat capacity using formula~(\ref{eq:heat_capacity}), as well as the value of the geometric intersection of clusters $U_n^{(W)}(t)$ on the lattice using expression~(\ref{eq:wolff_overlap}) and its variance. We present results averaged over the number of Wolff steps $N_{meas}$. 

\subsubsection{Ising model with Wolff updates}

\begin{figure}[H]
  \centering
  \includegraphics[width=0.48\textwidth]{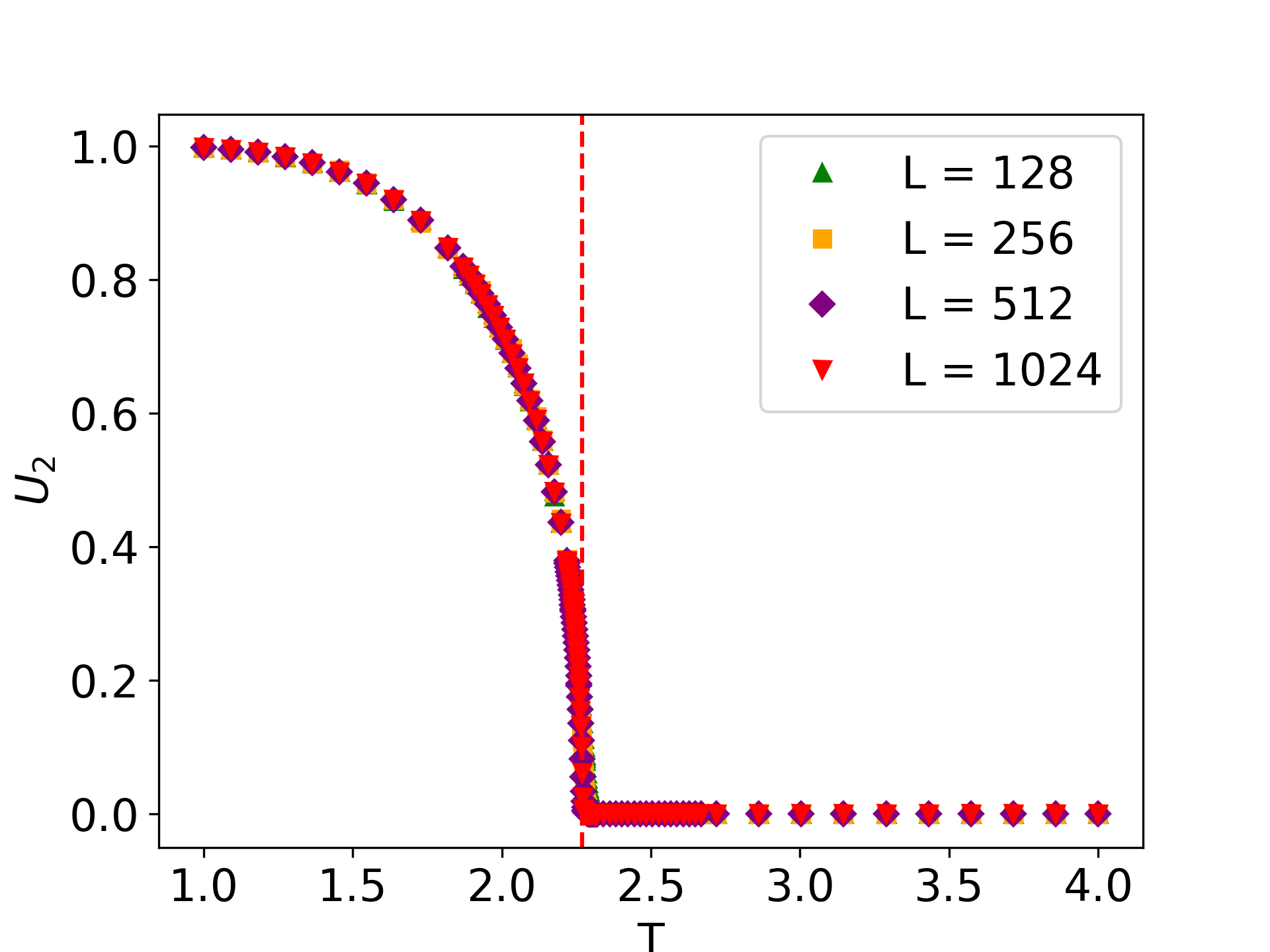}
  \caption{Dependence of $U_2^{(W)}$ on temperature for several values of size. The vertical dashed line indicates the critical temperature $T_c$.}
  \label{ising_U2_vs_T_wolff}
\end{figure}

Figure~\ref{ising_U2_vs_T_wolff} shows the dependence of $U_2^{(W)}$ on temperature for several lattice size values. In the high-temperature phase, the overlap $U_2^{(W)}$ is zero in the thermodynamic limit, since the cluster radius decreases rapidly with increasing temperature. 
In other words, the probability of intersection of two randomly located finite objects $C^{t}$ and $C^{t+n}$ on an infinite lattice tends to zero. 

In the low-temperature phase, the overlap $U_2^{(W)}$ is finite and vanishes at the critical point with an exponent approximately equal to $\mu^{(W)} = 0.428(3)$.
This exponent is estimated as follows.
We assume a power law near the critical point,  $U_2 \sim (T_c - T)^{\mu}$.
Taking the logarithm yields a linear relationship, so the slope on the logarithmic plot is the target exponent.  Also for each system size $L$, we estimate the finite-size effects using
\begin{equation}
y_L = y_{\infty} + \frac{a}{L}.
\label{eq:wolff_extrapolation}
\end{equation}
Extrapolating to $L \to \infty$ yields a smooth curve whose slope $\psi$ is the final estimate of the exponent.

\begin{figure}[H]
  \centering
  \includegraphics[width=0.48\textwidth]{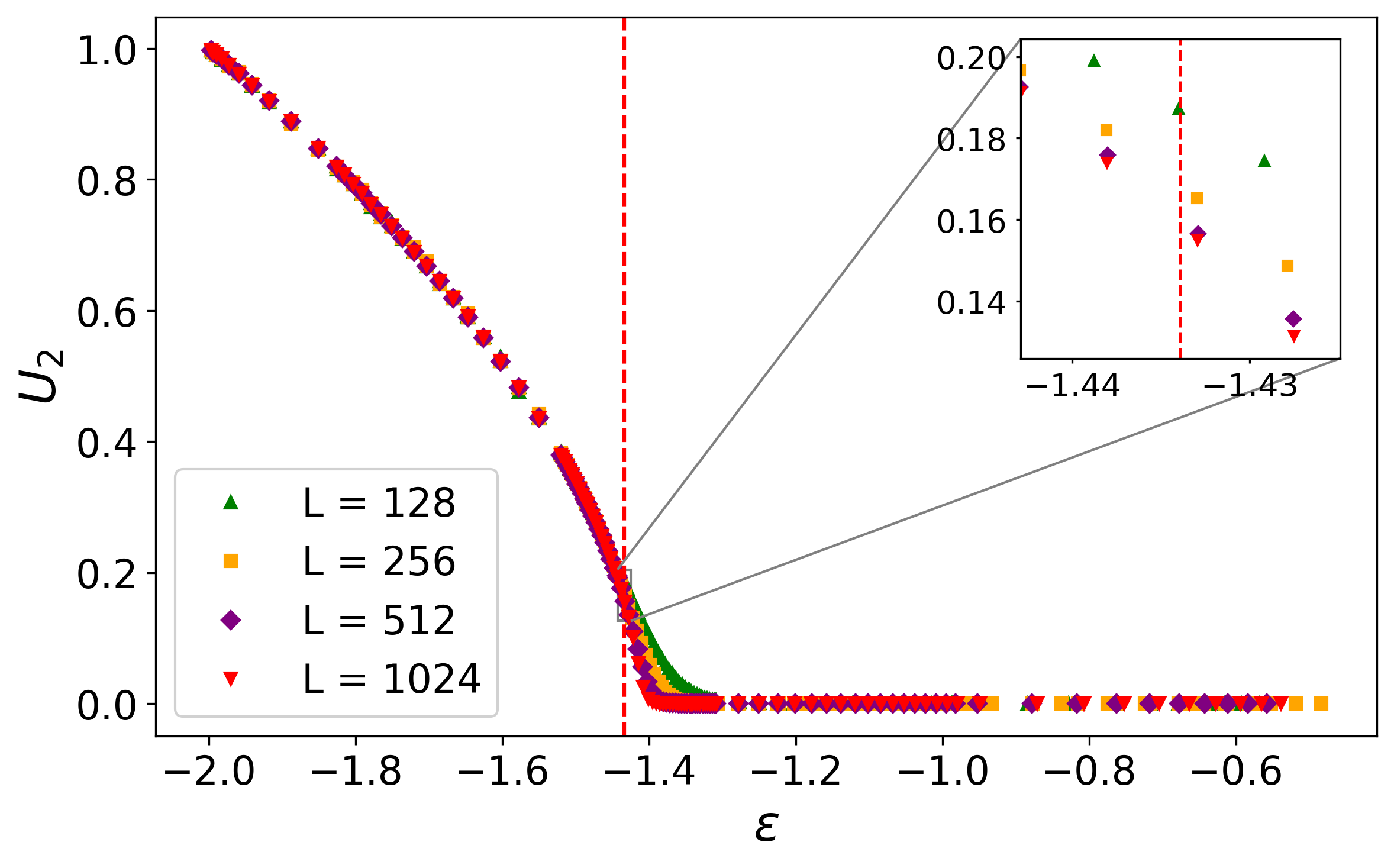}
  \caption{Wolff algorithm, Ising model. The average fraction of cluster overlap as a function of energy. The vertical dashed line indicates the critical energy corresponding to $T_c$. The inset shows a zoomed region near the critical point.}
  \label{fig:wolff-ising-mean}
\end{figure}

\begin{figure}[H]
  \centering
  \includegraphics[width=0.48\textwidth]{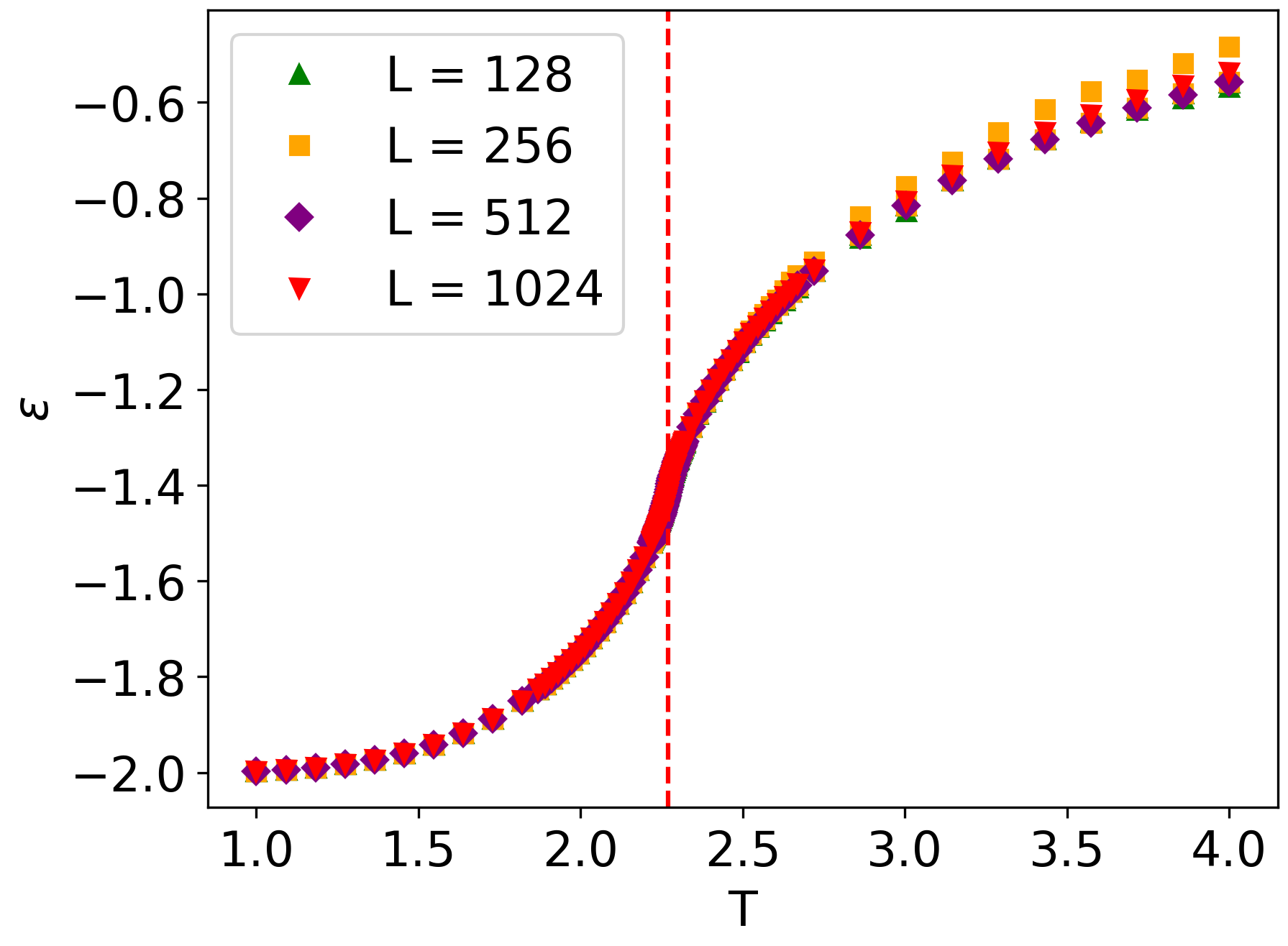}
  \caption{Wolff algorithm, Ising model. The energy per spin versus temperature. The vertical dashed line indicates the critical temperature $T_c$.}
  \label{ising_epsilon_vs_T_wolff}
\end{figure}

However, in the critical region, significant size-dependent effects are observed, which are clearly visible in the plot of the overlap $U_2^{(W)}$ versus energy in Figure~\ref{fig:wolff-ising-mean}. 
Plotting a graph of $U_2^{(W)}$ versus energy, initially calculated as a function of temperature, is possible due to the relationship between energy and temperature; see Figure~\ref{ising_epsilon_vs_T_wolff}.

\begin{figure}[H]
  \centering
  \includegraphics[width=0.48\textwidth]{"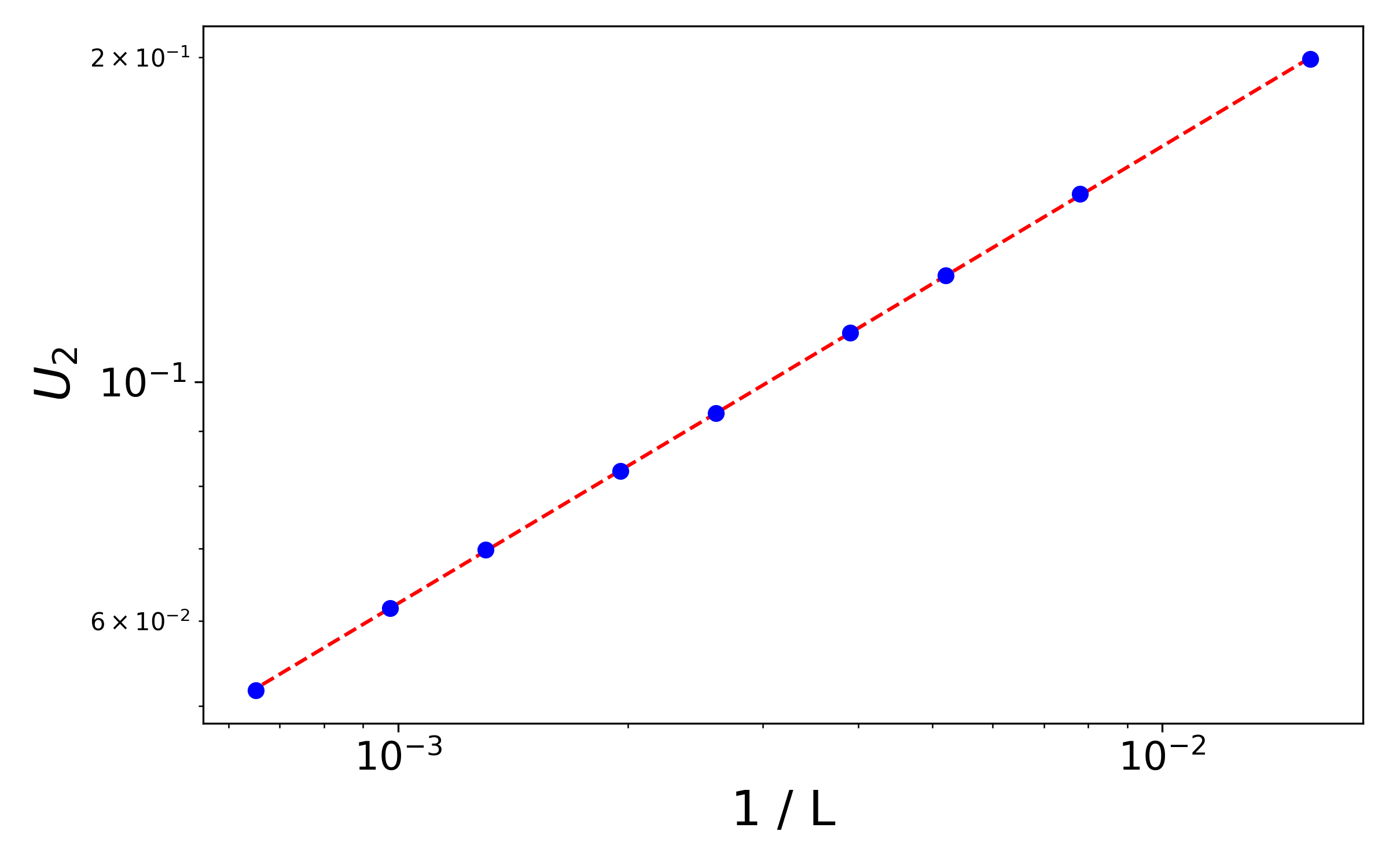"}
  \caption{Wolff updates, Ising model. Finite-size scaling of the average overlap at $T_c$.}
  \label{fig:wolff-ising-fss}
\end{figure}

We can gain additional insight into how the overlap value $U_2^{(W)}$ vanishes at the critical point as the lattice size changes. Figure~\ref{fig:wolff-ising-fss} shows the overlap value at the critical temperature $U_2^{(W)}(T_c)$ versus the reciprocal of the lattice size $L$ on a logarithmic scale along both axes for lattice sizes $L=64$, 128, 192, 256, 384, 512, 768, 1024, and 1536. Approximating this value with a nearly perfect straight line yields the exponent $\psi^{(W)}_{Ising}=0.4240(6)$, which is in good agreement with the value $\mu^{(W)}_{Ising}=0.428(3)$ estimated above.

We have no clear understanding of how this exponent might be related to the exponents within the Ising universality class. Moreover, this value coincides with the corresponding exponent estimated for the three-state Potts and four-state Potts models (see next subsections) within statistical errors. It is not yet clear whether this is simply a coincidence or a pattern of cluster overlap. 

\begin{figure}[H]
  \centering
  \includegraphics[width=0.48\textwidth]{"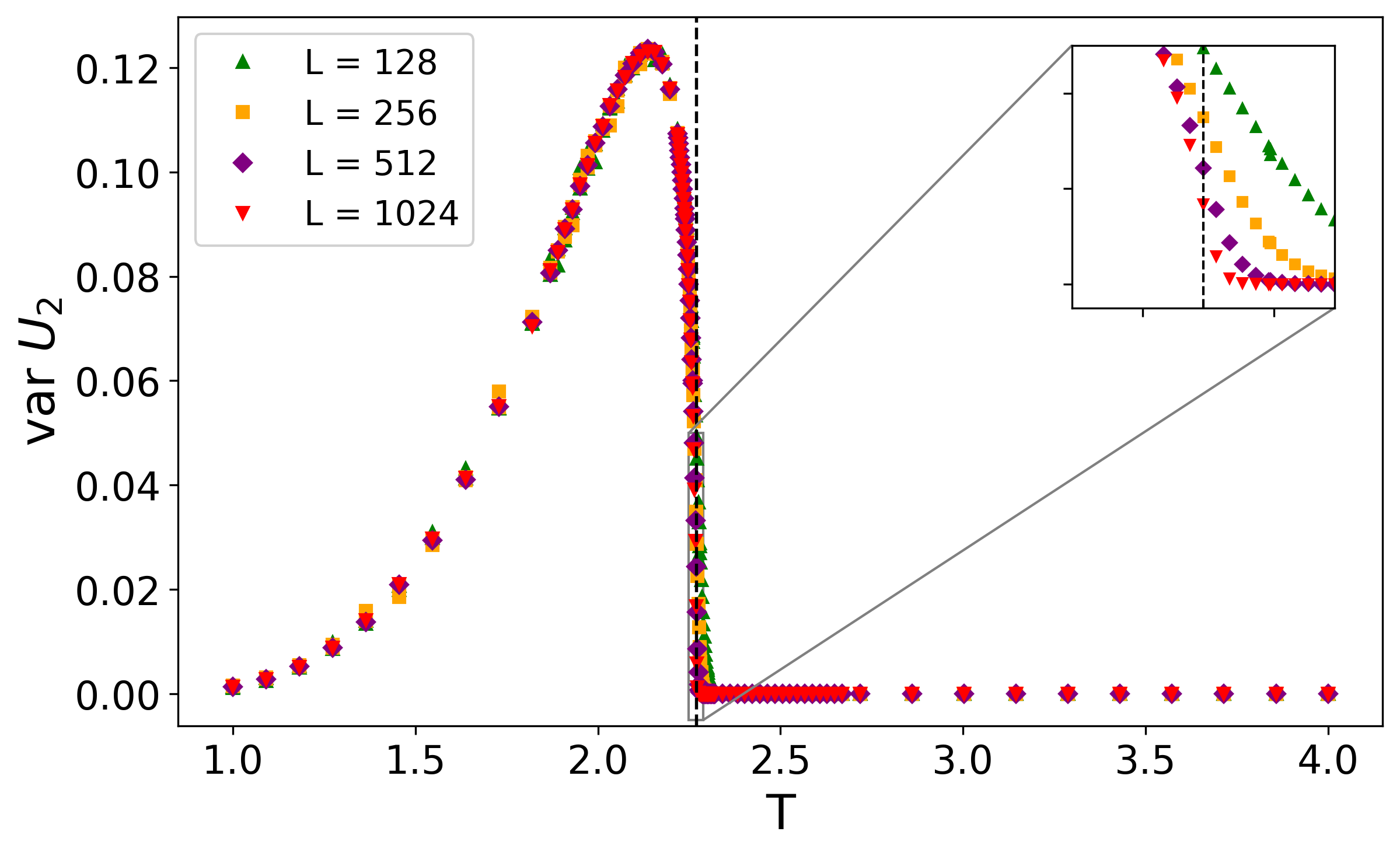"}
  \caption{Wolff updates, Ising model. The cluster overlap variance $\mathrm{Var}(U_2)$ as a function of temperature shows a maximum  slightly below the critical temperature. The vertical dashed line indicates the critical temperature $T_c$. The inset shows a zoomed region near the critical point.}
  \label{fig:wolff-ising-var}
\end{figure}

Another interesting quantity is the overlap variance, expr.~(\ref{eq:var_overlap}), which is shown in Figure~\ref{fig:wolff-ising-var}. This variance vanishes at the critical point in the thermodynamic limit, and its maximum shifts toward lower temperatures. 

In the spirit of the paper~\cite{lev2019}, we also plotted the dependence of the overlap variance on the specific heat capacity. This dependence has a limiting point at the peak associated with the critical region. 
\begin{figure}[H]
  \centering
  \includegraphics[width=0.48\textwidth]{"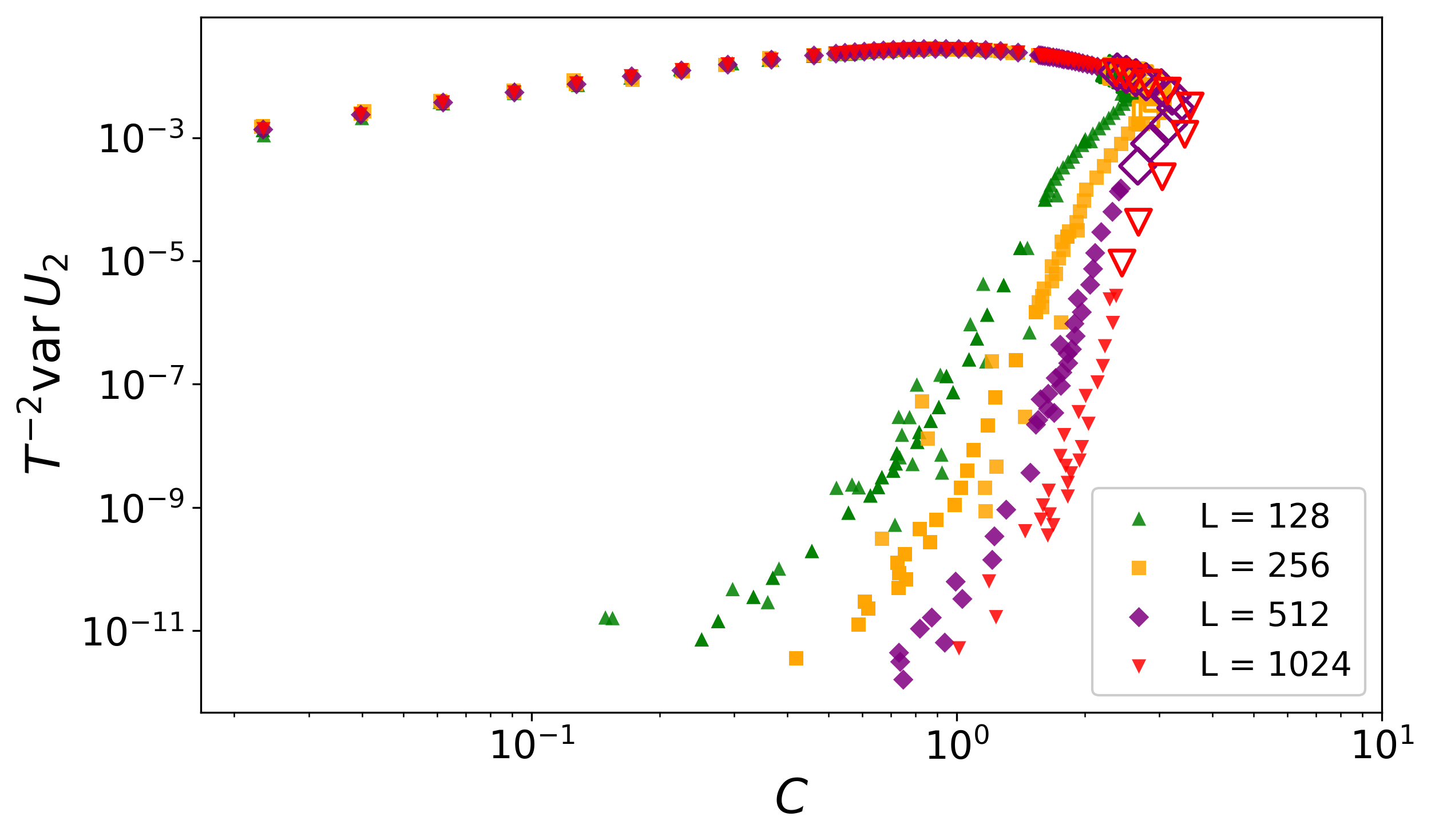"}
  \caption{Wolff updates, Ising model. Scaled variance $T^{-2}\mathrm{Var}(U_2)$ as a function of heat capacity, revealing the underlying thermodynamic singularity.}
  \label{fig:wolff-ising-var-cv}
\end{figure}

\subsubsection{Three-state Potts model with Wolff updates}

We present a similar analysis for the three-state Potts model, focusing on how the observed magnitude of cluster overlap behaves during the transition.

\begin{figure}[H]
  \centering
  \includegraphics[width=0.48\textwidth]{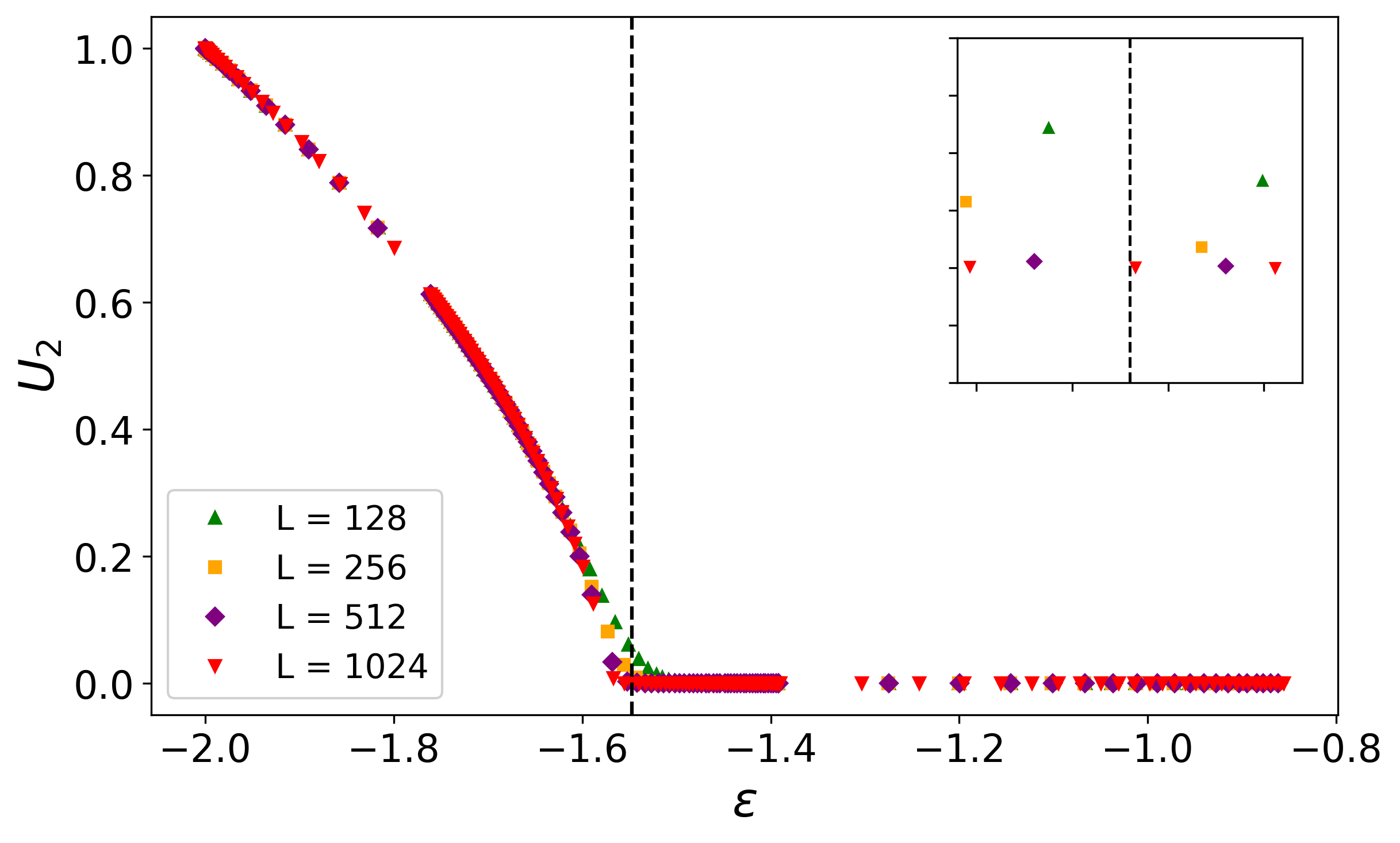}
  \caption{Wolff updates, three-state Potts model. Average cluster overlap fraction as a function of energy per spin. The vertical dashed line indicates the critical energy corresponding to $T_c$. The inset shows a zoomed region near the critical point.}
  \label{fig:wolff-potts-mean}
\end{figure}

Figure~\ref{fig:wolff-potts-mean} shows the dependence of the average overlap $U_2^{(W)}$ on the energy per spin. The transition is marked by a sharp drop in the overlap, which becomes increasingly sharp with system size, indicating the emergence of critical correlations.

\begin{figure}[H]
  \centering
  \includegraphics[width=0.48\textwidth]{"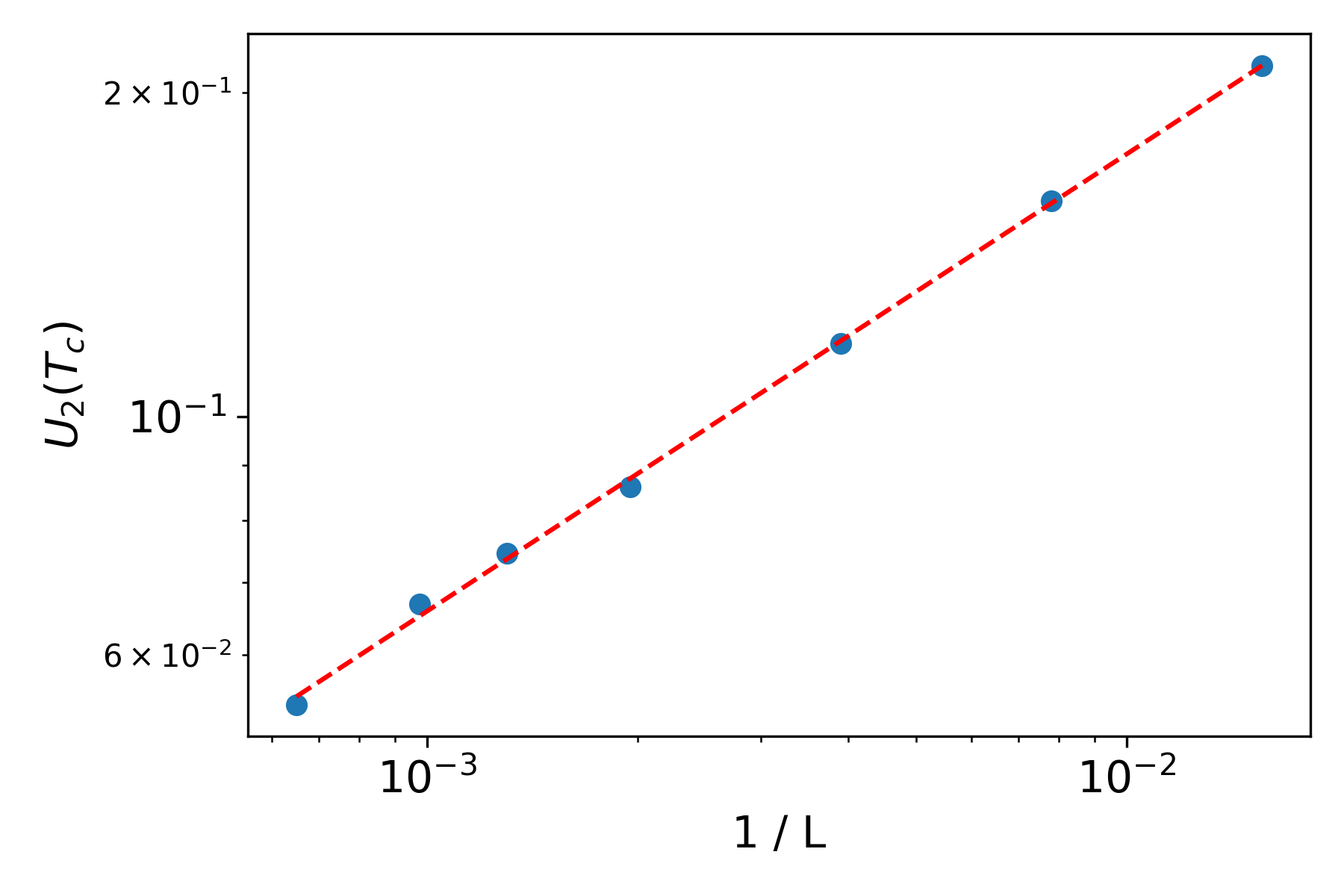"}
  \caption{Wolff updates, three-state Potts model. Scaling of the average overlap of finite-size clusters at the point $T_c$.}
  \label{fig:wolff-potts-fss}
\end{figure}

The dependence of the overlap on the lattice size at the critical point is shown in Fig.~\ref{fig:wolff-potts-fss}. The data follow a power-law decay in $L$, indicating that $U_2^{(W)}(T_c)$ vanishes in the thermodynamic limit. The corresponding value of the exponent is $\psi^{(W)}_{3-Potts} = 0.425(4)$, which is very close to the exponent estimated for the Ising model.

\begin{figure}[H]
  \centering
  \includegraphics[width=0.48\textwidth]{"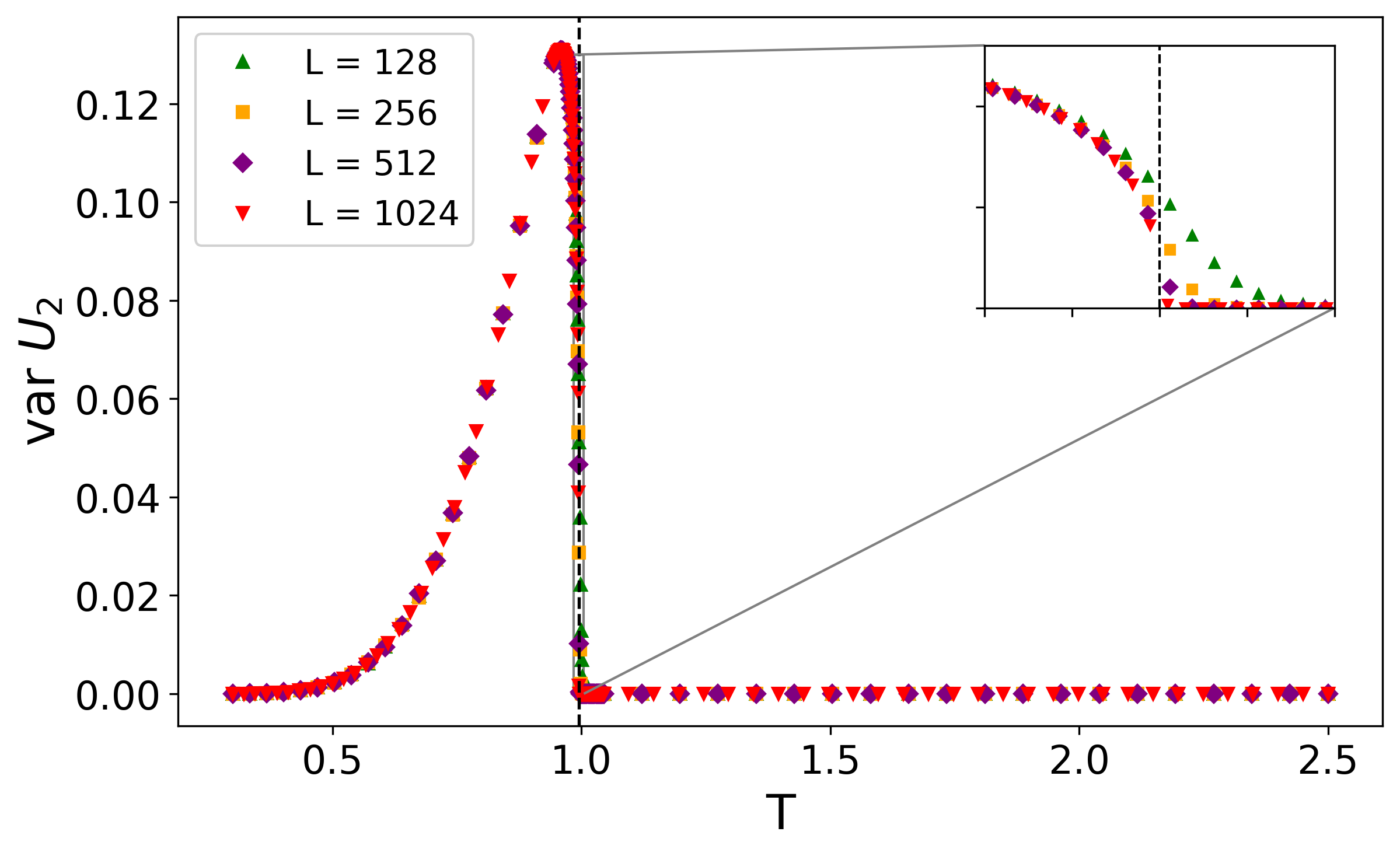"}
  \caption{Wolff updates, three-state Potts model. Cluster overlap variance $\mathrm{Var}(U_2)$ versus temperature. The vertical dashed line indicates the critical temperature $T_c$. The inset shows a zoomed region near the critical point.}
  \label{fig:wolff-potts-varT}
\end{figure}

The variance $\mathrm{Var}(U_2)$ forms a pronounced peak in the transition region (Fig.~\ref{fig:wolff-potts-varT}), with clear finite-size effects in both height and position.

\begin{figure}[H]
  \centering
  \includegraphics[width=0.48\textwidth]{"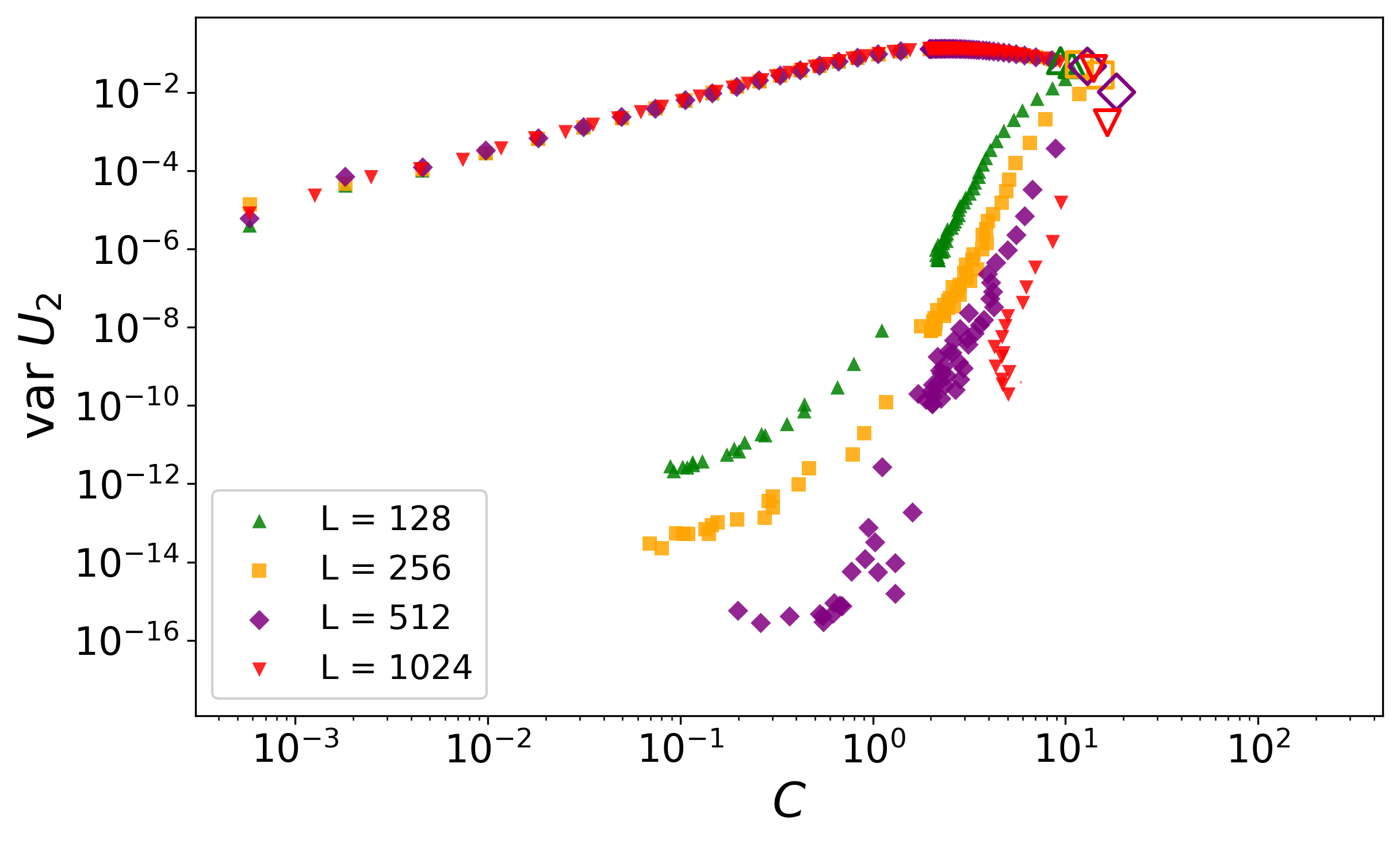"}
  \caption{Wolff updates, three-state Potts model. Rescaled variance $T^{-2}\mathrm{Var}(U_2)$ versus heat capacity.}
  \label{fig:wolff-potts-varC}
\end{figure}

 Just as in the Ising case  we can estimate the $\mu^{(W)}_{3-Potts}$ exponent of $U_2$ versus $T$. $\mu^{(W)}_{3-Potts} = 0.3525(5)$.
 We consider again the rescaled quantity $T^{-2}\mathrm{Var}(U_2)$ as a function of the heat capacity. The resulting curve (Fig.~\ref{fig:wolff-potts-varC}) exhibits a well-defined extremum aligned with the specific-heat peak.

\subsubsection{Four-state Potts model with Wolff updates}

We now extend the analysis to the four-state Potts model, This case provides a stringent test of whether the algorithmic-universality features identified for $q=2$ and $q=3$ (in particular the numerical coincidence of the Wolff overlap exponent $\psi^{(W)}$) persists).

\begin{figure}[H]
  \centering
  \includegraphics[width=0.48\textwidth]{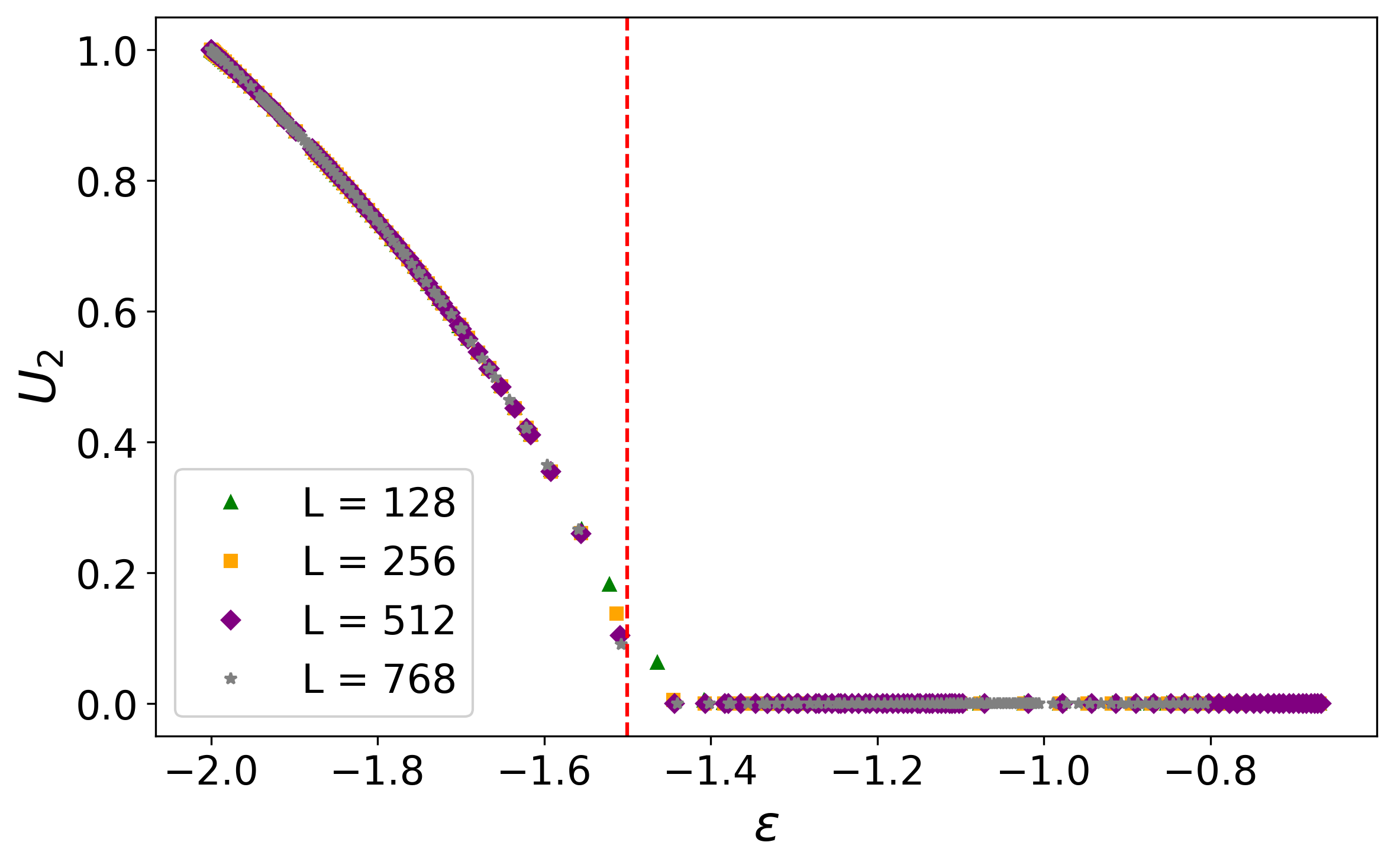}
  \caption{Wolff updates, four-state Potts model. Average cluster overlap fraction $U_2^{(W)}$ as a function of energy per spin, for several lattice sizes. The vertical dashed line indicates the critical energy corresponding to $T_c^{(4)} = 1/\ln 3$. }
  \label{fig:wolff-potts4-mean}
\end{figure}

Figure~\ref{fig:wolff-potts4-mean} shows the dependence of the average overlap $U_2^{(W)}$ on the energy per spin. As in the previous cases, the overlap drops through the transition region, and the curves become progressively sharper with system size.

\begin{figure}[H]
  \centering
  \includegraphics[width=0.48\textwidth]{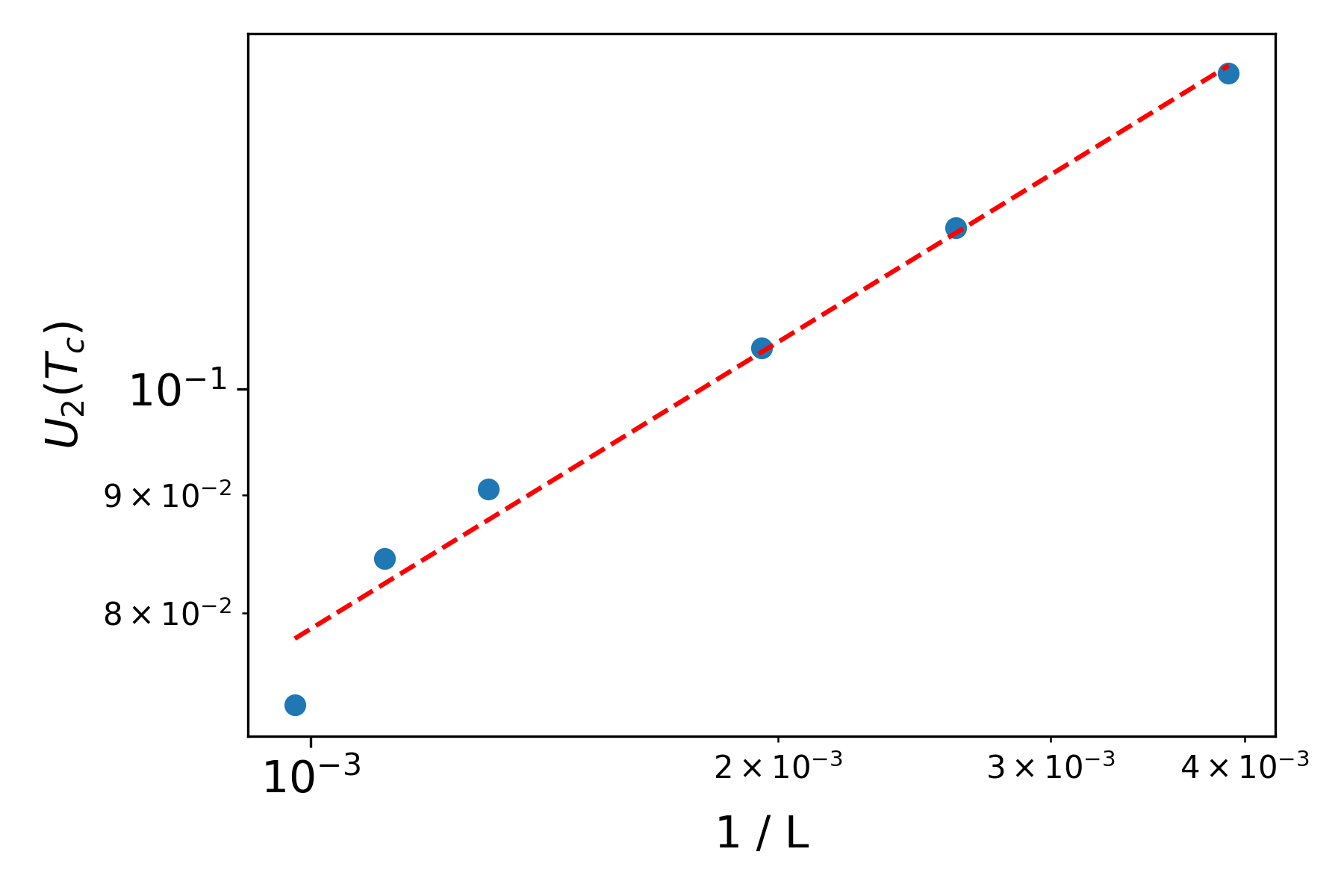}
  \caption{Wolff updates, four-state Potts model. Finite-size scaling of the average overlap at $T_c^{(4)}$, $U_2^{(W)}(T_c^{(4)})$ versus $1/L$ on log--log axes.}
  \label{fig:wolff-potts4-fss}
\end{figure}

The dependence of the overlap on the lattice size at the critical point is shown in Fig.~\ref{fig:wolff-potts4-fss}. A power-law decay in $L$ gives the exponent $\psi^{(W)}_{4-Potts} = 0.419(1)$ which is very similar to  $\psi^{(W)}_{Ising} = 0.428(3)$ and $\psi^{(W)}_{3-Potts} = 0.425(4)$ suggesting that this finite-size scaling exponent can be attributed to the algorithm-dependent cluster structure and not universality class itself.

\begin{figure}[H]
  \centering
  \includegraphics[width=0.48\textwidth]{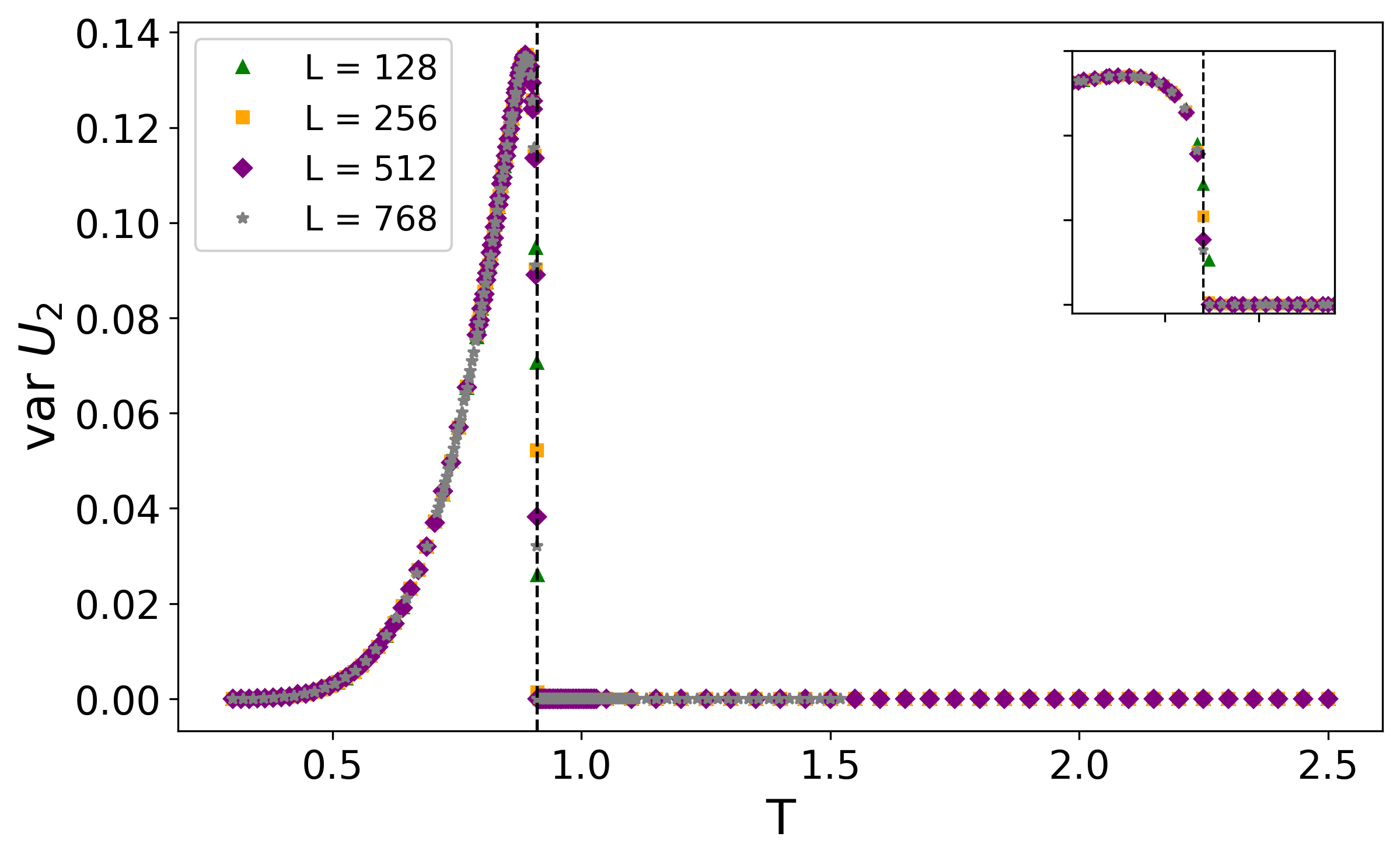}
  \caption{Wolff updates, four-state Potts model. Cluster overlap variance $\mathrm{Var}(U_2)$ versus temperature, for several lattice sizes. The vertical dashed line indicates the critical temperature $T_c^{(4)}$. The inset shows a zoomed region near the critical point.}
  \label{fig:wolff-potts4-varT}
\end{figure}

The variance $\mathrm{Var}(U_2)$  again forms a peak in the transition region (Fig.~\ref{fig:wolff-potts4-varT}), with finite-size effects in both height and position closely paralleling the Ising and three-state Potts cases. The exponent of $U_2$ versus $(T_c-T)$ is $\mu^{W}_{4-Potts} = 0.281(2)$ showing that $U_2$ drops sharper with $q$.

\begin{figure}[H]
  \centering
  \includegraphics[width=0.48\textwidth]{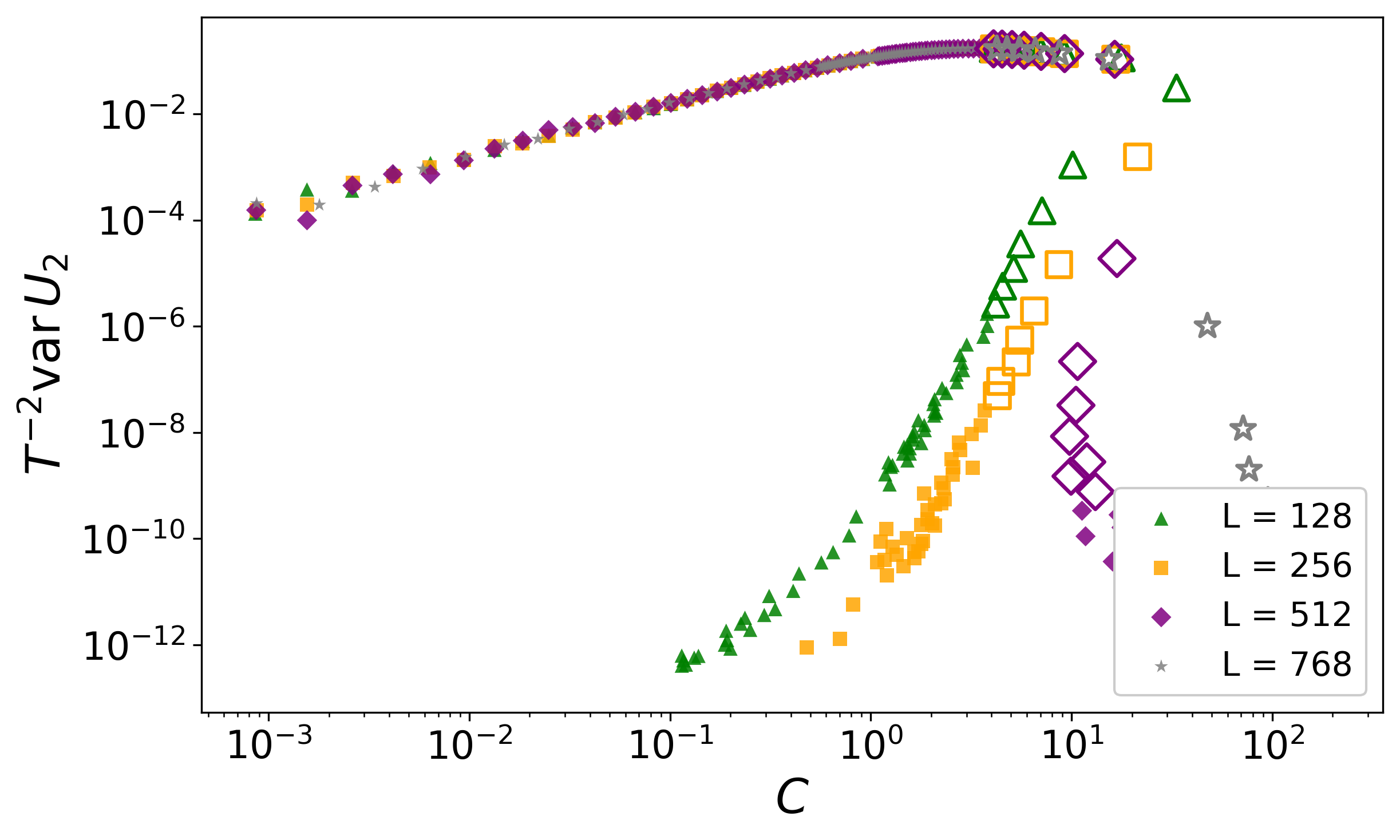}
  \caption{Wolff updates, four-state Potts model. Rescaled variance $T^{-2}\mathrm{Var}(U_2)$ versus heat capacity, for several lattice sizes.}
  \label{fig:wolff-potts4-varC}
\end{figure}

As in the Ising and three-state Potts cases, we will plot the rescaled quantity $T^{-2}\mathrm{Var}(U_2)$ against the heat capacity (Fig.~\ref{fig:wolff-potts4-varC}). The resulting curve exhibits an extremum aligned with the specific-heat peak, confirming that the Wolff overlap fluctuations track the underlying thermodynamic singularity also at the marginal value $q=4$.

\begin{figure}[H]
    \centering
    \includegraphics[width=0.48\textwidth]{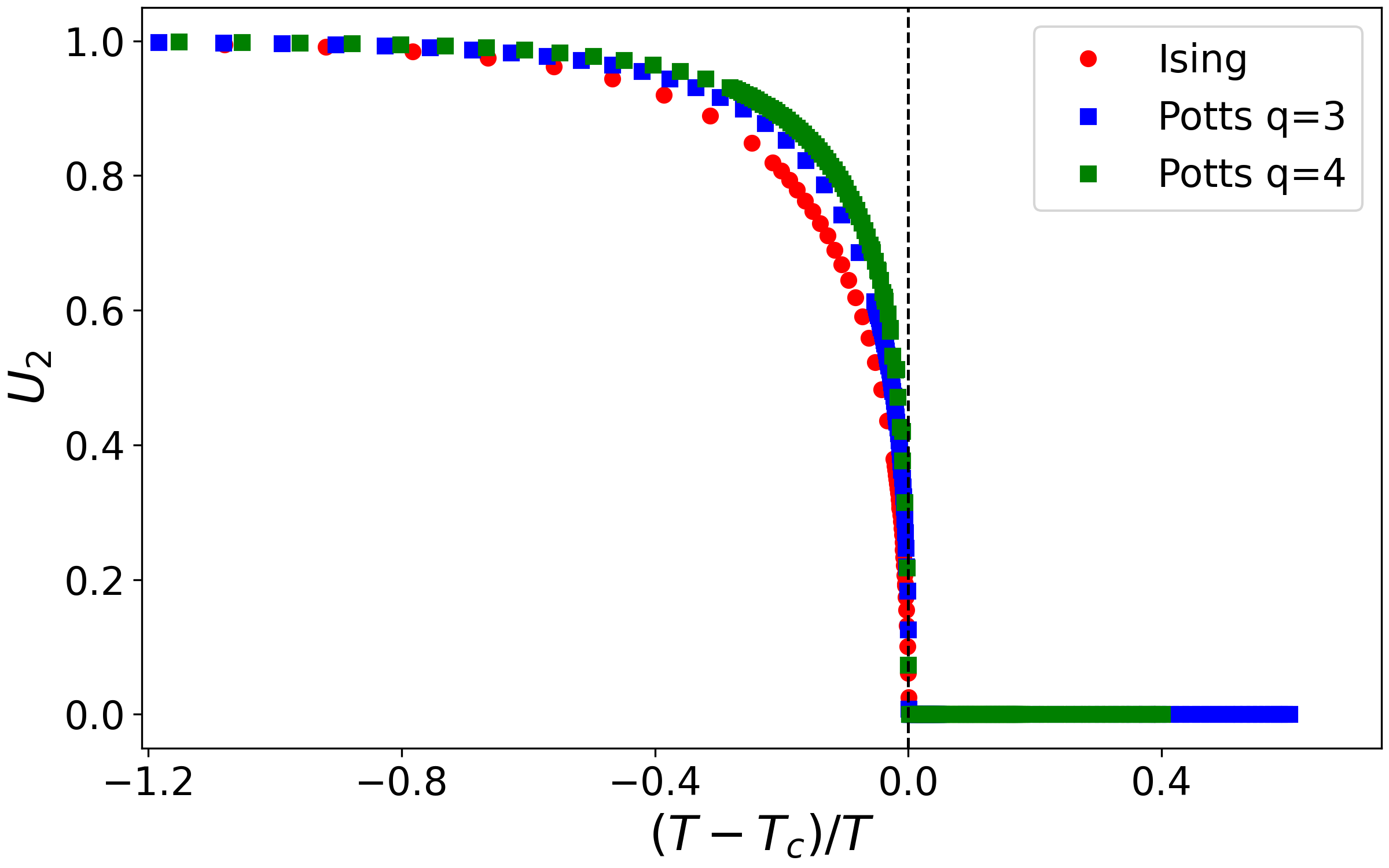}
    \caption{Overlap $U_2$ as a function of $(T - T_c)/T$ for the Ising and 3-state Potts models using the Wolff algorithm at $L=1024$. The vertical dashed line at $(T-T_c)/T = 0$ indicates the critical temperature $T_c$.}
    \label{fig:wolff_overlap_t2}
\end{figure}

To clarify the thermodynamic nature of the Wolff overlap  further, we have also plotted $U_2^{(W)}$ as a function of the reduced temperature variable $\tau=(T - T_c)/T$ and showed it in Figure~\ref{fig:wolff_overlap_t2}.

A sharp transition is observed in the critical region, separating the finite-overlap regime from the vanishing-overlap regime. The data for the Ising,  three- and four-state Potts models show a similar shape, getting sharper with $q$ , suggesting that the geometric overlap is determined primarily by universal properties of the cluster rather than by the microscopic details of spin symmetry.
The overlap $U_2$ can be interpreted as the order parameter for the Wolff algorithm.

% ===== Wolff: overlap vs (T - Tc)/T =====

\subsection{Overlaps in Swendsen--Wang updates}

\subsubsection*{Swendsen--Wang updates: Ising model}

Unlike the Wolff algorithm, Swendsen--Wang (SW) updates all Fortuin--Kasteleyn clusters simultaneously. This leads to strong decorrelation between successive configurations and significantly alters the behavior of the overlap observable.

\begin{figure}[H]
  \centering
  \includegraphics[width=0.48\textwidth]{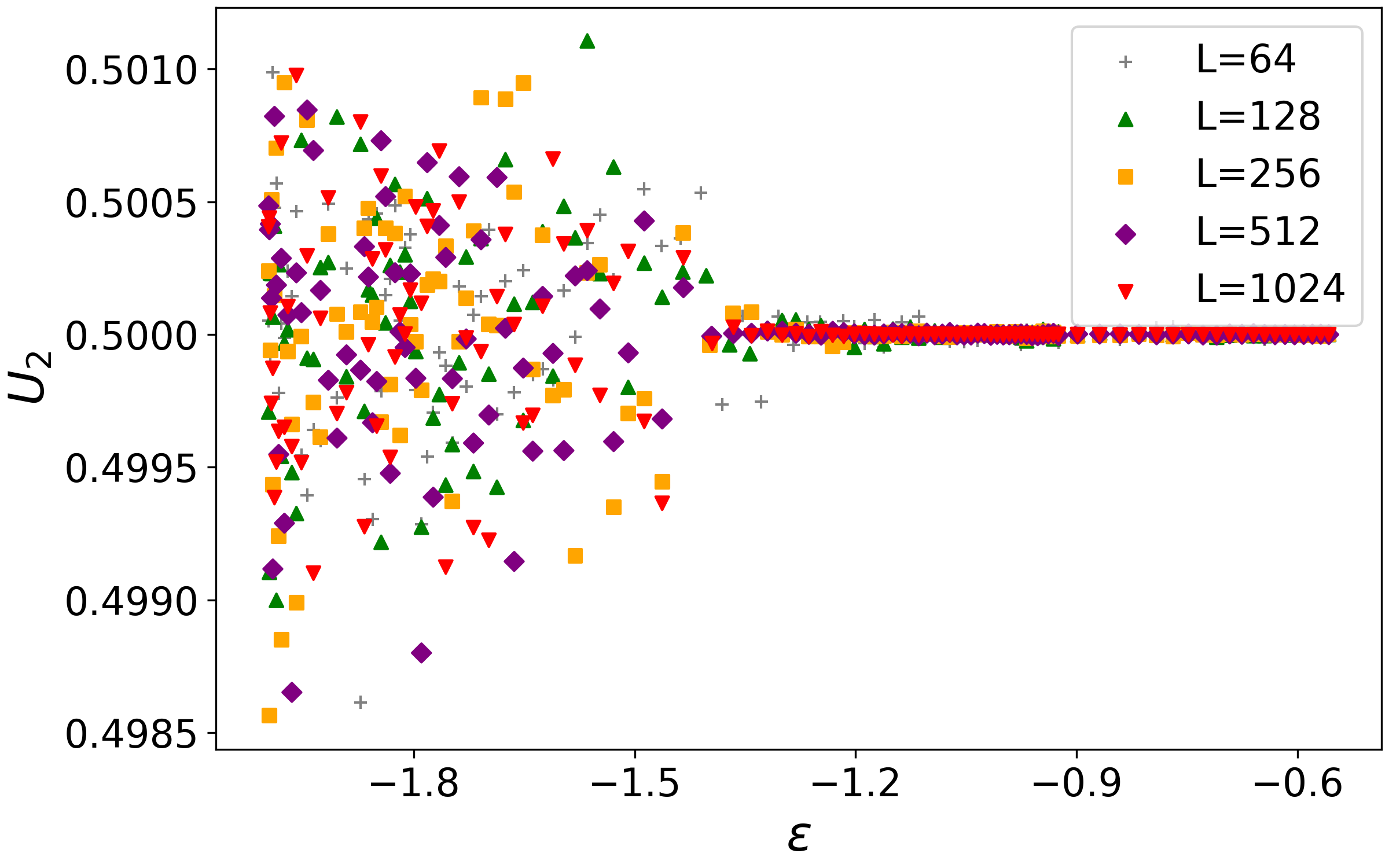}
  \caption{Swendsen--Wang updates, Ising model. Mean two-step overlap versus energy per spin.}
  \label{fig:sw-ising-mean}
\end{figure}

As shown in Fig.~\ref{fig:sw-ising-mean}, the average two-step overlap remains nearly constant (note the narrow range of the overlap scale) across the entire energy range, indicating that it does not carry direct thermodynamic information in this update scheme.

\begin{figure}[H]
  \centering
  \includegraphics[width=0.48\textwidth]{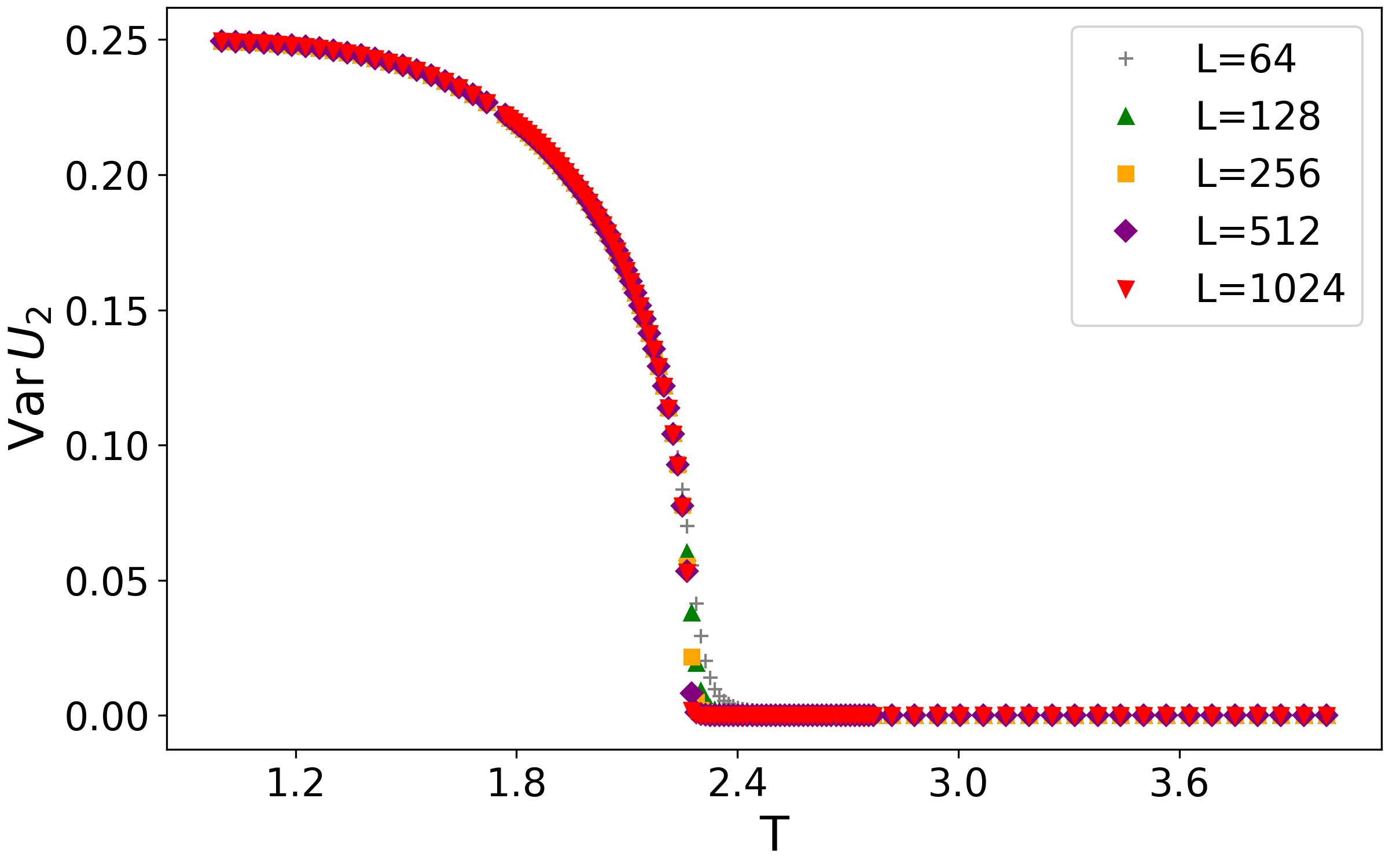}\\[4pt]
  \includegraphics[width=0.48\textwidth]{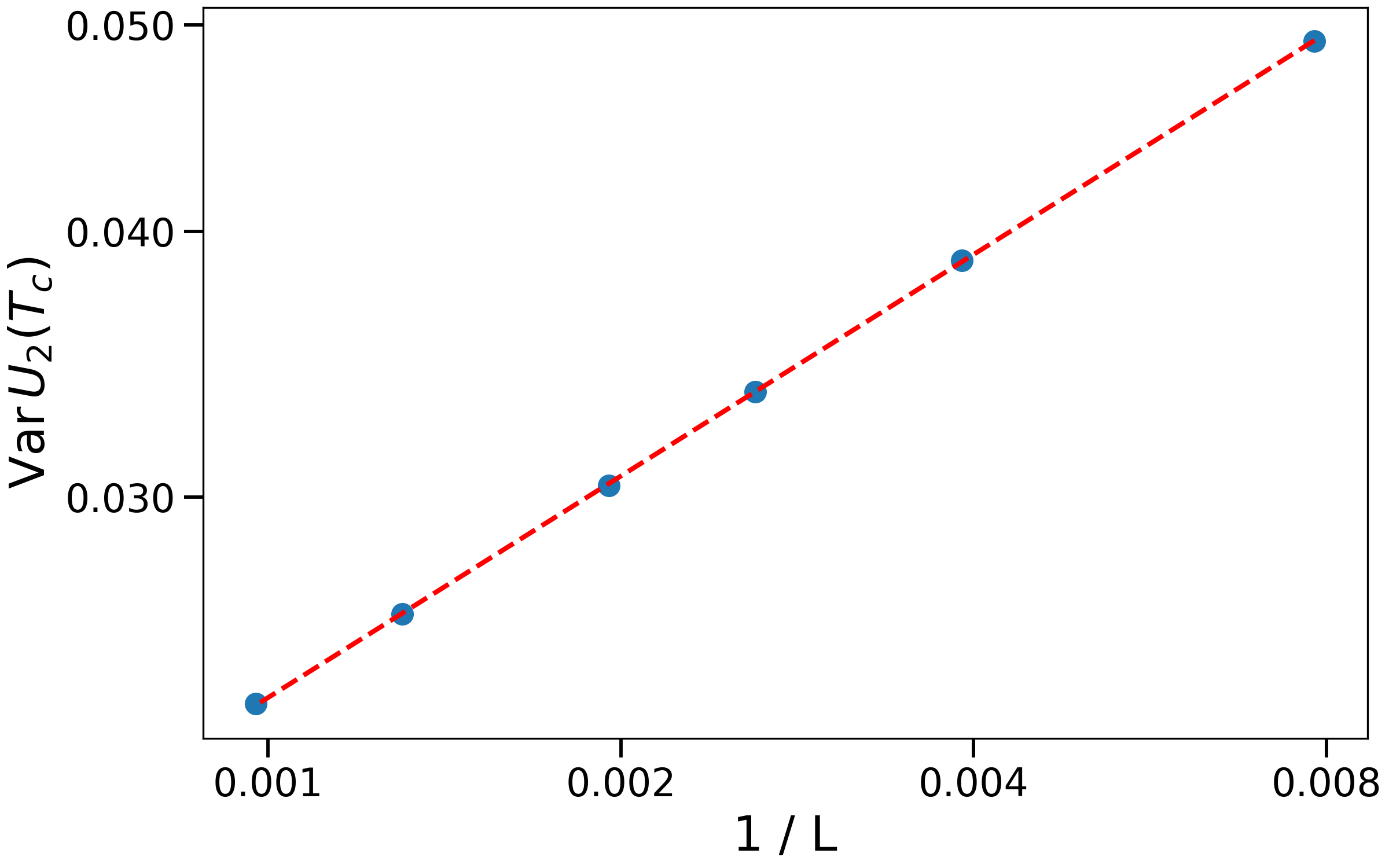}
  \caption{Swendsen--Wang updates, Ising model. Top: variance of the two-step overlap versus temperature. Bottom: finite-size scaling near $T_c$.}
  \label{fig:sw-ising-var-scale}
\end{figure}

The relevant information is encoded in the fluctuations. The overlap variance $\mathrm{Var}(U_2)$ is finite in the ordered phase and is rapidly suppressed near the critical temperature (Fig.~\ref{fig:sw-ising-var-scale}, top), while its behavior becomes increasingly sharper with increasing lattice size (Fig.~\ref{fig:sw-ising-var-scale}, bottom).

To quantify how $\mathrm{Var}(U_2)$ vanishes at the critical point, we repeat the same analysis as for the Wolff case, assuming a power form
\[
\mathrm{Var}(U_2) \sim (T_c - T)^{\mu^{(SW)}}.
\]
Extrapolating the effective exponent to the thermodynamic limit yields 
\[
\mu^{(SW)} \approx 0.348(4).
\]

Independently, the finite-size scaling of the variance at $T_c$ follows a power law in $L$, with slope $\psi^{(SW)}_{Ising} = 0.3458(9)$,
confirming that the suppression of fluctuations sharpens systematically with system size.

\begin{figure}[H]
  \centering
  \includegraphics[width=0.48\textwidth]{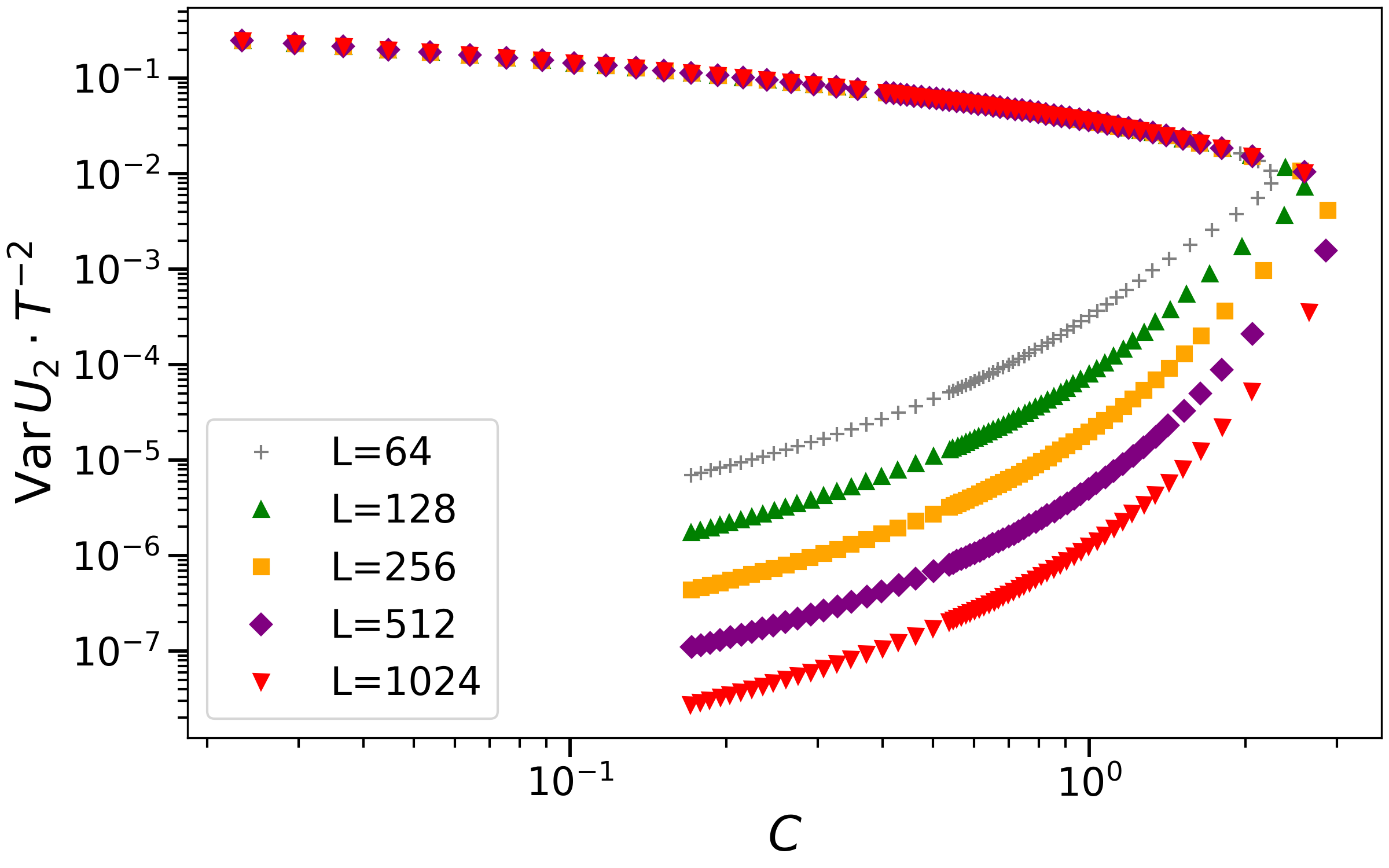}
  \caption{Swendsen--Wang updates, Ising model. Rescaled variance versus specific heat.}
  \label{fig:sw-ising-var-Cv}
\end{figure}

 In Fig.~\ref{fig:sw-ising-var-Cv} the rescaled variance $T^{-2}\mathrm{Var}(U_2)$ develops a cusp aligned with the specific-heat singularity.

\subsubsection*{Swendsen--Wang updates: three-state Potts model}

\begin{figure}[H]
  \centering
  \includegraphics[width=0.48\textwidth]{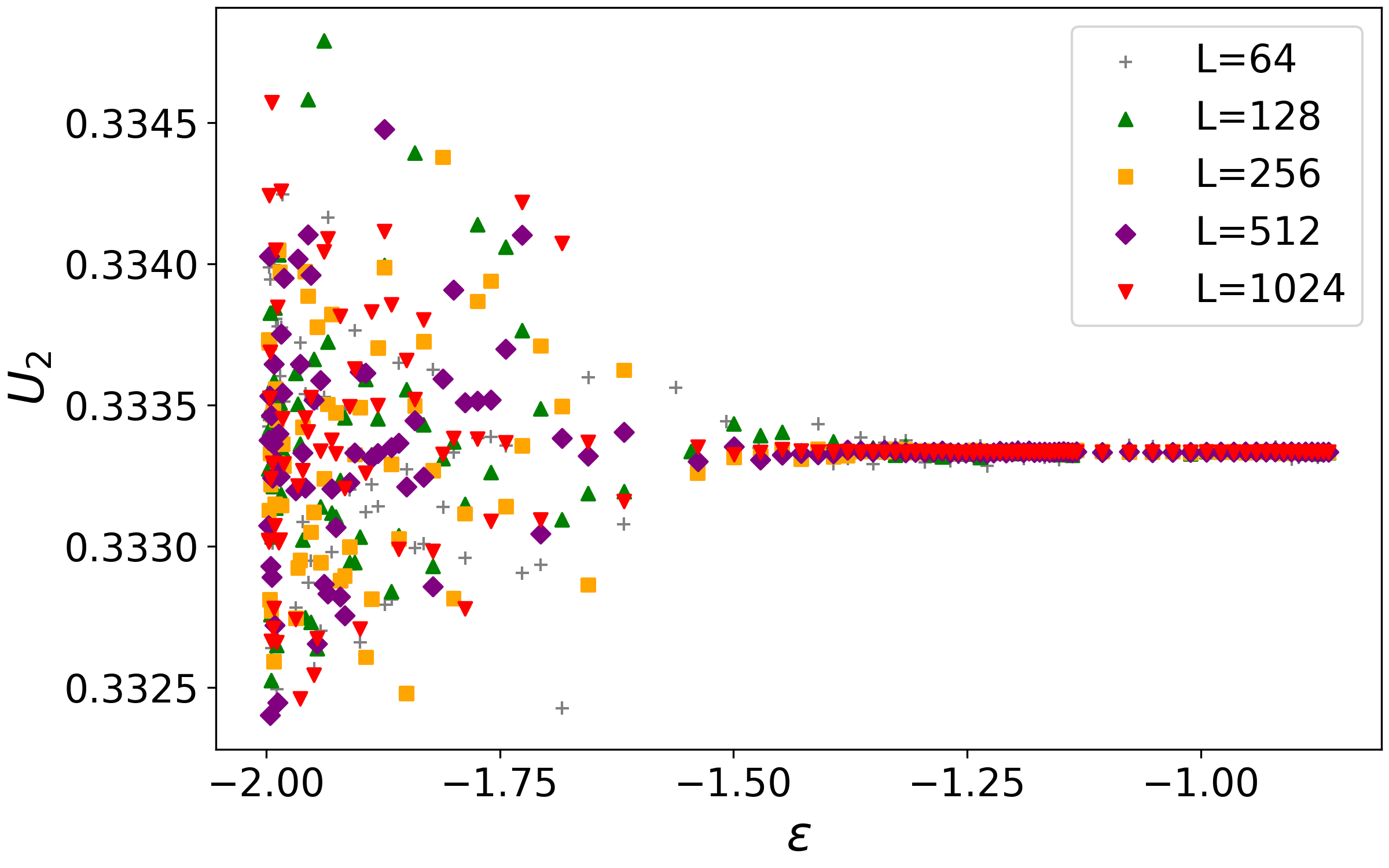}
  \caption{Swendsen--Wang updates, three-state Potts model. Mean two-step overlap versus energy.}
  \label{fig:sw-potts-mean}
\end{figure}

A similar qualitative picture holds for the three-state Potts model. The mean overlap (Fig.~\ref{fig:sw-potts-mean}) forms a nearly flat plateau close to the random value $1/3$, which is a consequence of the threefold degeneracy of the low-temperature phase, again showing no clear signature of the transition.

\begin{figure}[H]
  \centering
  \includegraphics[width=0.48\textwidth]{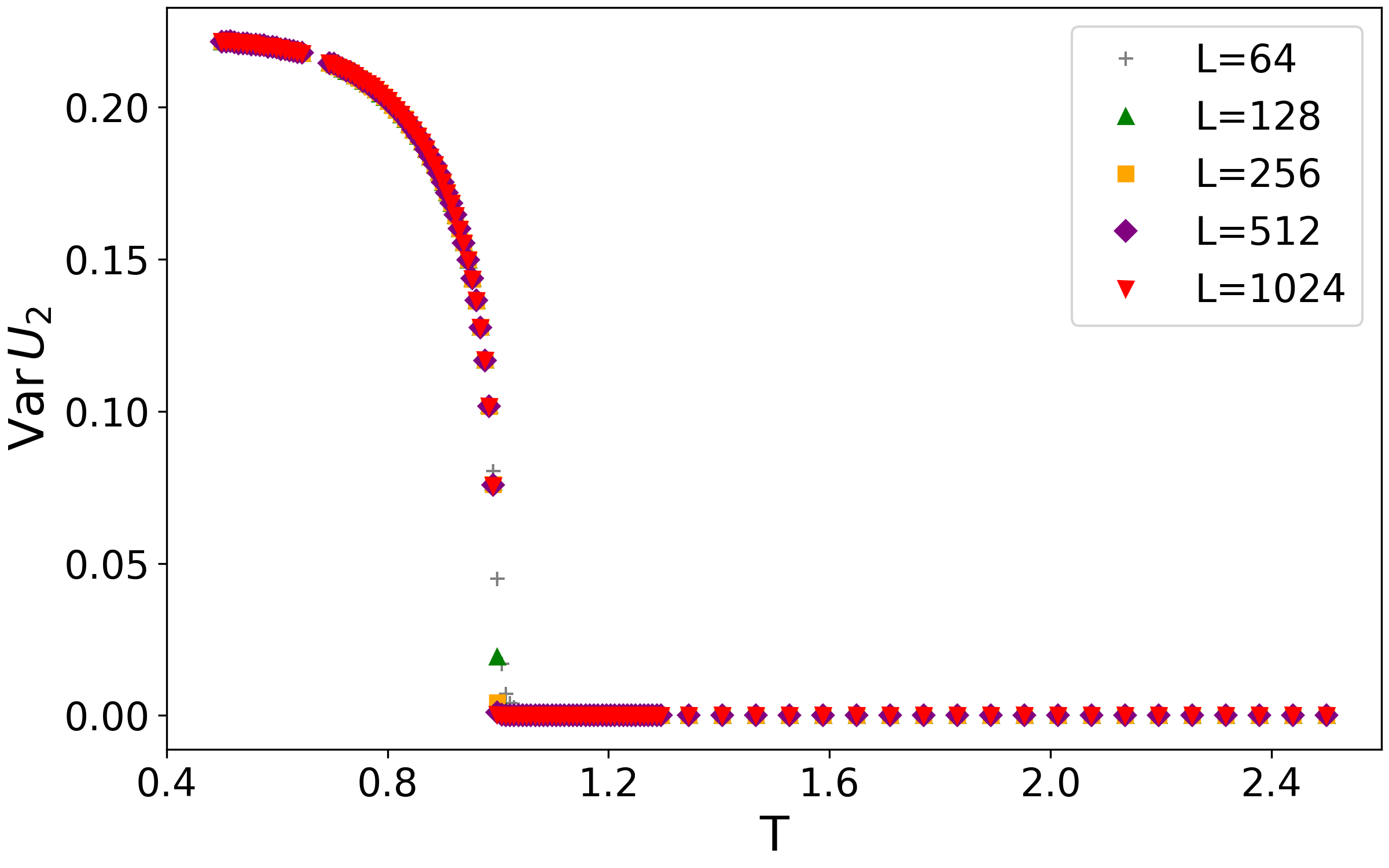}\\[4pt]
  \includegraphics[width=0.48\textwidth]{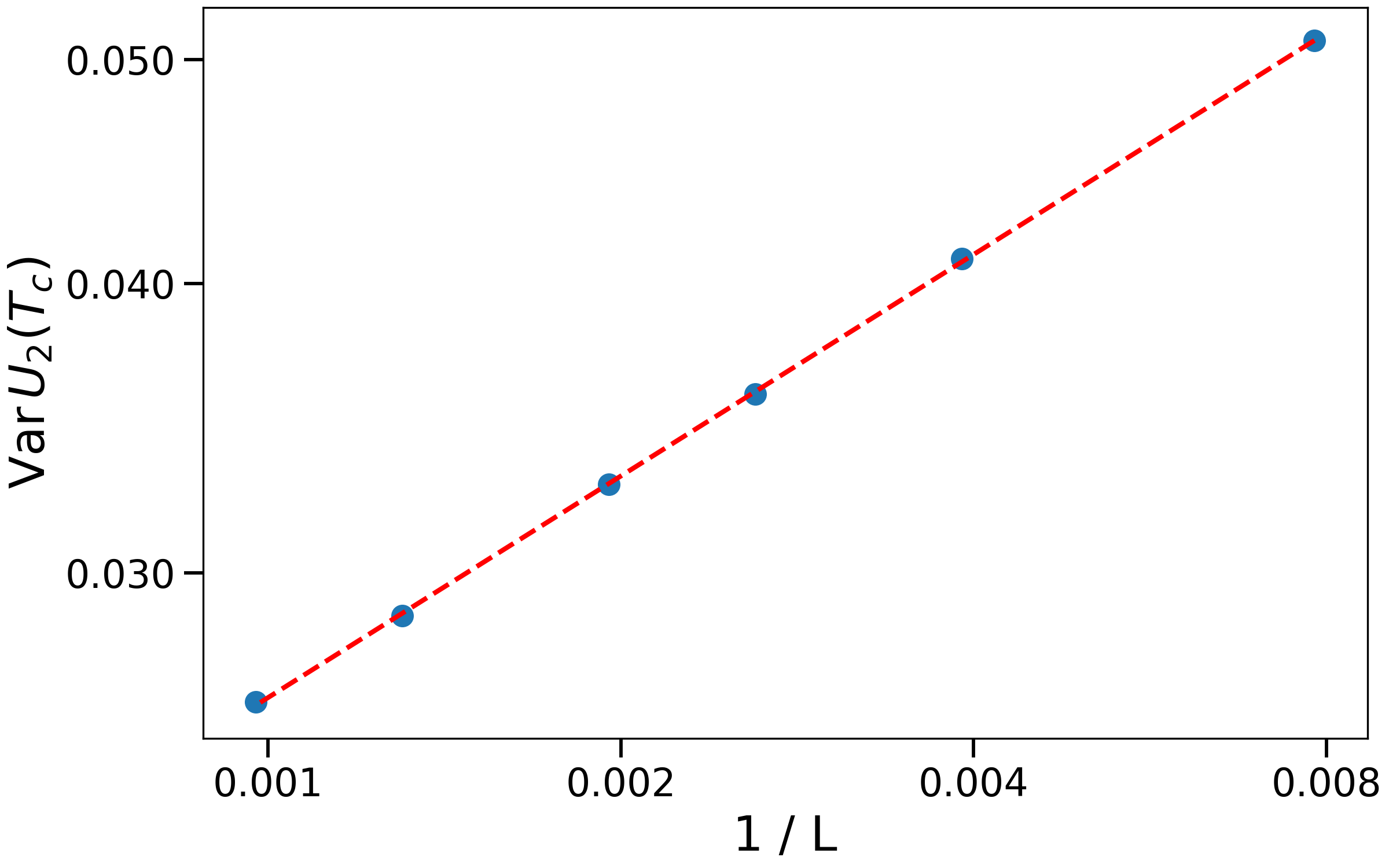}
  \caption{Swendsen--Wang updates, three-state Potts model. Top: variance versus temperature. Bottom: finite-size scaling near $T_c$.}
  \label{fig:sw-potts-var-scale}
\end{figure}

The variance, however, retains a strong sensitivity to criticality. It remains finite in the ordered phase and is rapidly suppressed near $T_c$ (Fig.~\ref{fig:sw-potts-var-scale}, top), with finite-size effects clearly visible (bottom).

Applying the same power-law analysis yields a thermodynamic-limit exponent
\[
\mu^{(SW)}_{3-Potts} \approx 0.266(3),
\]
which is significantly smaller than in the Ising case, indicating a slower vanishing of fluctuations near criticality.

The finite-size scaling at $T_c$ gives  $\psi^{(SW)}_{3-Potts} = 0.318(1)$,
demonstrating a stronger size dependence compared to the Ising model, consistent with the different universality class.

\begin{figure}[H]
  \centering
  \includegraphics[width=0.48\textwidth]{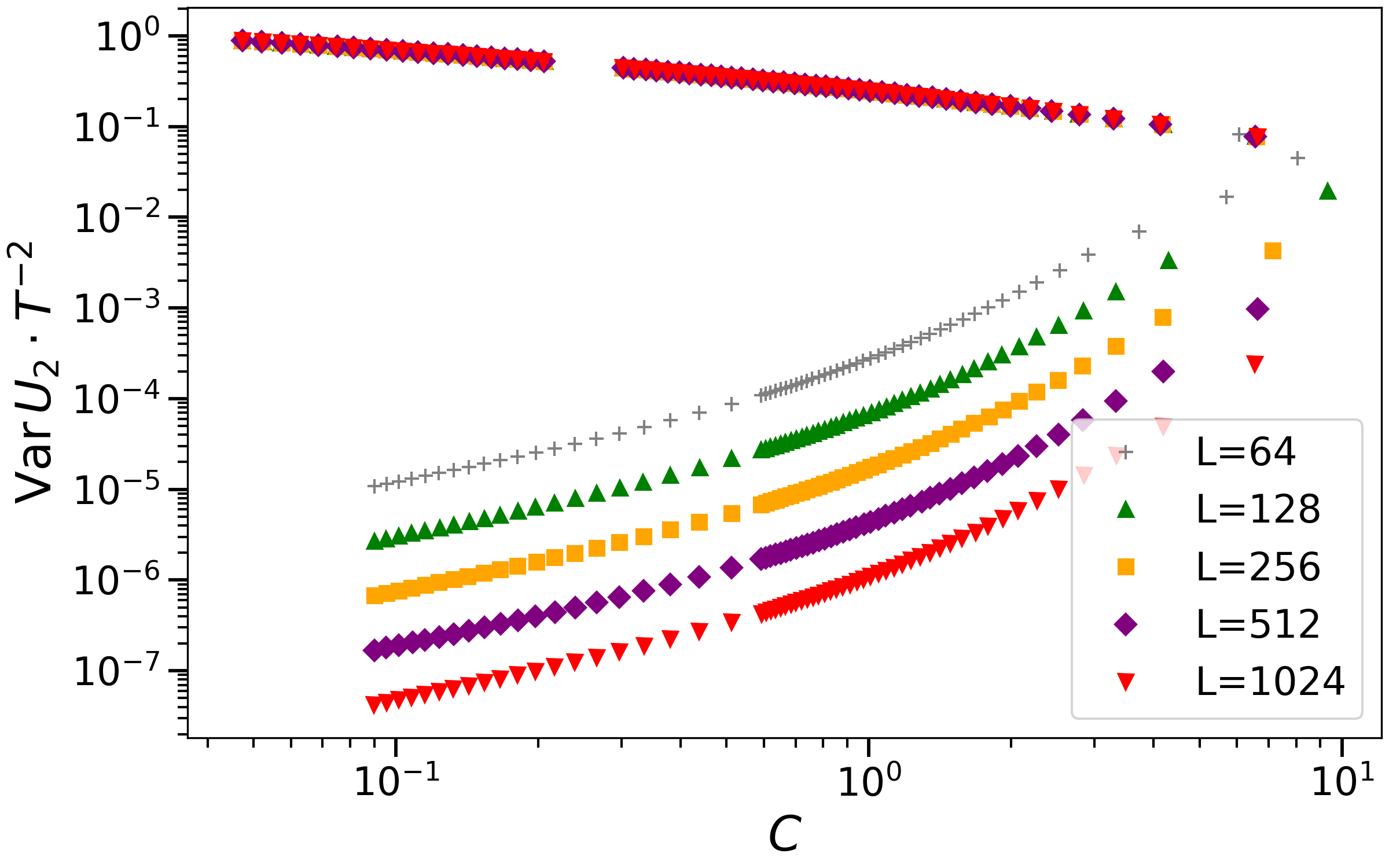}
  \caption{Swendsen--Wang updates, three-state Potts model. Rescaled variance versus specific heat.}
  \label{fig:sw-potts-var-Cv}
\end{figure}

Finally, when plotted against the heat capacity (Fig.~\ref{fig:sw-potts-var-Cv}), the rescaled variance again exhibits a cusp-like structure, confirming that the overlap fluctuations remain directly linked to the thermodynamic singularity.

\subsubsection*{Swendsen--Wang updates: four-state Potts model}

\begin{figure}[H]
  \centering
  \includegraphics[width=0.48\textwidth]{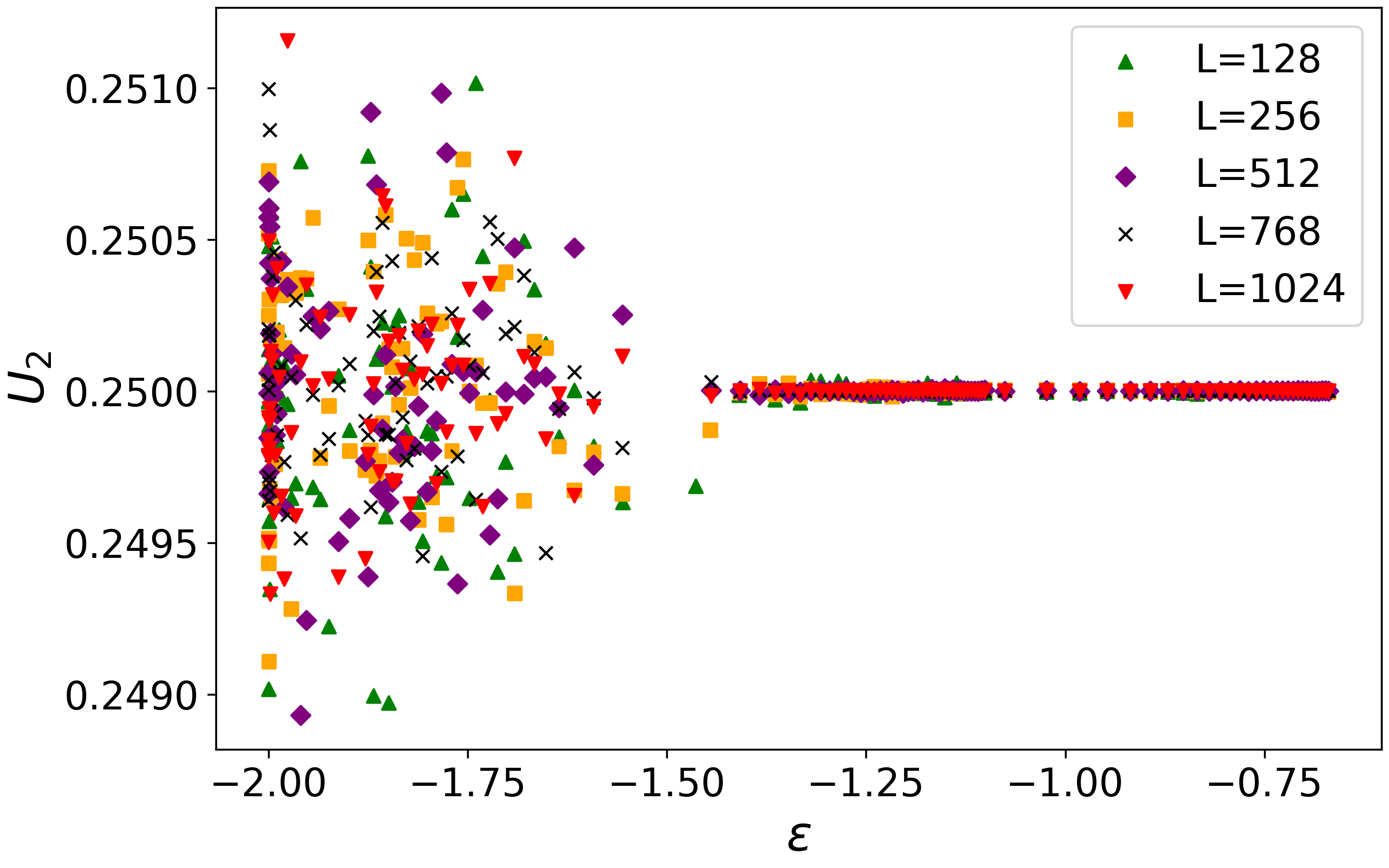}
  \caption{Swendsen--Wang updates, four-state Potts model. Mean two-step overlap versus energy per spin, for several lattice sizes.}
  \label{fig:sw-potts4-mean}
\end{figure}

We perform the same analysis for the four-state Potts model using Swendsen--Wang updates. The mean overlap (Fig.~\ref{fig:sw-potts4-mean}) forms a nearly flat plateau close to the random value $1/q = 1/4$, in agreement with the structure of the SW update: every FK cluster is independently recolored among $q=4$ states, so the spin-overlap baseline shifts from $1/3$ ($q=3$) to $1/4$ ($q=4$).

\begin{figure}[H]
  \centering
  \includegraphics[width=0.48\textwidth]{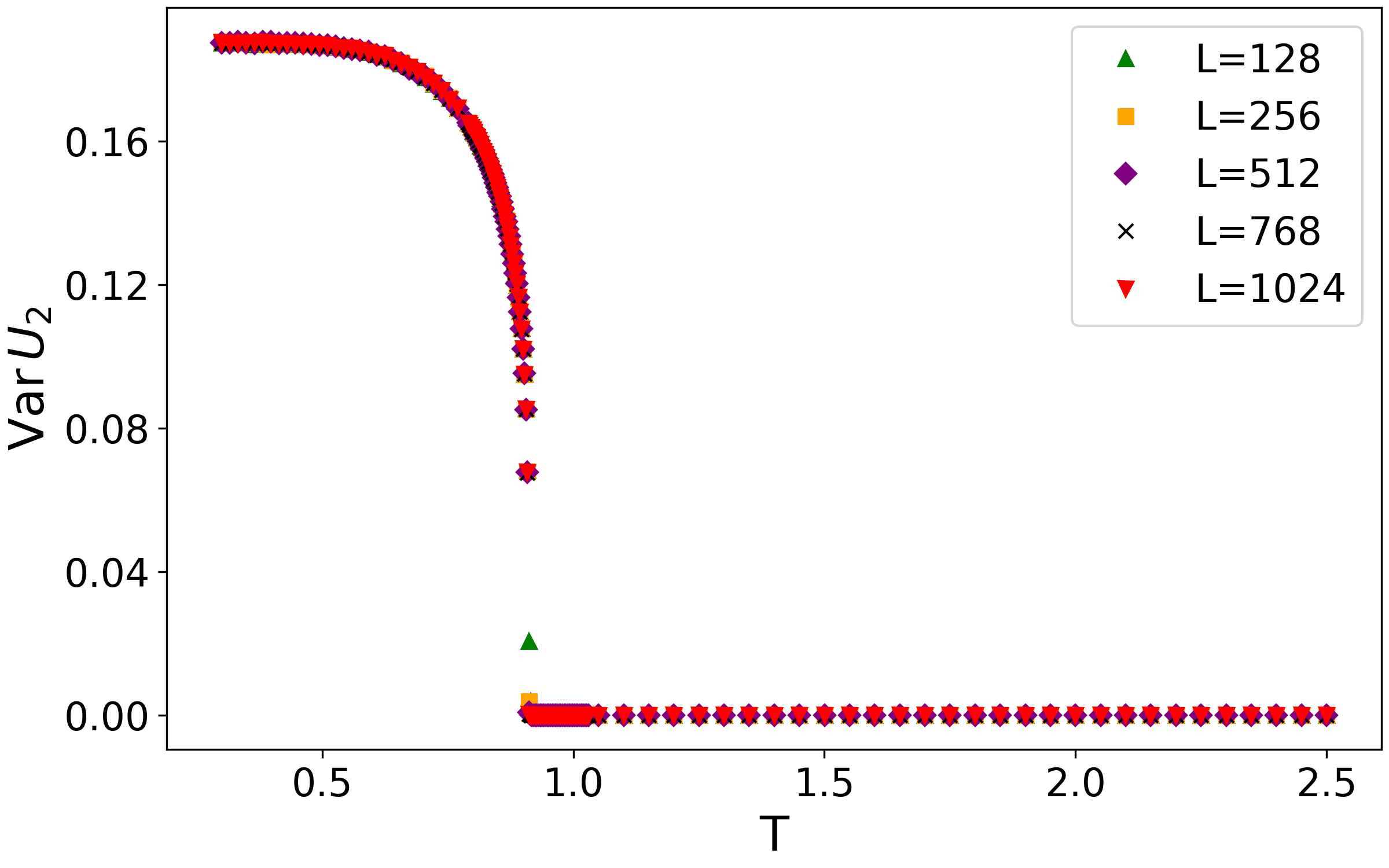}\\[4pt]
  \includegraphics[width=0.48\textwidth]{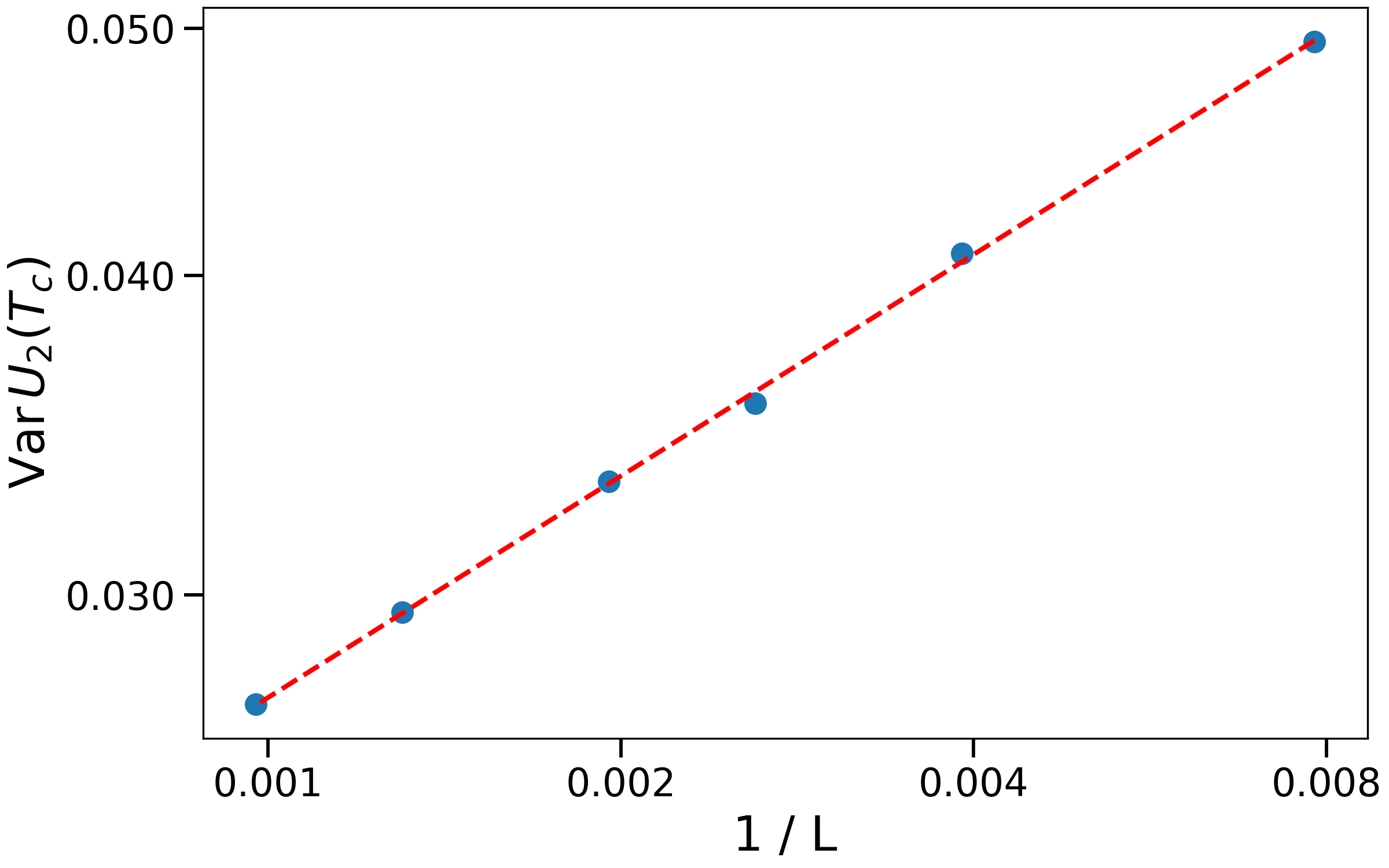}
  \caption{Swendsen--Wang updates, four-state Potts model. Top: variance of the two-step overlap versus temperature, for several lattice sizes. Bottom: finite-size scaling of the variance near $T_c^{(4)}$ on log--log axes.}
  \label{fig:sw-potts4-var-scale}
\end{figure}

The variance retains the same sensitivity to criticality as for $q=2$ and $q=3$: it remains finite in the ordered phase and is rapidly suppressed near $T_c^{(4)}$ (Fig.~\ref{fig:sw-potts4-var-scale}, top), with clear finite-size effects (bottom). Because $q=4$ is the marginal value of the Potts family, the size dependence is expected to be modified by multiplicative logarithmic corrections relative to the pure power-law behavior found for the Ising and three-state Potts cases.

Applying the same power-law analysis as in the Ising and three-state Potts cases, assuming a form $\mathrm{Var}(U_2)\sim (T_c-T)^{\mu^{(SW)}}$ and extrapolating the effective exponent to the thermodynamic limit, we obtain an estimate of $\mu^{(SW)}_{4-Potts} = 0.205(8)$ for the four-state Potts model. The structural argument given around Eq.~(\ref{eq:Usw_variance}) --- the recoloring factor $1/q^2 = 1/16$ in the variance and the spectral modification of the SW transfer operator at the marginal point --- suggests that $\mu^{(SW)}$ should continue the monotone trend $\mu^{(SW)}_{Ising}=0.358(4) > \mu^{(SW)}_{3-Potts}=0.266(3) > \mu^{(SW)}_{4-Potts}=0.205(8)$, although a quantitative prediction requires the full spectral data of the SW transfer matrix and is left for future work.

 For the finite size scaling exponent, the exponent is $\psi^{(SW)}_{4-Potts}=0.288(4)$ following the same relative positions depending on $q$ as the $\mu^{(SW)}$.

\begin{figure}[H]
  \centering
  \includegraphics[width=0.48\textwidth]{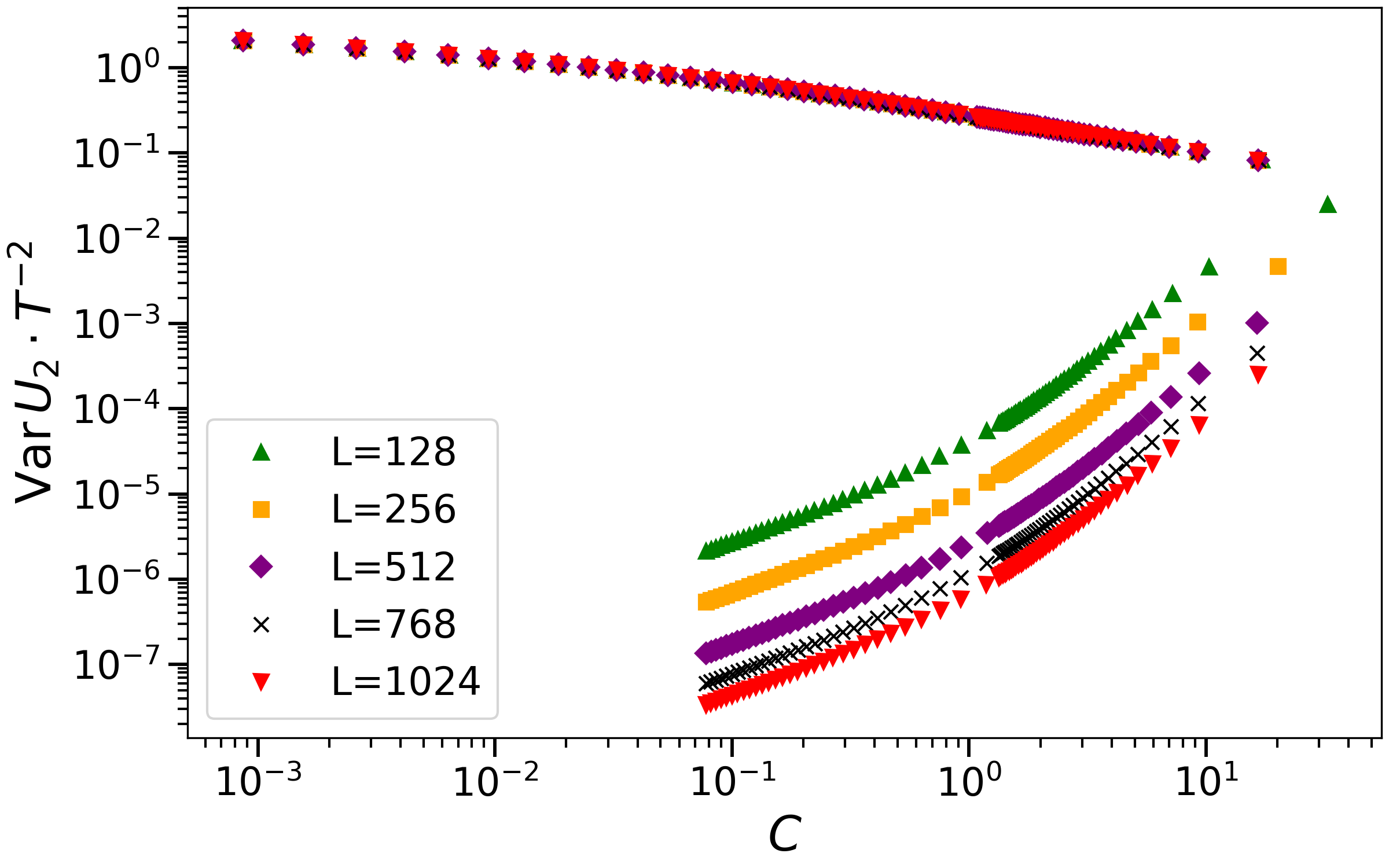}
  \caption{Swendsen--Wang updates, four-state Potts model. Rescaled variance $T^{-2}\mathrm{Var}(U_2)$ versus specific heat, for several lattice sizes.}
  \label{fig:sw-potts4-var-Cv}
\end{figure}

 For the four-state Potts model we plot the rescaled variance $T^{-2}\mathrm{Var}(U_2)$  against the heat capacity again (Fig.~\ref{fig:sw-potts4-var-Cv}). For this case the cusp-like structure is still present, in parallel with the Ising and three-state Potts cases.

\begin{figure}[H]
    \centering
    \includegraphics[width=0.48\textwidth]{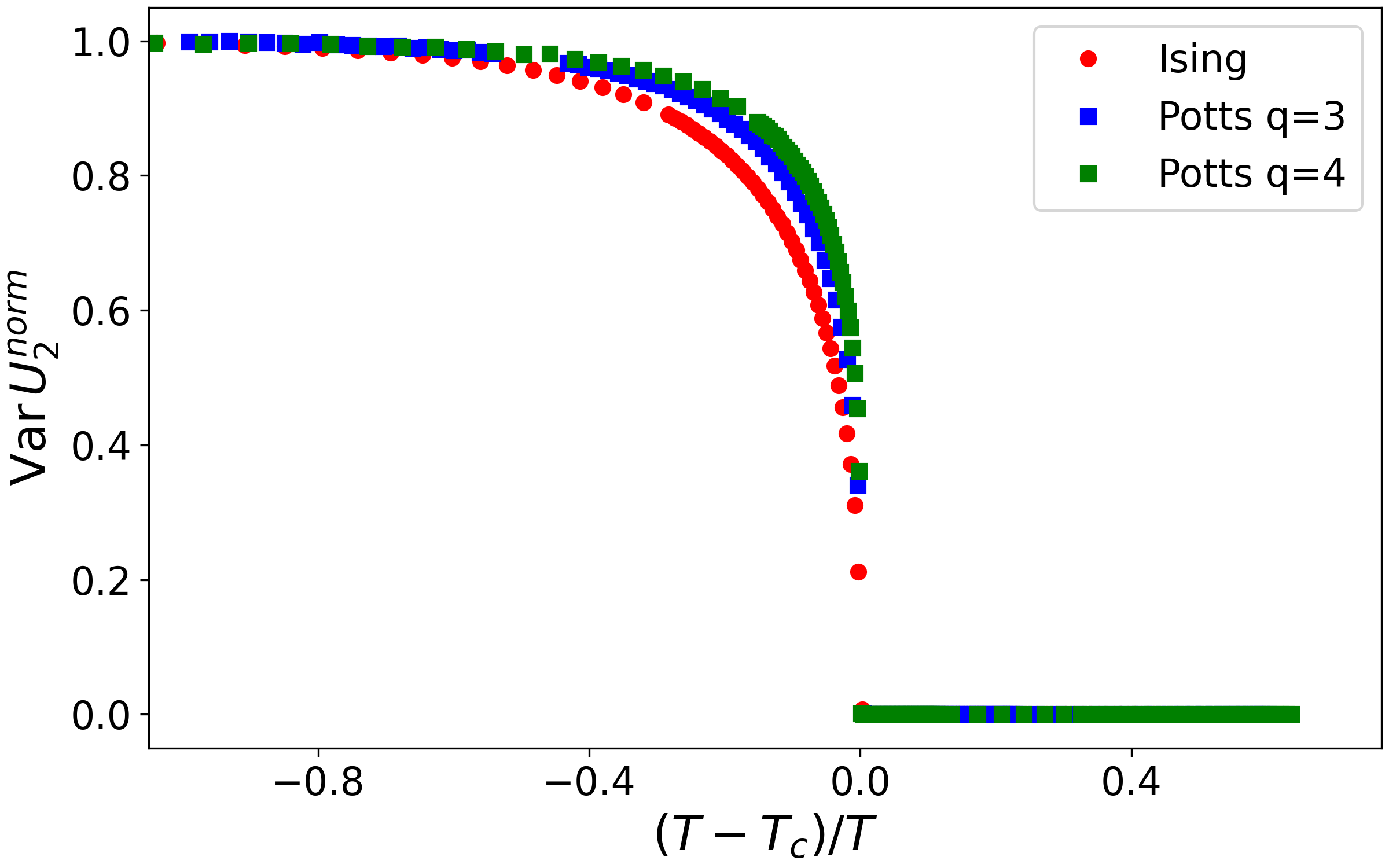}
    \caption{$\mathrm{Var}\,U^{norm}_2$ as a function of $(T - T_c)/T$ for the Ising, 3-state Potts and 4-state Potts models using the Swendsen--Wang algorithm at $L=1024$. The vertical dashed line at $(T-T_c)/T = 0$ indicates the critical temperature $T_c$.}
    \label{fig:sw_var_t2}
\end{figure}

The thermodynamic nature of the overlap fluctuations in Swendsen--Wang updates  can also be shown in terms of the reduced temperature variable ( Figure~\ref{fig:sw_var_t2}). Note that the vertical axis is normalized by the overlap value at low temperatures, $\mathrm{Var}\,U^{norm}_2 = \mathrm{Var}\,U_2(T)/\mathrm{Var}\,U_2(-\infty)$.

Below $T_c$, fluctuations are large due to the coexistence of large clusters, whereas above $T_c$ they decay rapidly as clusters become finite and uncorrelated. The similarity of the curves for Ising and Potts models indicates that the overlap variance in the Swendsen--Wang algorithm also reflects universal geometric features of cluster percolation and critical fluctuations, encoding the system's response to stochastic updates of the configuration.

% ===== SW: variance vs (T - Tc)/T =====

 \subsection{Configuration overlaps in the Metropolis algorithm}

For Metropolis updates, the configuration overlap $U_2$ measures the fraction of spins remaining unchanged after two full steps of the algorithm. One step is defined as $N=L^2$ local spin-flip attempts, see Eq.~(\ref{eq:overlap_general}). 

\subsubsection*{Metropolis updates: Ising model}

\begin{figure}[H]
  \centering
  \includegraphics[width=0.48\textwidth]{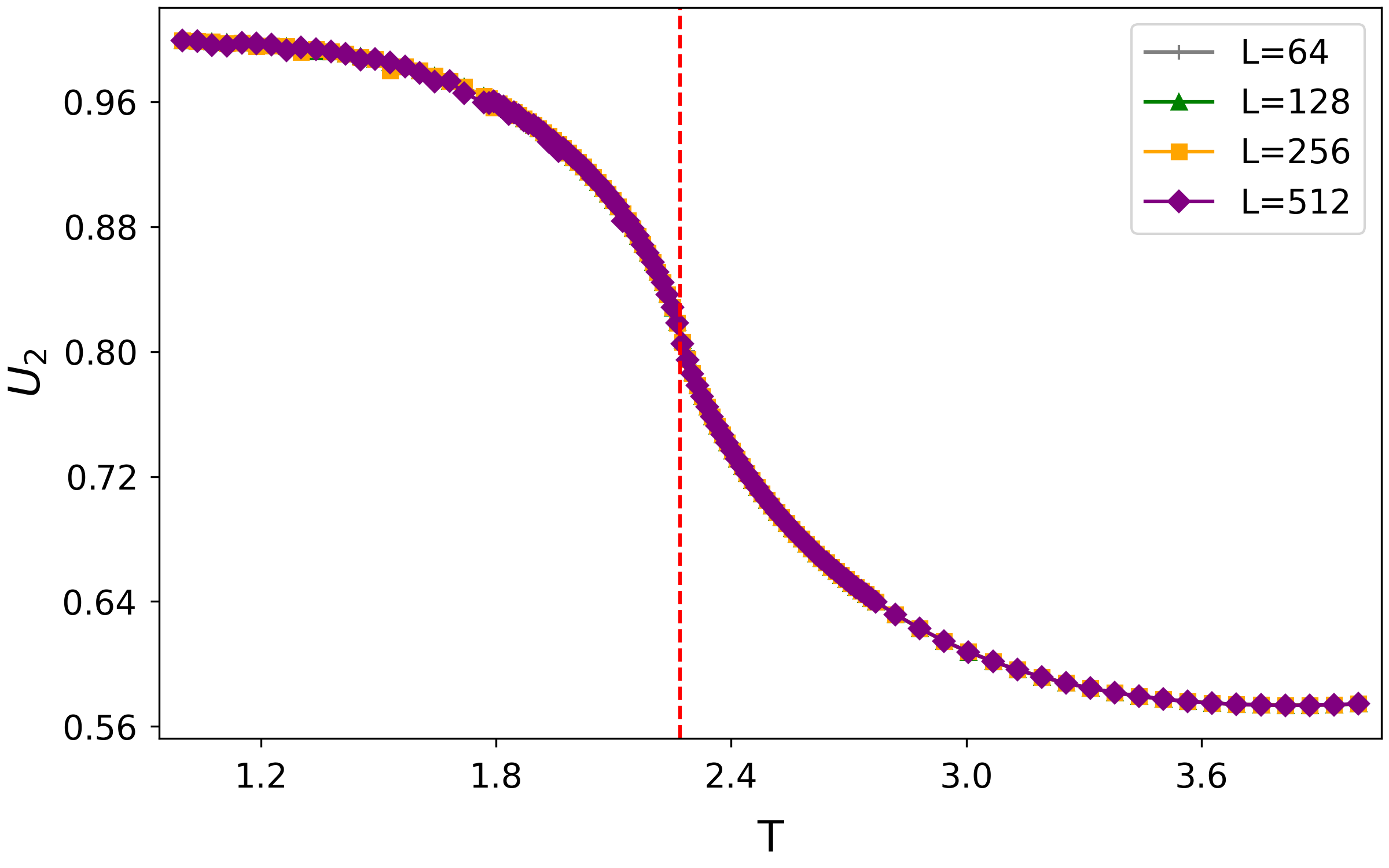}\\[4pt]
  \includegraphics[width=0.48\textwidth]{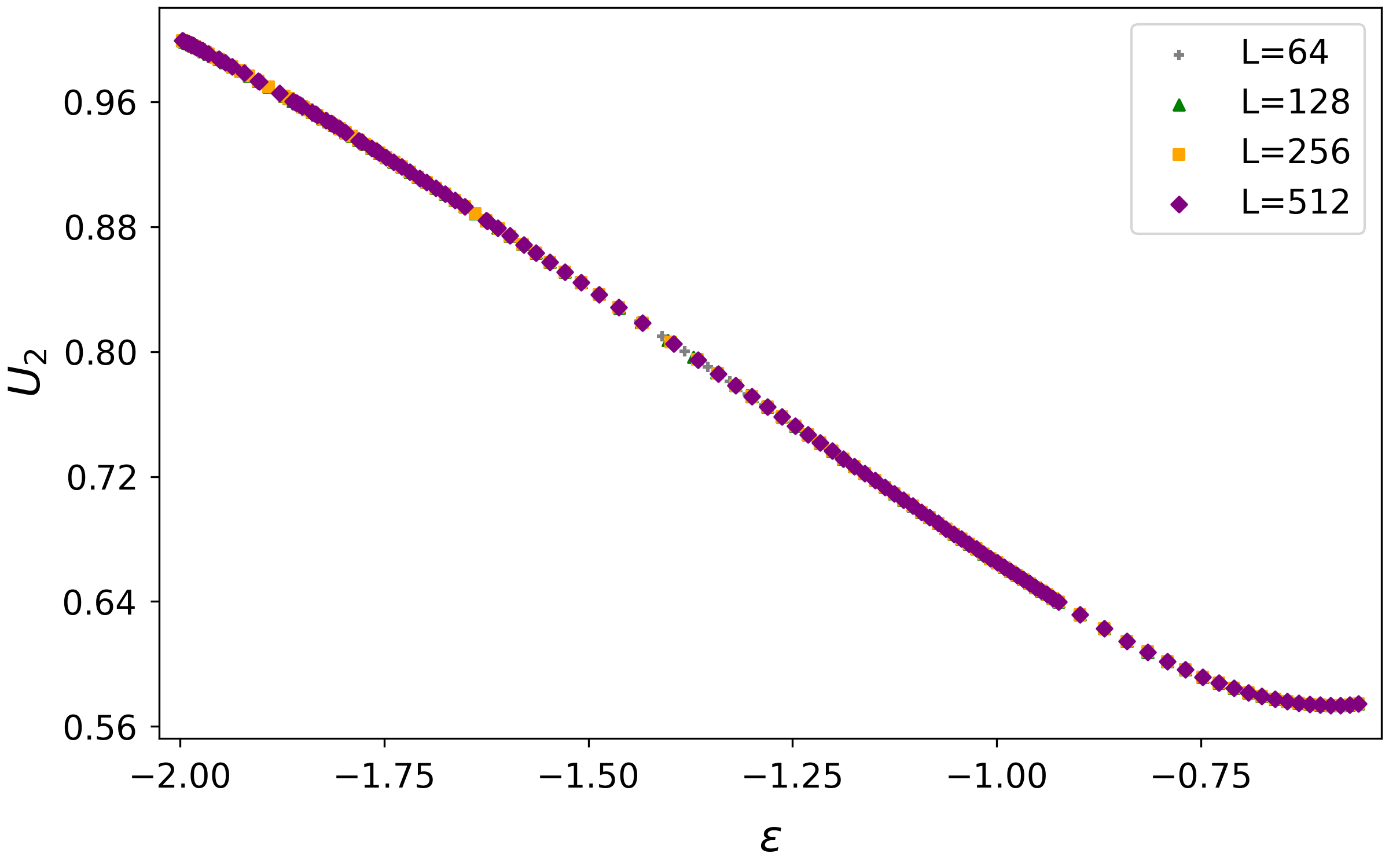}
  \caption{Metropolis updates, Ising model. Top: two-step overlap as a function of temperature. 
  Bottom: the same observable as a function of energy per spin. The vertical dashed line in the top panel indicates the critical temperature $T_c$.}
  \label{fig:metro-ising-mean}
\end{figure}

In the two-dimensional Ising model, the mean overlap decreases smoothly with increasing temperature and forms a nearly size-independent curve when plotted against temperature or energy density (Fig.~\ref{fig:metro-ising-mean}). This behavior confirms that the overlap acts as a thermodynamic function of state, rather than as a purely dynamic algorithmic quantity.

\begin{figure}[H]
  \centering
  \includegraphics[width=0.48\textwidth]{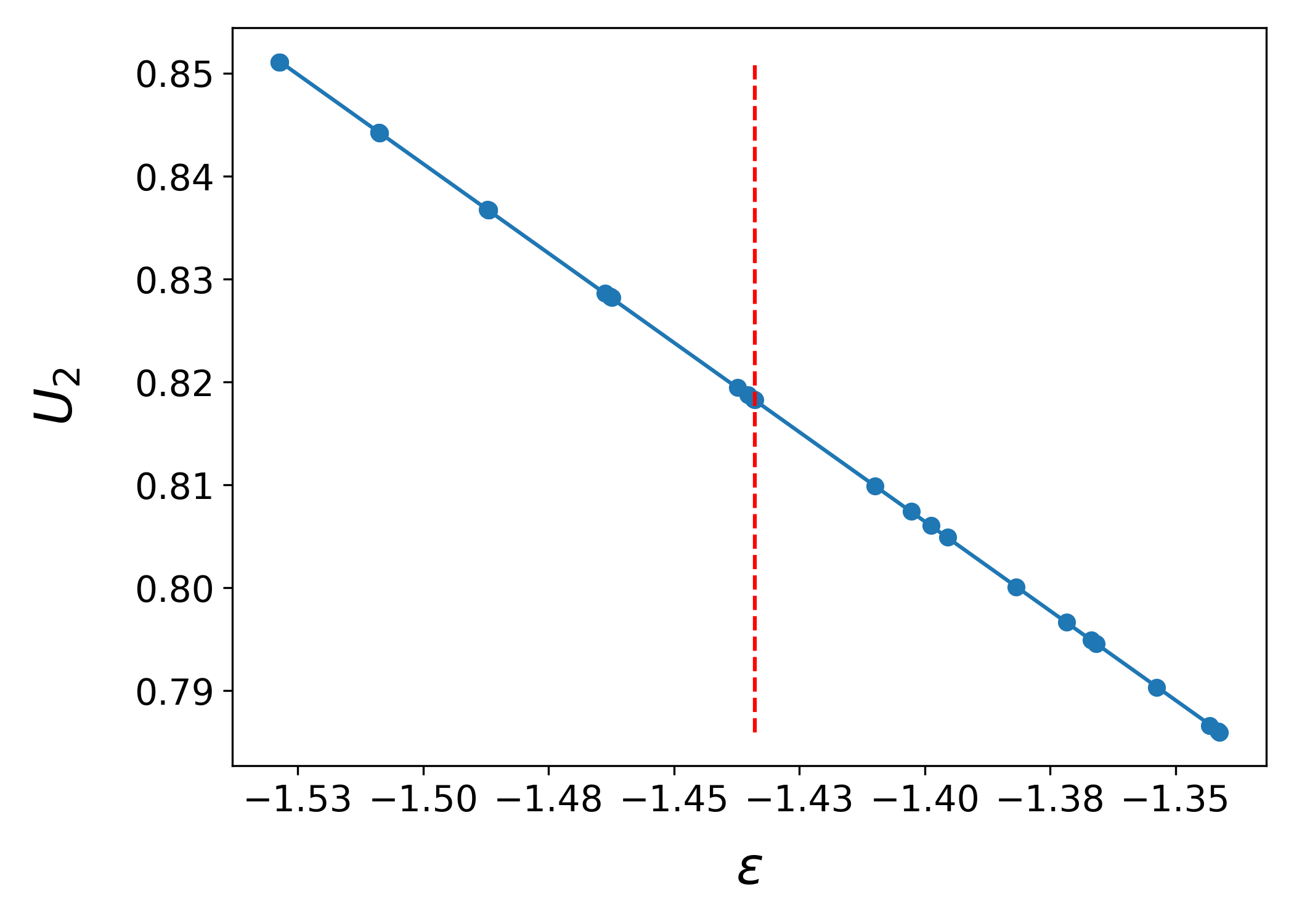}
  \caption{Metropolis updates, Ising model. Local linear fit of the mean overlap as a function of the energy density near $E_c$. The vertical dashed line indicates the critical energy $E_c$ corresponding to $T_c$.}
  \label{fig:metro-ising-lin}
\end{figure}

Near the critical region, the average overlap is approximately linear in energy. Figure~\ref{fig:metro-ising-lin} shows a local linear approximation near $E_c$, demonstrating that the overlap remains differentiable during the phase transition and responds regularly to thermodynamic changes. The slope is $\xi = -0.347(3)$.

\begin{figure}[H]
  \centering
  \includegraphics[width=0.48\textwidth]{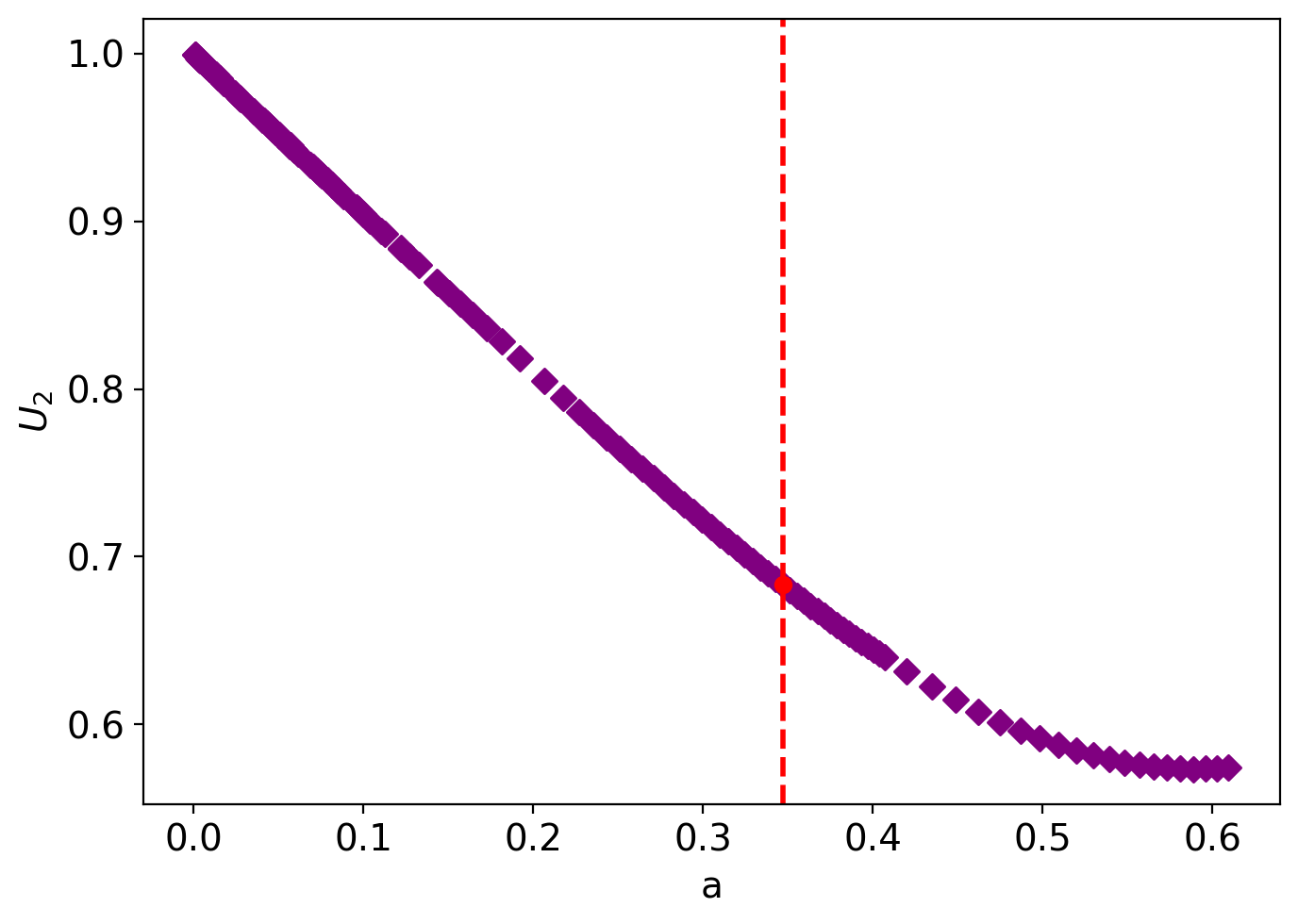}
  \caption{Metropolis updates, Ising model. $U_2$ versus the Metropolis acceptance rate, $L = 512$. The red dashed vertical line indicates the value of the acceptance rate at the critical temperature $T_c$ ($a = 0.347$), where $U_2(a)$ starts to deviate from the linear dependence in the high-temperature phase.}
  \label{fig:metro-ising-acceptance}
\end{figure}

\begin{figure}[H]
  \centering
  \includegraphics[width=0.48\textwidth]{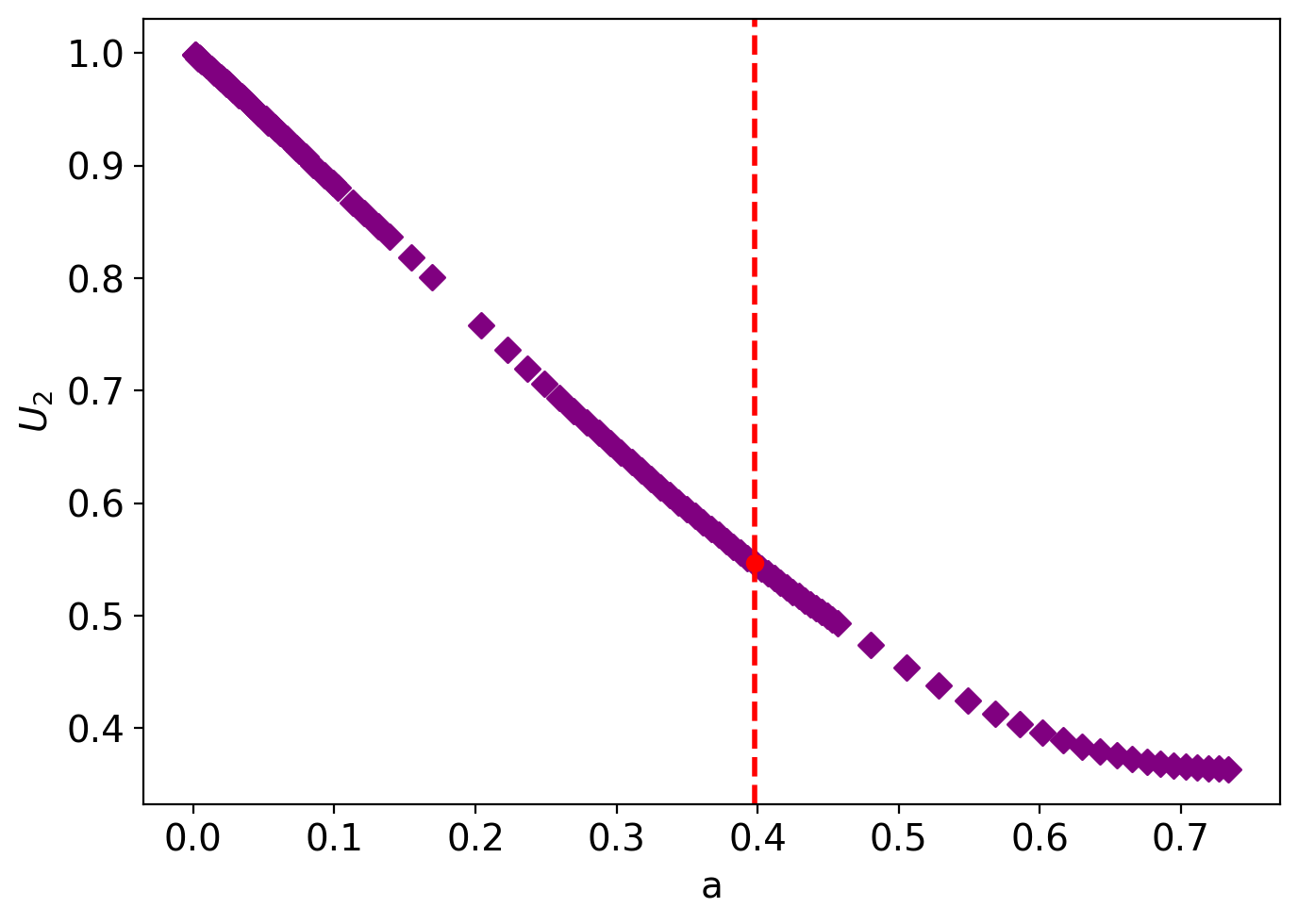}
  \caption{Metropolis updates, Potts model. $U_2$ versus the Metropolis acceptance rate, $L = 512$. The red dashed vertical line indicates the value of the acceptance rate at the critical temperature $T_c$ ($a = 0.397$), where $U_2(a)$ starts to deviate from the linear dependence.}
  \label{fig:metro-potts-acceptance}
\end{figure}

% {\color{red} \Large add comparison with spin frequency}
The relation between the average overlap $U_2$ and the Metropolis acceptance rate~\cite{barash2017b} provides further insight into the thermodynamic meaning of the overlap (Figs.~\ref{fig:metro-ising-acceptance} and~\ref{fig:metro-potts-acceptance}). The acceptance rate at the critical point is $a=0.347$ for the Ising model and $a=0.397$ for the three-state Potts model~\cite{barash2017b}. Note the similarity of the curves in Figures~\ref{fig:metro-ising-mean} and~\ref{fig:metro-ising-acceptance} for the Ising model, and of the pairs of curves in Figures~\ref{fig:metro-potts-mean} and~\ref{fig:metro-potts-acceptance} for the three-state Potts model.

In both models, $U_2$ exhibits a nearly linear dependence on the average acceptance probability at lower temperatures. For higher values of $T$, and thus of the acceptance rate, $U_2$ seems to retain some between-states dependence, and thus $U_2(a)$ deviates from linearity.

\begin{figure}[H]
  \centering
  \includegraphics[width=0.48\textwidth]{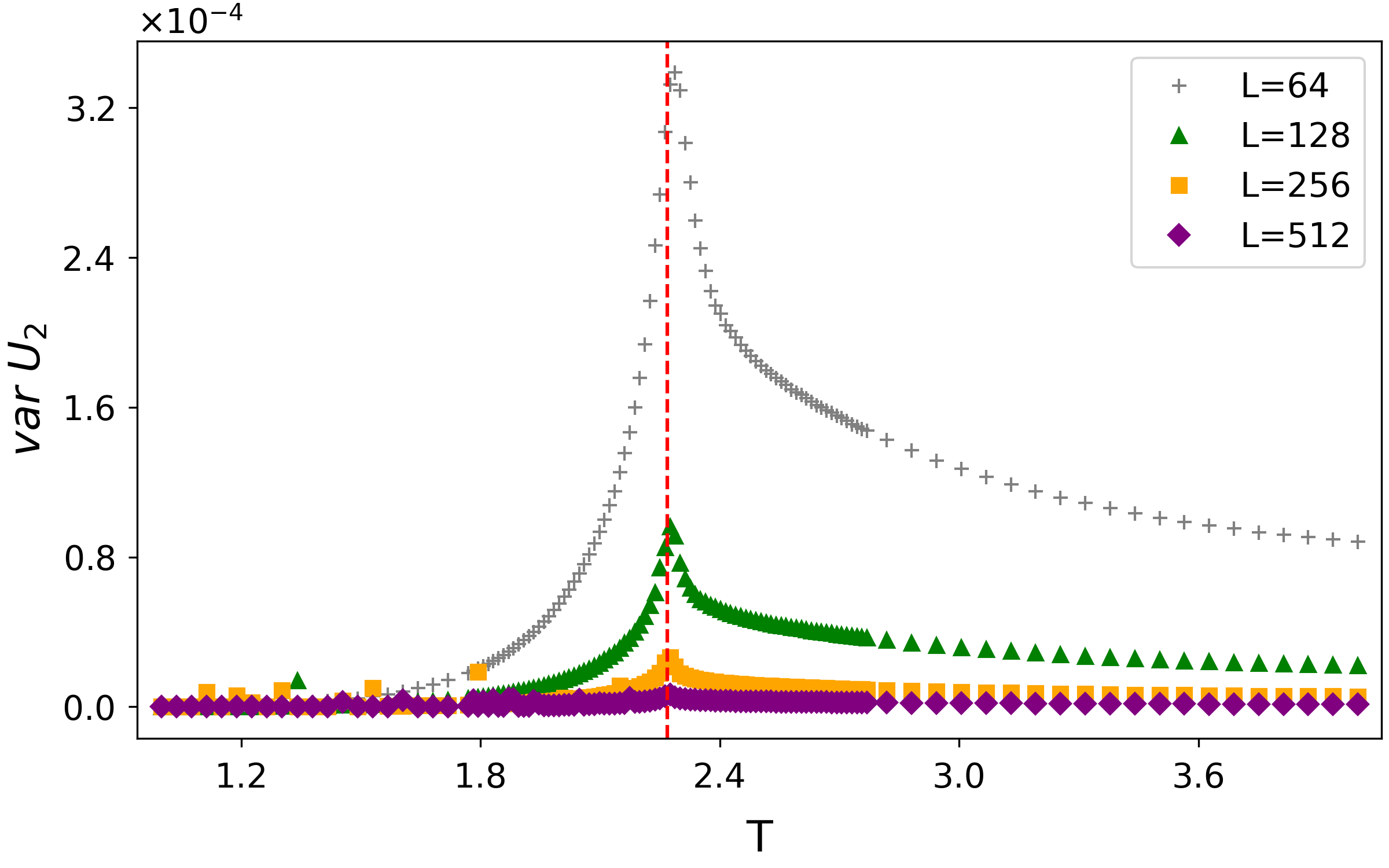}\\[4pt]
  \includegraphics[width=0.48\textwidth]{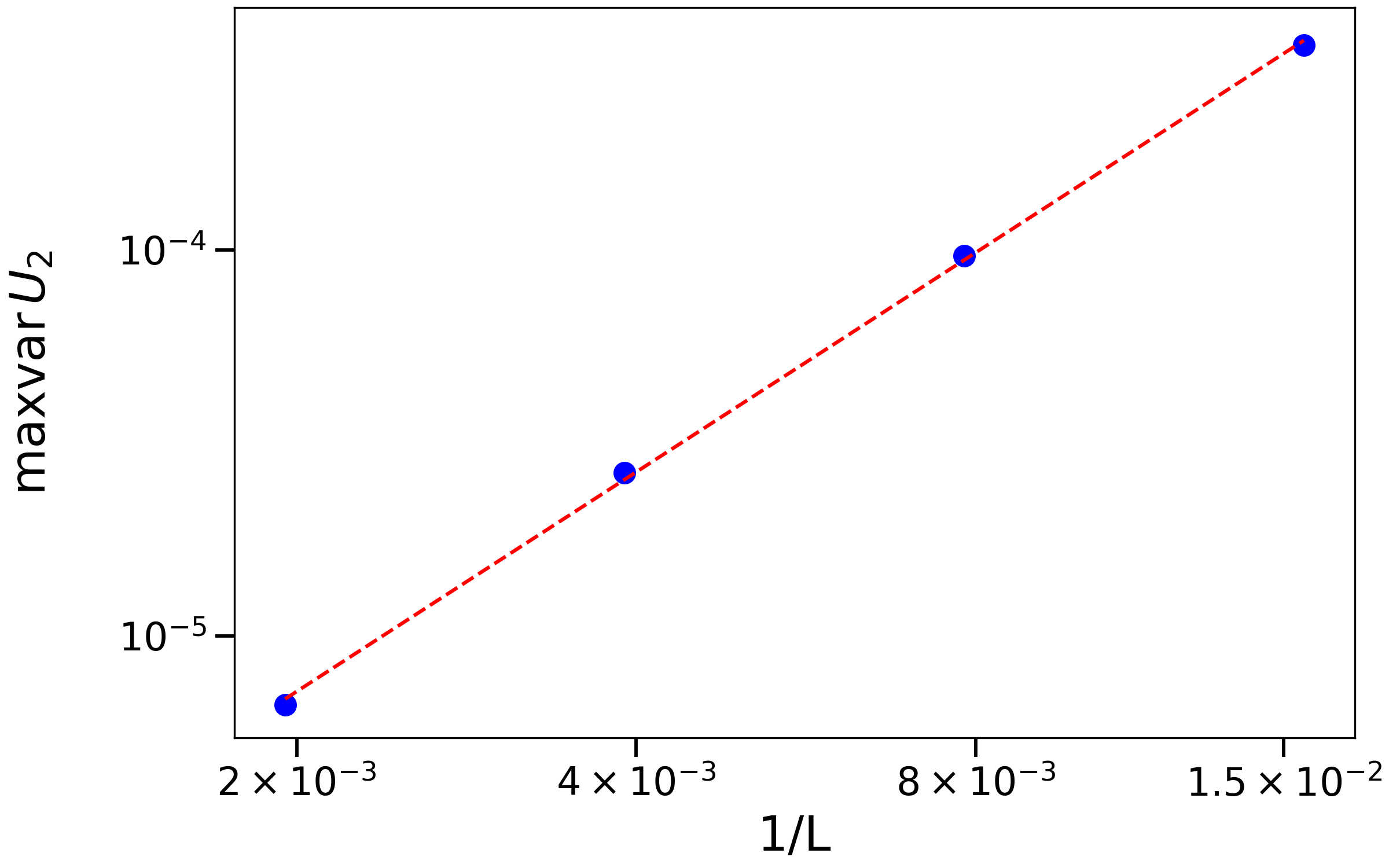}
  \caption{Metropolis updates, Ising model. Top: the variance of the two-step overlap as a function of 
  temperature. Bottom: the scaling of the peak variance with system size. The vertical dashed line in the top panel indicates the critical temperature $T_c$.}
  \label{fig:metro-ising-var}
\end{figure}

The overlap variance exhibits a peculiarity in the critical region: at $T_c$, a pronounced peak is observed whose height decreases with the system size (Fig.~\ref{fig:metro-ising-var}, top). The scaling of the peak (Fig.~\ref{fig:metro-ising-var}, bottom) follows a power law in $L$. The magnitude of the power-law decay is approximately 2, which corresponds to the lattice dimension. Therefore, it is a geometrical finite-size effect. 

\subsubsection*{Metropolis updates: three-state Potts model}

\begin{figure}[H]
  \centering
  \includegraphics[width=0.48\textwidth]{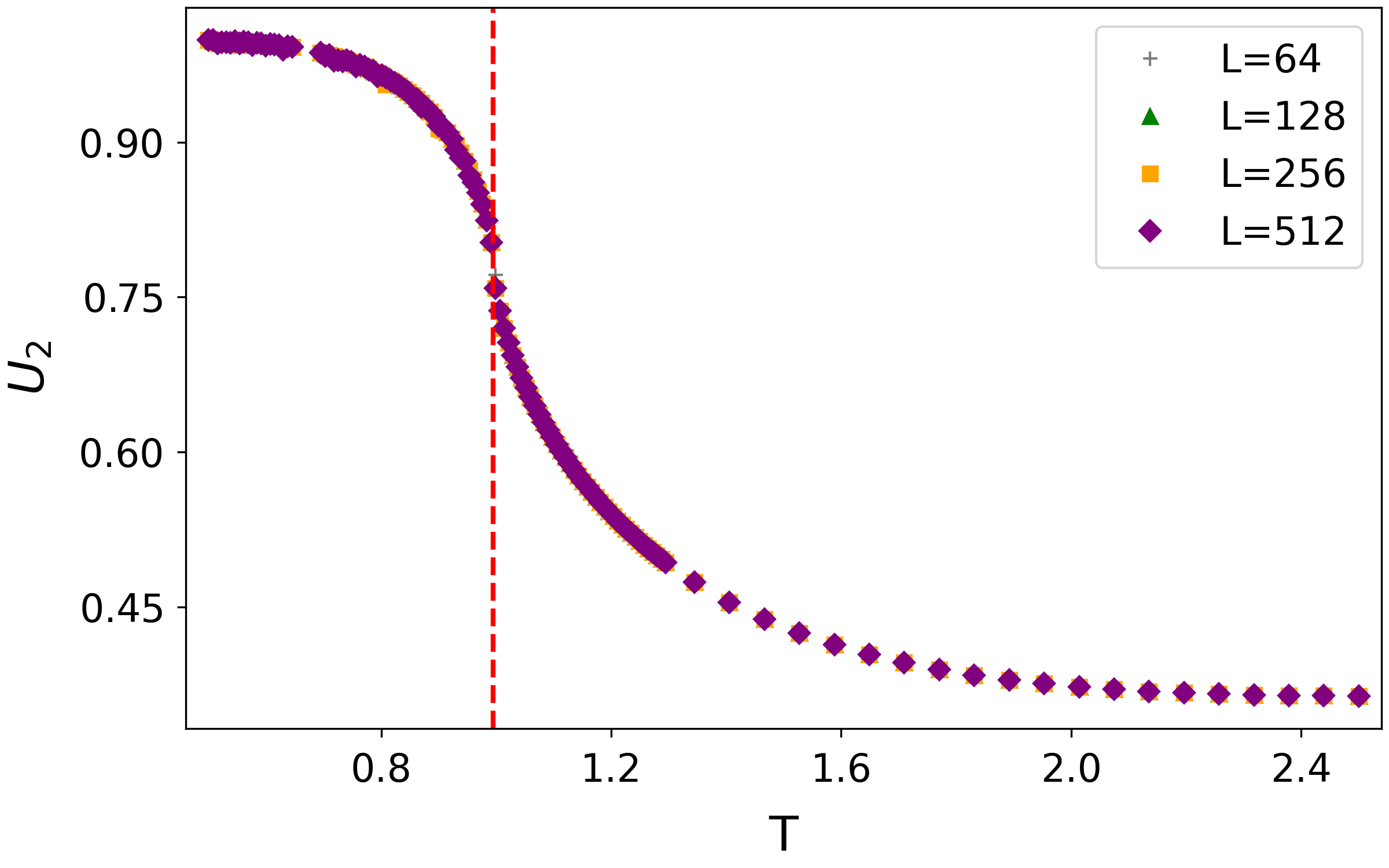}\\[4pt]
  \includegraphics[width=0.48\textwidth]{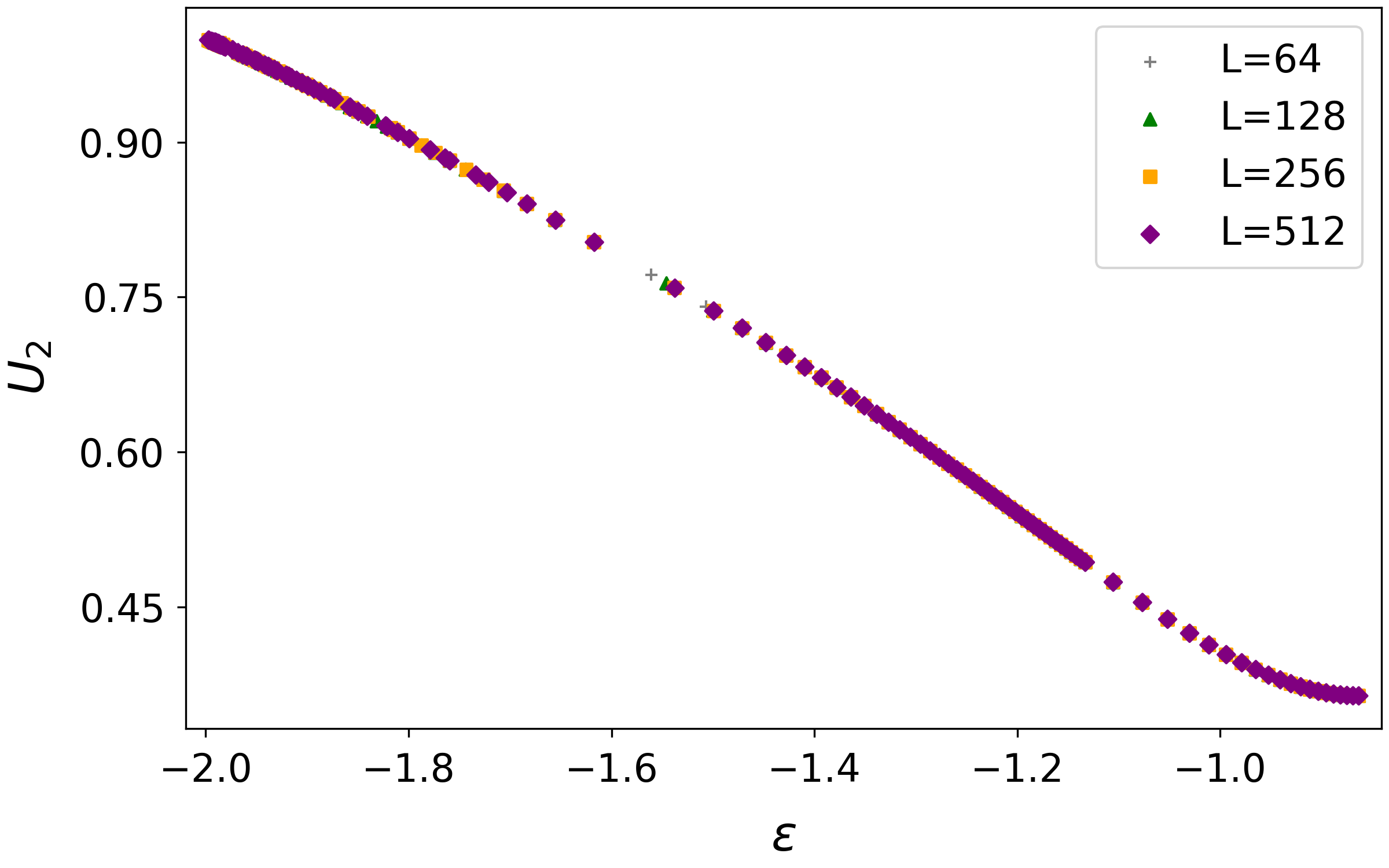}
  \caption{Metropolis updates, three-state Potts model. Top: the average two-step overlap as a function of temperature. Bottom: the average overlap as a function of energy density. The vertical dashed line in the top panel indicates the critical temperature $T_c$.}
  \label{fig:metro-potts-mean}
\end{figure}

Applying the same analysis to the three-state Potts model yields similar qualitative behavior. The average overlap again varies smoothly with temperature and vanishes when plotted versus energy (Fig.~\ref{fig:metro-potts-mean}), supporting its interpretation as a thermodynamic quantity.


\begin{figure}[H]
  \centering
  \includegraphics[width=0.48\textwidth]{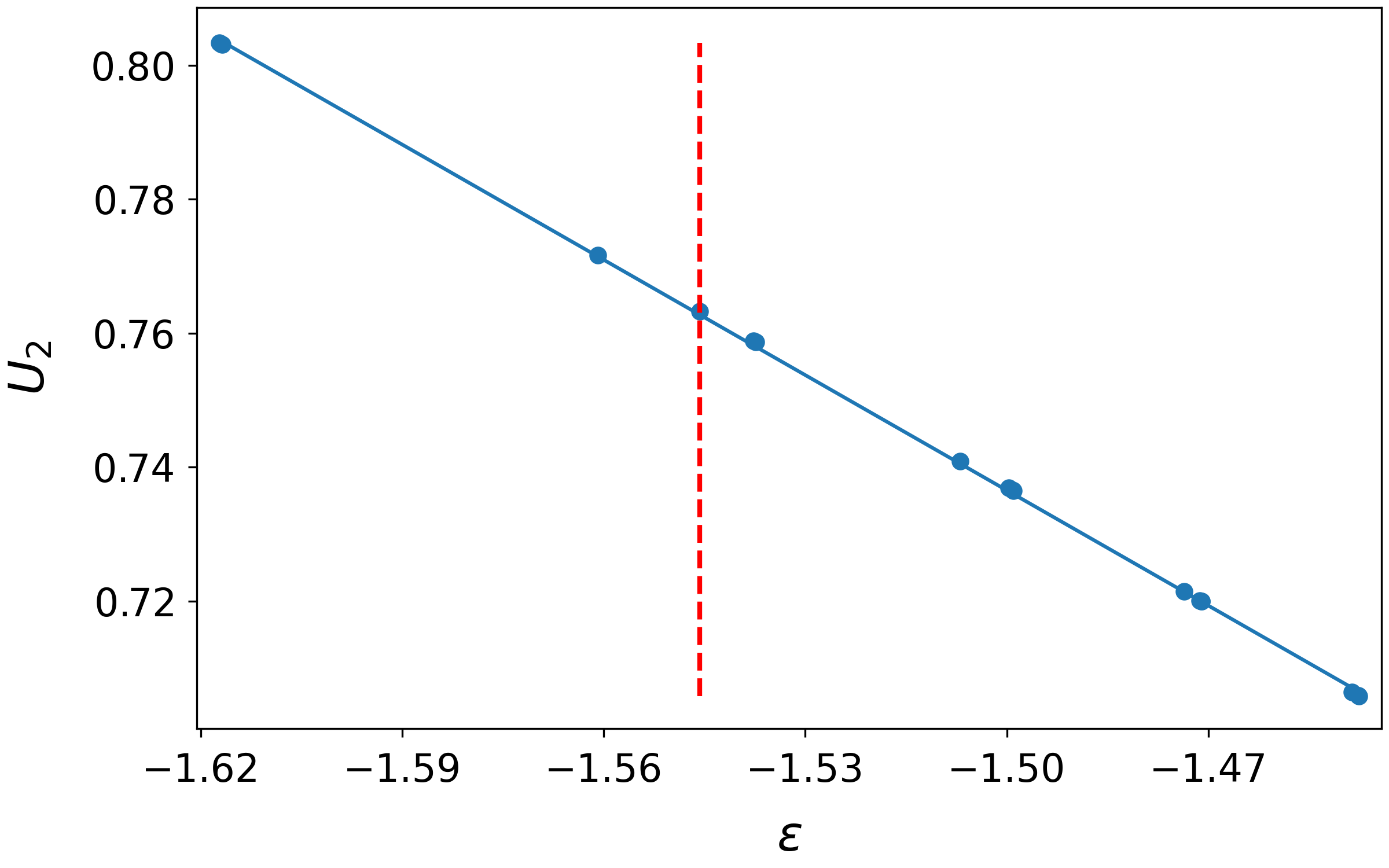}
  \caption{Metropolis updates, three-state Potts model. Local linear dependence of the mean overlap on the energy density near the critical region. The vertical dashed line indicates the critical energy $E_c$ corresponding to $T_c$.}
  \label{fig:metro-potts-lin}
\end{figure}

Near the transition, the energy dependence is again locally linear, but with a slope of $-0.573(5)$ (Fig.~\ref{fig:metro-potts-lin}), which coincides with the slope for the Ising model.

\begin{figure}[h]
  \centering
  \includegraphics[width=0.48\textwidth]{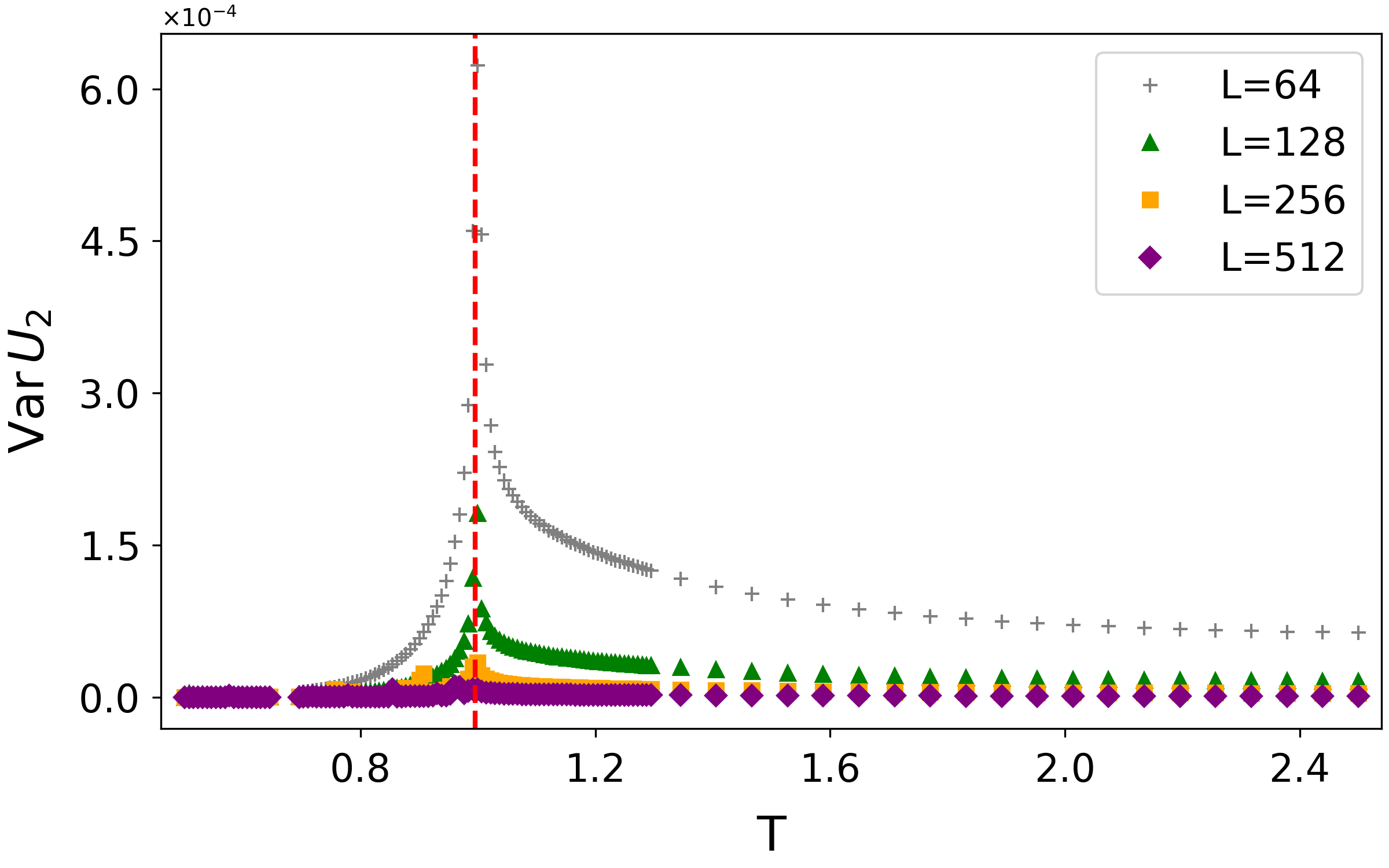}\\[4pt]
  
  \caption{Metropolis updates, three-state Potts model. Variance of the two-step overlap versus temperature. The vertical dashed line in the top panel indicates the critical temperature $T_c$.}
  \label{fig:metro-potts-var}
\end{figure}

The variance exhibits a singularity in the critical region (Fig.~\ref{fig:metro-potts-var}, top). The maximum value decreases inversely proportional to the square of the linear lattice size, just as in the Ising model.

\begin{figure}[h]
  \centering
\includegraphics[width=0.48\textwidth]{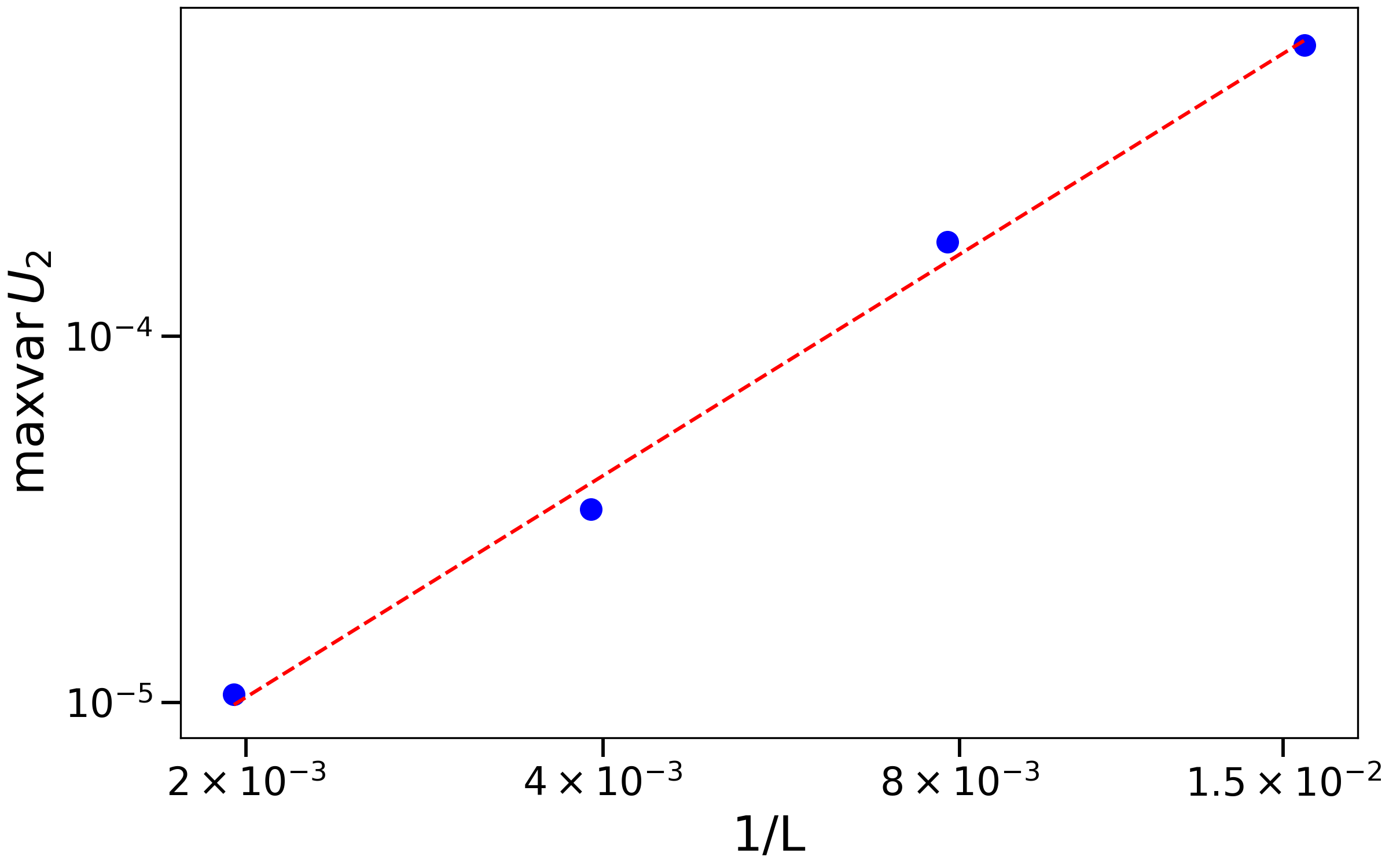}
 \caption{ Finite-size scaling of the peak variance, differing from the Ising case due to the Potts heat-capacity exponent. }
  \label{fig:metro-potts-var-atTc}
\end{figure}

\section*{Multistep overlaps}

To verify that the observed thermodynamic behavior is not specific to the two-step overlap $U_2$, we analyzed multistep overlaps $U_n$ for several values of $n$ in both Swendsen--Wang and Wolff updates for the Ising and three-state Potts models. In all cases, the qualitative behavior remains unchanged: the temperature dependence sharpens with increasing system size, and the critical region is clearly identifiable from either the mean (Wolff) or the variance (Swendsen--Wang).

For the Wolff algorithm, the mean geometric overlap $\langle U_n\rangle$ decreases systematically with increasing $n$, but preserves the same critical structure as $U_2$. The family of curves collapses toward a limiting curve as $n$ increases, indicating convergence to a stationary overlap value. The approach to the infinite-step limit is well described by an exponential form
\begin{equation}
U_n(T) = U_\infty(T) + A(T)e^{-n/\tau(T)},
\label{eq:extrapolation}
\end{equation}
allowing reliable extrapolation to $n\to\infty$. The resulting infinite-step overlaps exhibit the same singular structure near $T_c$ as the finite-$n$ data.

\begin{figure}[ht]
  \centering
  \includegraphics[width=0.95\columnwidth,clip=false]{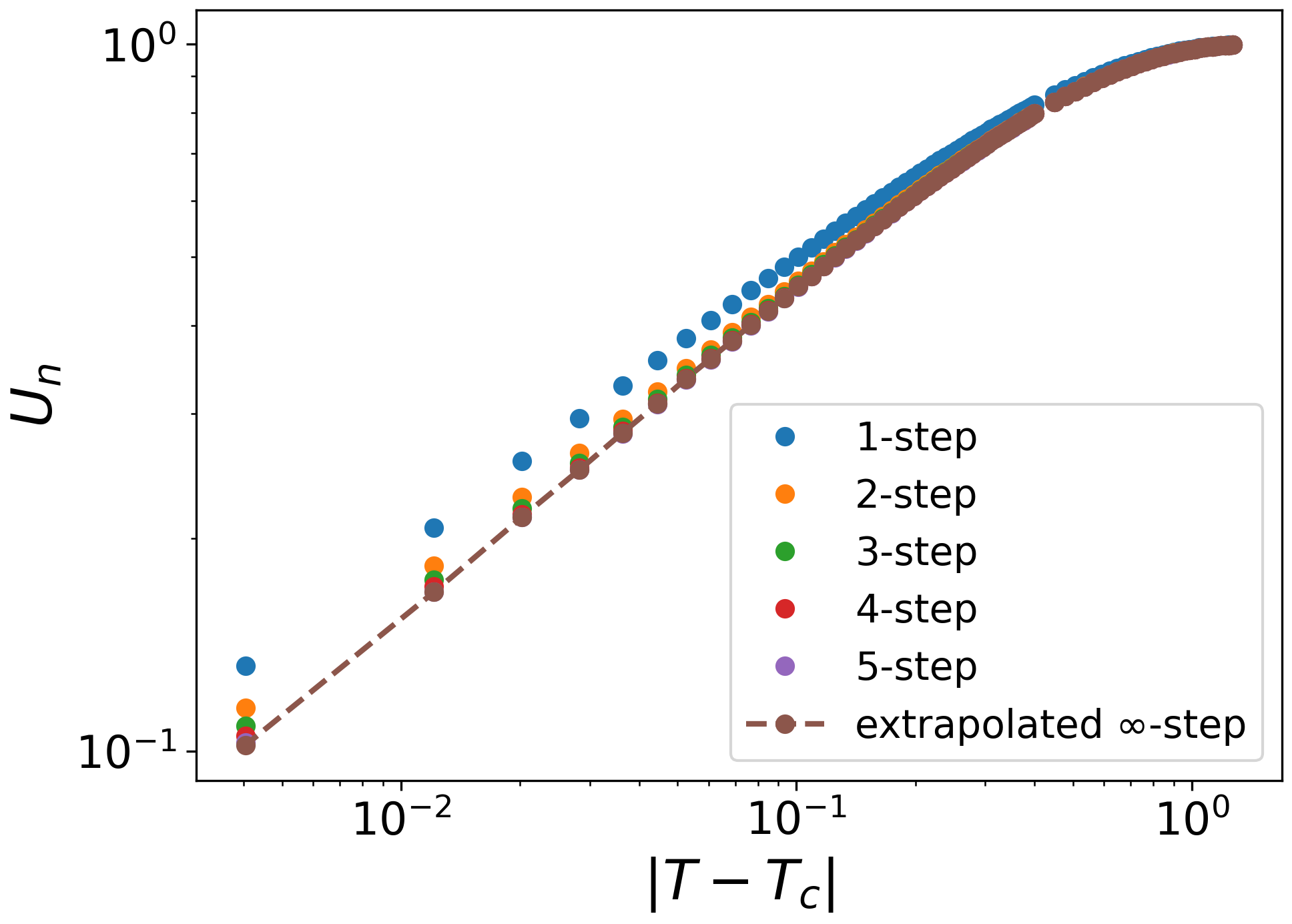}
  \caption{Wolff updates, Ising model. Multistep cluster overlaps $\langle U_n\rangle$ versus temperature for several $n$. The curves converge exponentially toward an infinite-step limit while preserving the critical structure.}
  \label{fig:wolff-ising-multistep}
\end{figure}

\begin{figure}[H]
  \centering
  \includegraphics[width=0.95\columnwidth,clip=false]{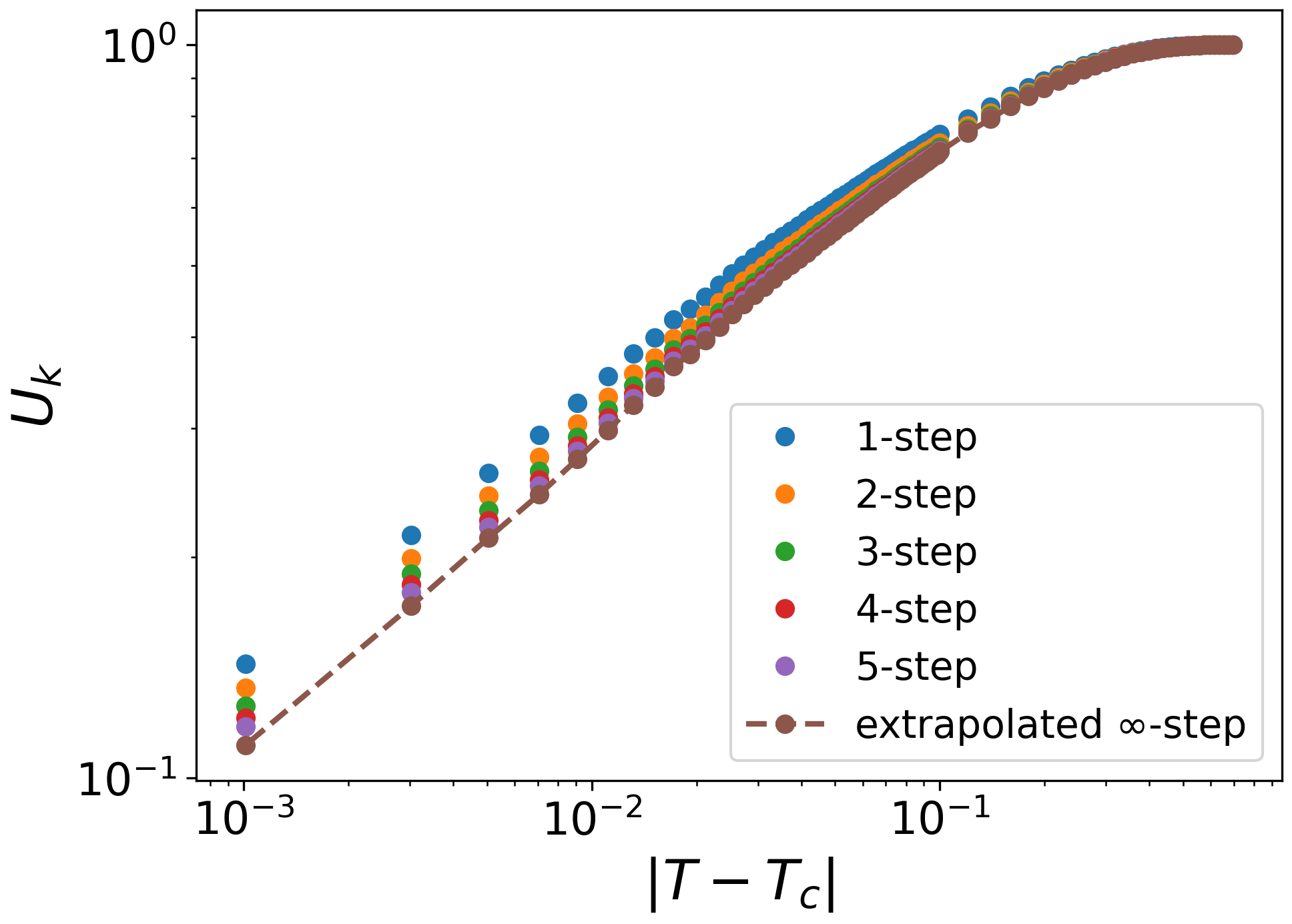}
  \caption{Wolff updates, three-state Potts model. Multistep cluster overlaps $\langle U_n\rangle$ versus temperature. Exponential convergence toward $U_\infty(T)$ is observed across the full thermodynamic range.}
  \label{fig:wolff-potts-multistep}
\end{figure}

For Swendsen--Wang updates, where the mean overlap remains near its random value, the variance $\mathrm{Var}(U_n)$ continues to act as the thermodynamic indicator. As $n$ increases, the variance curves approach a limiting profile, again following an exponential convergence in $n$. The suppression of fluctuations above $T_c$ and their enhancement below $T_c$ persist for all $n$, demonstrating that the susceptibility-like character of overlap fluctuations is robust under multistep generalization.

\begin{figure}[H]
  \centering
  \includegraphics[width=0.95\columnwidth,clip=false]{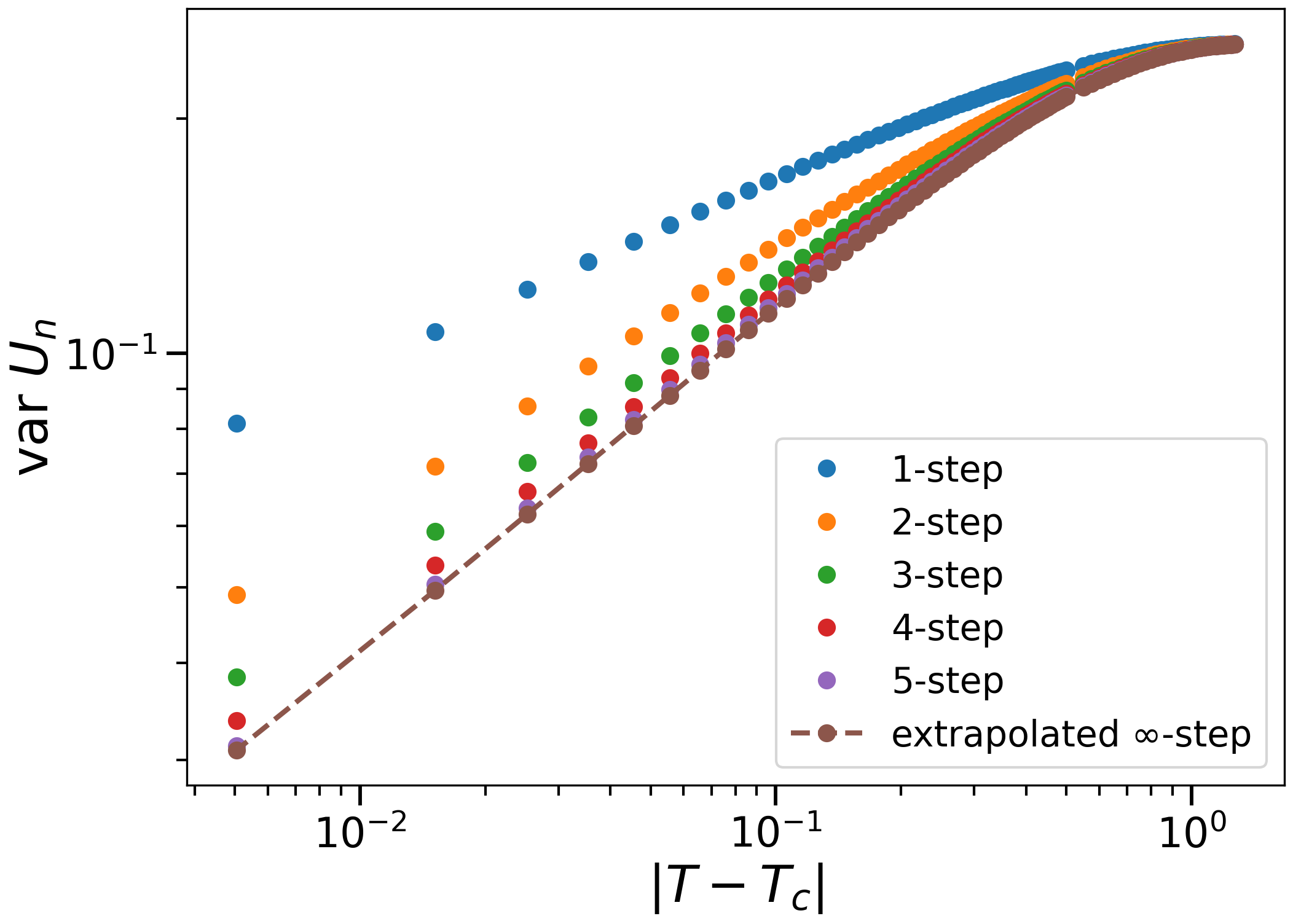}
  \caption{Swendsen--Wang updates, Ising model. Variance of the multistep overlap $\mathrm{Var}(U_n)$ versus temperature. Increasing $n$ leads to exponential convergence toward a limiting curve while preserving the sharp critical suppression at $T_c$.}
  \label{fig:sw-ising-multistep}
\end{figure}

\begin{figure}[H]
  \centering
  \includegraphics[width=0.95\columnwidth,clip=false]{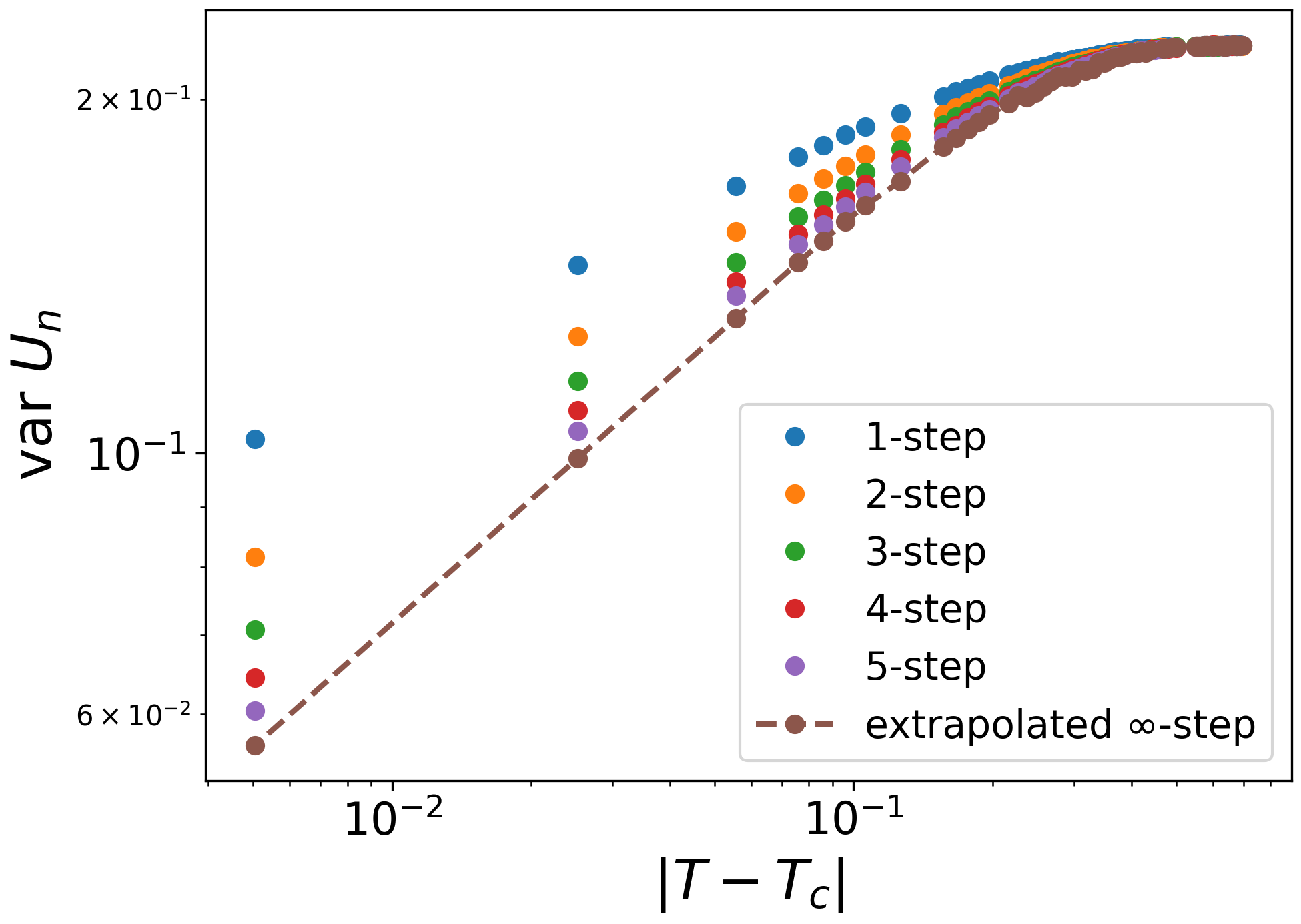}
  \caption{Swendsen--Wang updates, three-state Potts model. Multistep overlap variance $\mathrm{Var}(U_n)$ versus temperature. The thermodynamic signature of the transition remains clearly visible for all $n$.}
  \label{fig:sw-potts-multistep}
\end{figure}

Overall, multistep overlaps exhibit the same thermodynamic structure as the two-step observable $U_2$. The exponential approach to an infinite-step limit demonstrates that the overlap is not merely a short-time dynamical artifact but a well-defined stationary quantity of the Markov chain. Both the limiting mean (Wolff) and the limiting variance (Swendsen--Wang) retain the same universality-class--dependent singular behavior identified for $U_2$, reinforcing the interpretation of algorithmic overlaps as genuine thermodynamic observables.

\section*{Discussion}

\begin{table*}[t]
\caption{\label{tab:exponents}
Critical exponents of the two-dimensional Ising, three-state and four-state Potts models together with the algorithmic exponents extracted in this work from the finite-size scaling of the two-step algorithmic overlap $U_2$.}
\begin{ruledtabular}
\begin{tabular}{lccc}

 & $q=2$ (Ising) & $q=3$ Potts & $q=4$ Potts \\
\hline
\multicolumn{4}{l}{\emph{Standard static exponents (2D Potts, exact)}}\\
$\alpha$           & $0$ (log)   & $1/3$    & $2/3$         \\
$\beta$            & $1/8$       & $1/9$    & $1/12$        \\
$\gamma$           & $7/4$       & $13/9$   & $7/6$         \\
$\nu$              & $1$         & $5/6$    & $2/3$         \\
$\eta$             & $1/4$       & $4/15$   & $1/4$         \\
$\delta$           & $15$        & $14$     & $15$          \\
$\beta/\nu$        & $1/8$       & $2/15$   & $1/8$         \\
$\gamma/\nu$       & $7/4$       & $26/15$  & $7/4$         \\
$d_F = 2-\beta/\nu$& $15/8$      & $28/15$  & $15/8$ (log)  \\
\hline
\multicolumn{4}{l}{\emph{This work --- Wolff updates}}\\
$\mu^{(W)}$ from $U_2 \sim (T_c-T)^{\mu^{(W)}}$ & $0.428(3)$  & $0.352(5)$ & $0.281(2)$ \\
$\psi^{(W)}$ from FSS of $U_2(T_c)$ vs. $1/L$   & $0.4240(6)$ & $0.425(4)$ & $0.419(1)$ \\
\hline
\multicolumn{4}{l}{\emph{This work --- Swendsen--Wang updates}}\\
$\mu^{(SW)}$ from $\mathrm{Var}(U_2)\sim(T_c-T)^{\mu^{(SW)}}$ & $0.348(4)$  & $0.266(3)$ & $0.205(8)$ \\
$\psi^{(SW)}$ from FSS of $\mathrm{Var}(U_2)$ at $T_c$ vs. $1/L$ & $0.3458(9)$ & $0.318(1)$ & $0.288(4)$ \\

\end{tabular}
\end{ruledtabular}
\end{table*}

We have formulated the problem of determining the geometric properties of the dynamics of three Monte Carlo algorithms for simulating spin systems. All three methods --- the local Metropolis method~\cite{metropolis1953}, the Swendsen--Wang cluster method~\cite{swendsen1987}, and the single-cluster Wolff method~\cite{wolff1989} --- satisfy detailed balance~\cite{landau2005}. At the same time, the relaxation time exponent $z$ differs and decreases in the following order: Metropolis, Swendsen--Wang, Wolff (see, for example, the book~\cite{landau2005}). The traditional explanation is that cluster methods have an advantage in the critical region because they take into account the effect of the growing correlation length. However, the significantly smaller value of the critical exponent remains not fully understood. A possible explanation should be sought in the geometric aspects of the relaxation dynamics of configurations.

In the present paper, we examined the relaxation dynamics by studying the spatial overlap of successive configurations of the systems. We studied the relaxation dynamics of successive configurations for the two-dimensional Ising model and, to avoid peculiarities specific to this model, also for the three-state Potts model. This allowed us to identify some characteristic features of the geometric properties, and we can state that the Swendsen--Wang method, when analyzing such properties, occupies an intermediate position between the local Metropolis method and the single-cluster Wolff method. This is related to the fact that the Wolff method uses only a single cluster in relaxation, whose average size in the critical region behaves similarly to the magnetic susceptibility. In contrast, in the Metropolis method and the Swendsen--Wang method, all spins of the system are updated at each Monte Carlo step.

The overlap functions $U_n$ that we propose behave differently in the three algorithms studied. At the same time, these functions carry information both about the phase transition and about the specific features of the construction of the overlap functions $U_n$. The most interesting is the behavior of the overlap function at two successive steps, $U_2$. For the Wolff algorithm, this function changes from 1 at low temperatures to 0 at the phase transition point (in the thermodynamic limit --- infinite system size). Interestingly, the ``critical exponents'' associated with the singularity of the $U_2$ function are indistinguishable within statistical uncertainty, suggesting that the critical dynamics described by the geometric properties of the intersection of two fractal structures at the transition point does not depend on the universality class of the phase transition. Details are given in the subsection ``Overlaps in Wolff updates'' --- see Figure~\ref{fig:wolff_overlap_t2} and its discussion, as well as the estimates of these exponents using formula~(\ref{eq:wolff_extrapolation}).

For the Wolff algorithm we extracted the overlap exponents in two independent ways: as the slope $\mu^{(W)}$ of $U_2 \sim (T_c-T)^{\mu^{(W)}}$ on approach to $T_c$, and as the finite-size-scaling slope $\psi^{(W)}$ of $U_2(T_c)$ versus $1/L$. For the Ising model these two estimates nearly coincide, $\mu^{(W)}_{\text{Ising}} = 0.428(3)$ and $\psi^{(W)}_{\text{Ising}} = 0.4240(6)$, agreeing up to their combined uncertainties. The finite-size exponent $\psi^{(W)}$ is, moreover, the same for all three models within errors, $\psi^{(W)}_{\text{Ising}} = 0.4240(6)$, $\psi^{(W)}_{3\text{-Potts}} = 0.425(4)$ and $\psi^{(W)}_{4\text{-Potts}} = 0.419(1)$, even though the static ratio $\beta/\nu$ differs between $q=2$ and $q=3$. The naive factorized estimate $\langle U_2^{(W)}\rangle|_{T_c}\sim L^{-2\beta/\nu}$ would give $1/4$, $4/15$ and $1/4$ for the three models and would not produce a common value, so the shared $\psi^{(W)}$ reflects the geometry of how two successive Wolff clusters intersect at criticality rather than a static FK quantity.

The thermal Wolff exponent $\mu^{(W)}$ behaves differently and depends on $q$, decreasing as $\mu^{(W)}_{\text{Ising}} = 0.428(3)$, $\mu^{(W)}_{3\text{-Potts}} = 0.352(5)$ and $\mu^{(W)}_{4\text{-Potts}} = 0.281(2)$, so the overlap drops more steeply on approach to $T_c$ for larger $q$. We read this as the thermal slope mixing in the model-dependent rate at which the system orders away from $T_c$, whereas the finite-size slope $\psi^{(W)}$ isolates the critical cluster geometry set by the algorithm. The Swendsen--Wang exponents depend on $q$ in both channels: $\mu^{(SW)}_{\text{Ising}} = 0.358(4)$, $\mu^{(SW)}_{3\text{-Potts}} = 0.266(3)$, $\mu^{(SW)}_{4\text{-Potts}} = 0.205(8)$, and $\psi^{(SW)}_{\text{Ising}} = 0.3458(9)$, $\psi^{(SW)}_{3\text{-Potts}} = 0.318(1)$, $\psi^{(SW)}_{4\text{-Potts}} = 0.288(4)$. As discussed around Eq.~(\ref{eq:Usw_variance}), $q$ enters the SW variance both through the random-cluster weight $q^{k(\omega)}$ and through the recoloring factor $1/q^2$, and the SW transfer operator carries a $q$-dependent spectrum, so this dependence is expected. The contrast between the two cluster algorithms is then that the single-cluster Wolff overlap yields a finite-size exponent common to all three models, while the multi-cluster SW overlap fluctuations remain tied to the model-specific FK connectivity. We also note that for SW the thermal and finite-size estimates are close for $q=2$ ($0.358(4)$ versus $0.3458(9)$) but separate for the Potts cases, with the gap largest at $q=4$.

The two-step overlap $U_2$ in the Swendsen--Wang algorithm is approximately $1/2$, but the fluctuations of this function are stronger in the ordered region and increase with decreasing temperature. Interestingly, the variance $\mathrm{Var}\,U_2$ of the two-step overlap behaves qualitatively similarly to the function $U_2$ in the Wolff algorithm, as can be seen from Figure~\ref{fig:wolff_overlap_t2} for the Wolff algorithm and from Figure~\ref{fig:sw_var_t2} for the Swendsen--Wang algorithm. However, the values of the effective exponents differ for these functions.

The behavior of the function $U_2$ when simulated by the Metropolis method differs significantly from the case of the cluster algorithms. The overlap function in the critical region behaves more like the spin-flip probability studied in work~\cite{lev2019}. This connection is shown in Figure~\ref{fig:metro-ising-acceptance}. The spin-flip probability is nothing but the frequency of spin flips in Glauber dynamics~\cite{glauber1963,lev2019}. The variance $\mathrm{Var}\,U_2$ in this case decreases in the critical region with system size. This may indicate a faster growth of critical slowing down in the local algorithm compared to the cluster algorithms.

The data that support the findings of this article are openly available at Ref.~\cite{data}.

\acknowledgments

IP and LS are supported by the Russian Science Foundation (grant 25-11-00158). YD acknowledges the support of the National Natural Science Foundation of China (NSFC) under Grant No. 12275263, as well as the Innovation Program for Quantum Science and Technology (under Grant No. 2021ZD0301900). YD is also supported by the Natural Science Foundation of Fujian Province of China (Grant No. 2023J02032).

\appendix

\section{Derivation of Eq.~(\ref{eq:Usw_variance})}
\label{app:SW-variance-derivation}

Taking the square of the configuration overlap of Eq.~(\ref{eq:overlap_general}) and
then taking the stationary expectations one gets:
\begin{equation}
\mathrm{Var}\!\big(U_n^{(\mathrm{SW})}\big)
= \frac{1}{N^2}\sum_{i,j}
\Big\{
\Pr\!\big[s_i^{(t)}{=}s_i^{(t+n)}\wedge s_j^{(t)}{=}s_j^{(t+n)}\big]
- p_1^2
\Big\},
\label{eq:var_raw}
\end{equation}
with $p_1 = \Pr[s_i^{(t)}=s_i^{(t+n)}]$ independent of $i$ bytranslation invariance. Then let's define the two-time FK connectivity event
\begin{equation}
A_{ij}^{(n)}
= \big\{\,i\leftrightarrow j\text{ in }\omega^{(t)}
\text{ and in }\omega^{(t+n)}\,\big\}.
\label{eq:Aij}
\end{equation}
The SW recoloring rule has two consequences. First is that on $A_{ij}^{(n)}$ the sites $i,j$ are recolored together at every step and start with the same color, so $s_i^{(t)}{=}s_j^{(t)}$ and $s_i^{(t+n)}{=}s_j^{(t+n)}$ deterministically, and the joint agreement event collapses to the single-site agreement of the pair of probability $p_1$. Second is that on the complement $(A_{ij}^{(n)})^c$ at least one independent uniform recoloring separates the trajectories of $i$ and $j$; conditional on this independence the agreement events at $i$ and $j$ factorize, giving $p_1^2$. Hence
\begin{align}
\Pr\!\big[s_i^{(t)}{=}s_i^{(t+n)}&\wedge s_j^{(t)}{=}s_j^{(t+n)}\big]
\nonumber\\
&= p_1\,\Pr[A_{ij}^{(n)}]
+ p_1^2\,\big(1-\Pr[A_{ij}^{(n)}]\big).
\label{eq:joint_agree}
\end{align}
Inserting Eq.~(\ref{eq:joint_agree}) into Eq.~(\ref{eq:var_raw}) and
subtracting $p_1^2$ yields the exact identity
\begin{equation}
\mathrm{Var}\!\big(U_n^{(\mathrm{SW})}\big)
= \frac{p_1(1-p_1)}{N^2}\sum_{i,j}
\Pr_{\mu_{\mathrm{ES}}}\!\big[A_{ij}^{(n)}\big],
\label{eq:var_exact}
\end{equation}
with the diagonal $i=j$ contributing $p_1(1-p_1)/N$. Above $T_c$ the
single-site agreement probability approaches the random value
$p_1\to 1/q$, so $p_1(1-p_1)\to (q-1)/q^2$ and
Eq.~(\ref{eq:var_exact}) reduces to
Eq.~(\ref{eq:Usw_variance}) up to the trivial diagonal correction.

\section{Snapshot illustrations of the algorithmic overlaps}
\label{app:snapshots}

This appendix collects lattice-level snapshots of the algorithmic
overlaps introduced in the main text, for the three thermodynamic
regimes $T>T_c$, $T<T_c$, and $T=T_c$. The figures are drawn on the two-dimensional Ising model at $L=24$ and are intended to make the geometric content of the SW configuration overlap and the Wolff cluster overlap directly visible.

\begin{figure*}[ht]
  \centering
  \includegraphics[width=0.92\textwidth]{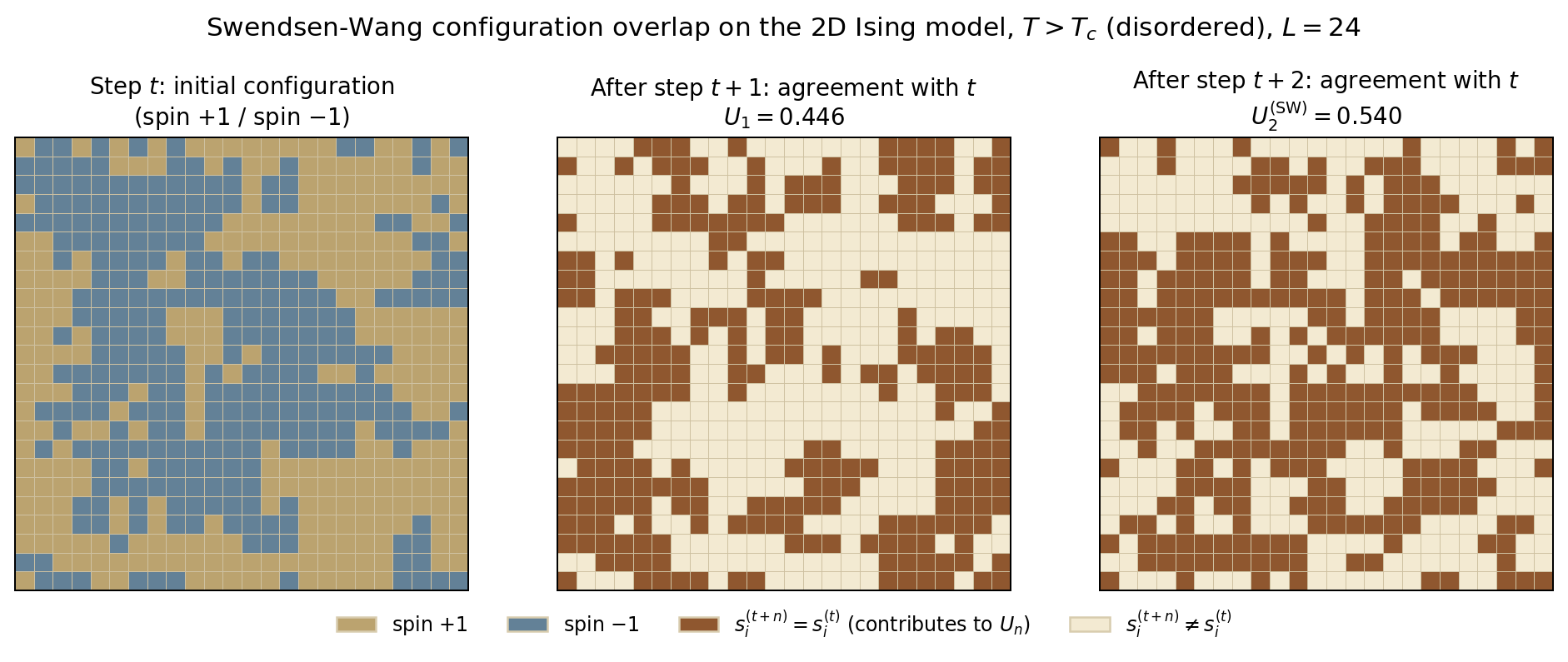}\\[2pt]
  \includegraphics[width=0.92\textwidth]{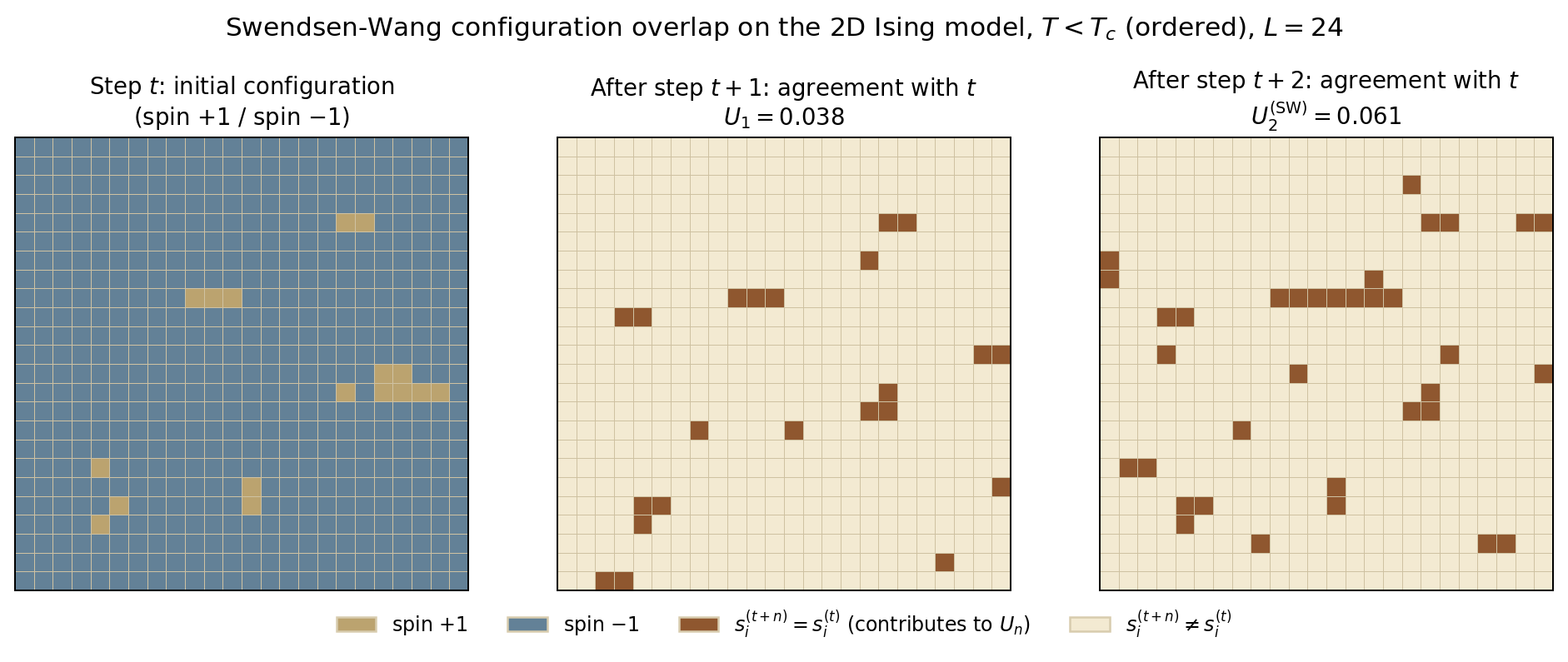}\\[2pt]
  \includegraphics[width=0.92\textwidth]{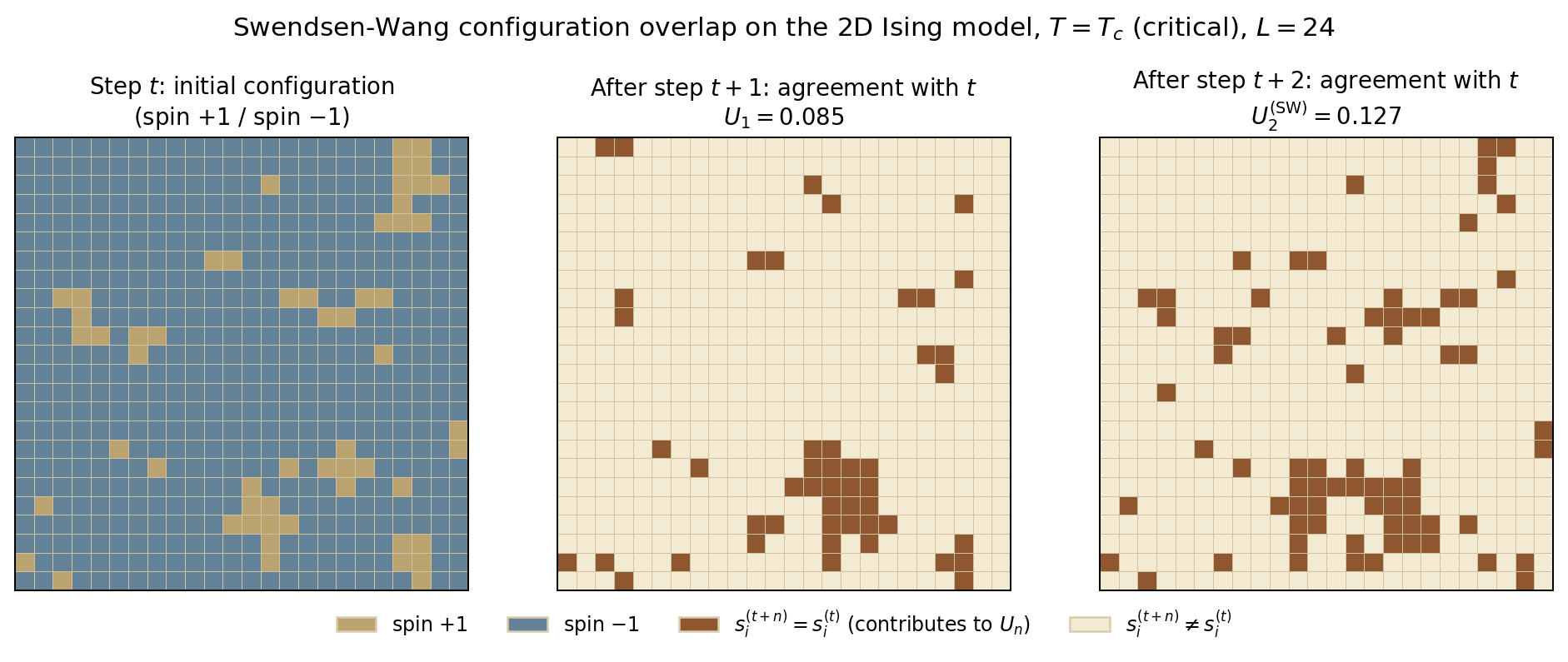}
  \caption{Swendsen--Wang configuration overlap on the 2D Ising
  model, $L=24$. Top: $T>T_c$ (disordered). Middle: $T<T_c$
  (ordered). Bottom: $T=T_c$ (critical). Left: spin configuration
  $\{s_i^{(t)}\}$. Center and right: after one and two SW updates,
  sites where $s_i^{(t+n)}=s_i^{(t)}$ (green) and
  $s_i^{(t+n)}\neq s_i^{(t)}$ (red). The values of $U_1$ and
  $U_2^{(\mathrm{SW})}$ shown above each panel are for the specific configuration displayed.}
  \label{fig:sw-fk-snapshots}
\end{figure*}

\paragraph*{Swendsen--Wang algorithm.}
Figure~\ref{fig:sw-fk-snapshots} shows three rows, one per regime. In each row the left plot is the spin configuration $\{s_i^{(t)}\}$ (orange = $+1$, blue = $-1$). The center and right plots show, after one and two SW updates respectively, the sites at which $s_i^{(t+n)}=s_i^{(t)}$ (green, contributing to $U_n$) and $s_i^{(t+n)}\neq s_i^{(t)}$ (red). Above $T_c$ the FK clusters are small and the independent recoloring decorrelates sites across the lattice, so $U_n\!\approx\!1/q$. Below $T_c$ a giant FK cluster carries a large fraction of the spins. With probability $1/q$ per SW step that cluster is recolored to the same value (yielding $U_1\!\approx\!1$, $U_2\!\approx\!0^{+}$) and with probability $1-1/q$ to a different value (yielding $U_1\!\approx\!0$, $U_2\!\approx\!1$), producing the one-step bistability and larger variance $\mathrm{Var}(U_2)$ explicitly shown in Figs.~\ref{fig:sw-ising-var-scale} and~\ref{fig:sw-potts-var-scale}. At $T_c$ the FK clusters span all scales and the green/red overlap pattern itself acquires a fractal spatial structure.

\begin{figure*}[ht]
  \centering
  \includegraphics[width=0.92\textwidth]{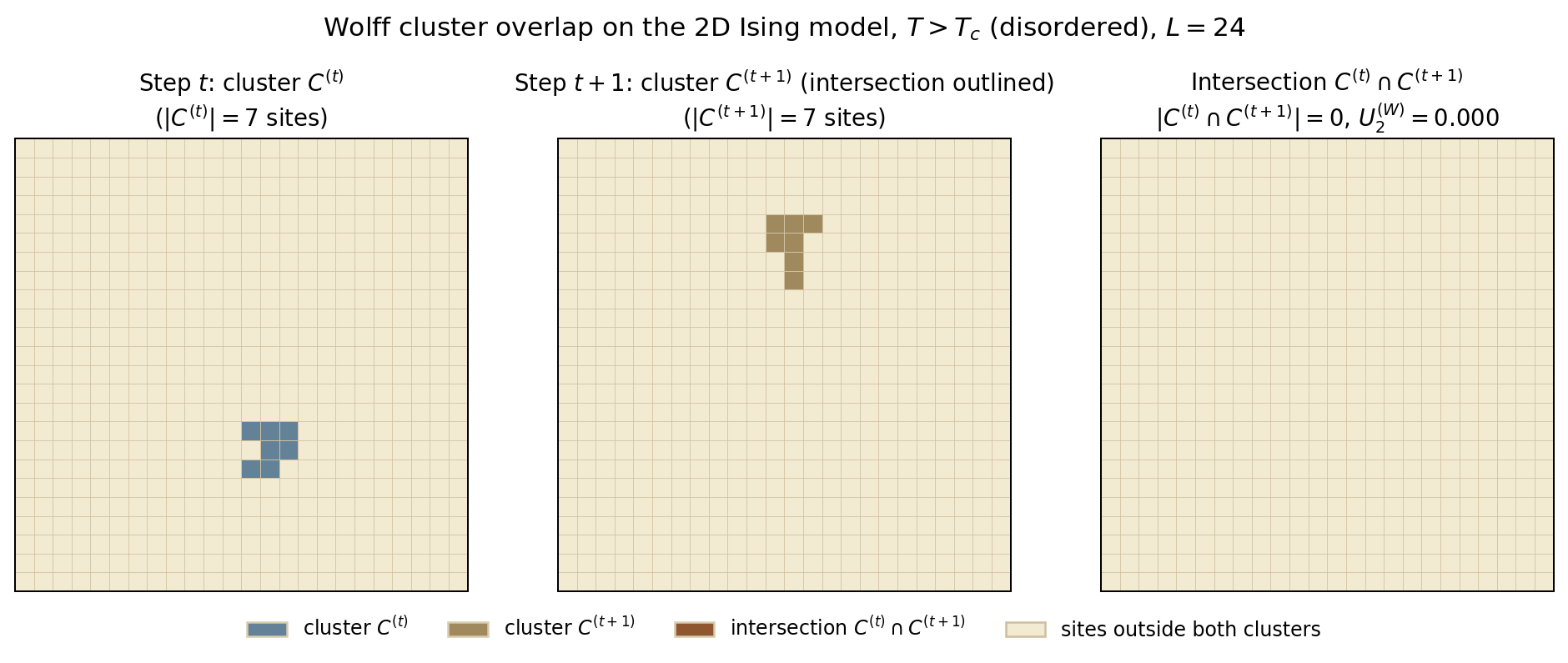}\\[2pt]
  \includegraphics[width=0.92\textwidth]{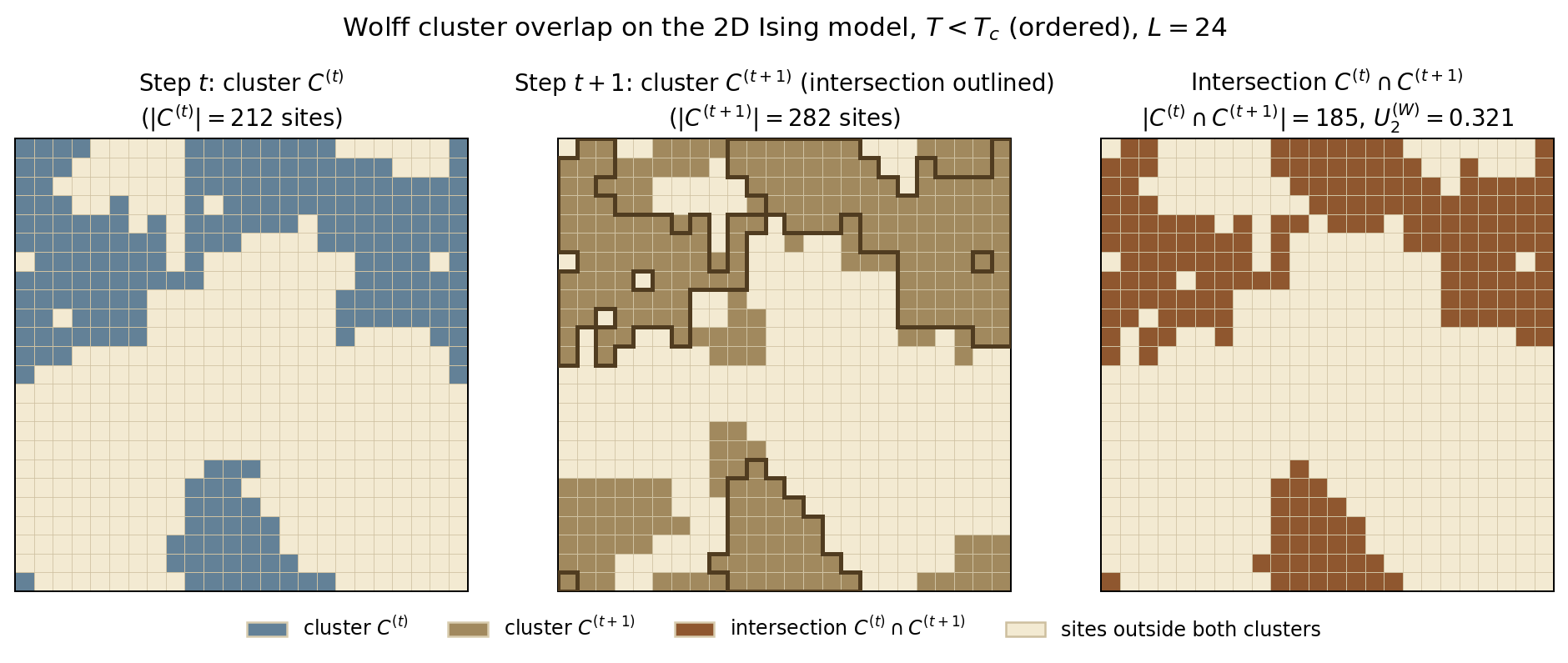}\\[2pt]
  \includegraphics[width=0.92\textwidth]{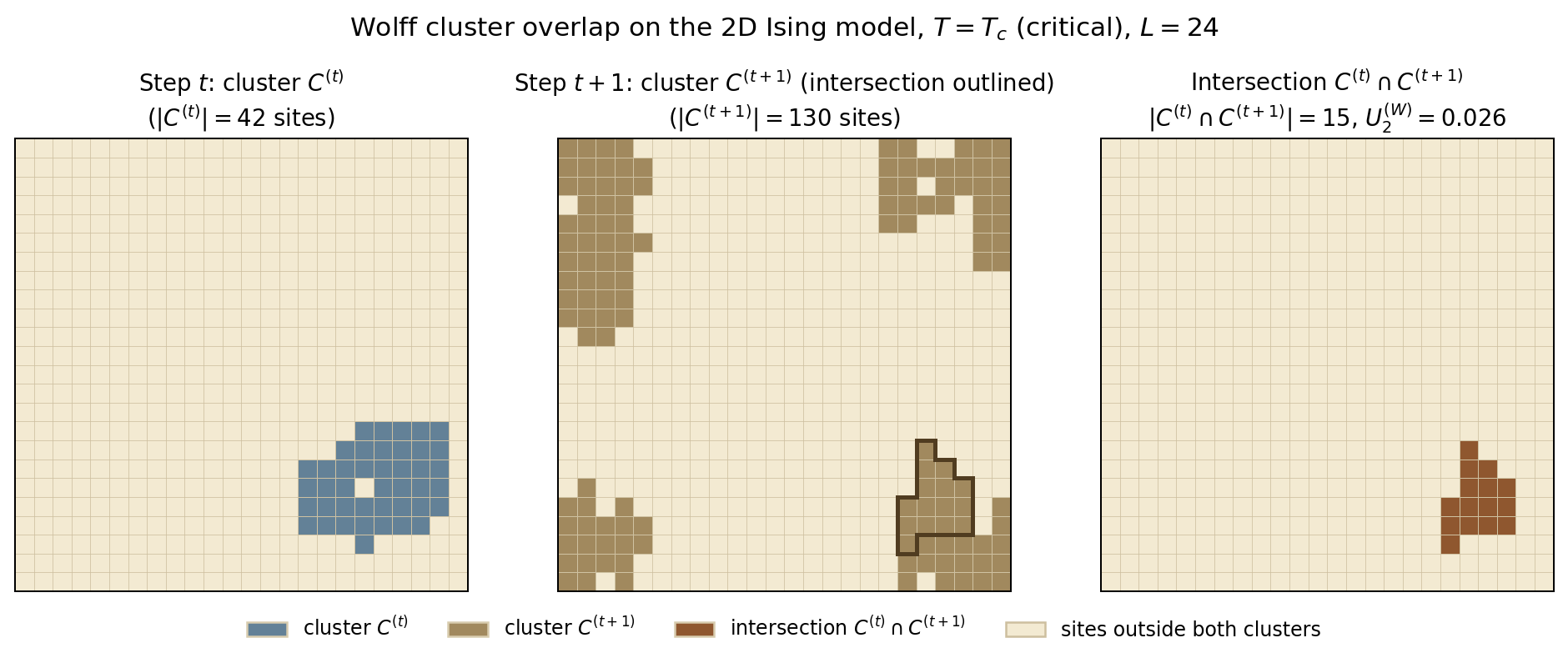}
  \caption{Wolff cluster overlap on the 2D Ising model, $L=24$. Top:
  $T>T_c$. Middle: $T<T_c$. Bottom: $T=T_c$. Left: $C^{(t)}$ (blue).
  Center: $C^{(t+1)}$ (red, intersection outlined). Right: the
  intersection $C^{(t)}\cap C^{(t+1)}$ (green). Cluster sizes and
  the value $U_2^{(W)}$ shown above each panel are for the specific
  pair of clusters displayed.}
  \label{fig:wolff-fk-snapshots}
\end{figure*}

\paragraph*{Wolff algorithm.}
Figure~\ref{fig:wolff-fk-snapshots} shows, for each of the three
regimes, the FK cluster $C^{(t)}$ at the seed site $i_0^{(t)}$
(blue), the next-step cluster $C^{(t+1)}$ at the new seed
$i_0^{(t+1)}$ (red, with the intersection $C^{(t)}\cap C^{(t+1)}$
outlined), and the intersection itself (green). Above $T_c$ both
clusters are finite and almost surely disjoint, so
$U_2^{(W)}\!\to\!0$ in the thermodynamic limit. Below $T_c$ each
Wolff cluster occupies a macroscopic fraction of the lattice and
$U_2^{(W)}$ is finite. At $T_c$ both clusters are scale-free
fractals of dimension $d_F=2-\beta/\nu$; their typical intersection is itself fractal, and its fractional measure $|C^{(t)}\cap C^{(t+1)}|/N$ vanishes algebraically with $L$. This is the geometric content of the singularity in
$\langle U_2^{(\mathrm{W})}\rangle$ presented in
Figs.~\ref{fig:wolff-ising-fss} and~\ref{fig:wolff-potts-fss}.

\clearpage
\bibliography{references}
\end{document}